\documentclass[]{pasj02} 
\usepackage[switch,mathlines]{lineno}
\jyear{2024}
\Received{2024/12/17}
\Accepted{2025/03/21}
\Published{}
\usepackage{amsmath}
\usepackage{url}
\usepackage{pdflscape}
\usepackage{longtable}
\usepackage{arydshln}

\begin{document} 
\title{ALMA 2D Super-resolution Imaging Survey of Ophiuchus Class I/Flat Spectrum/II Disks - I: Discovery of New Disk Substructures}
\author{
Ayumu \textsc{Shoshi},
\altaffilmark{1,2}
\altemailmark\orcid{0000-0001-6580-6038} 
\email{shoshi.ayumu.660@s.kyushu-u.ac.jp} 
Masayuki \textsc{Yamaguchi},
\altaffilmark{2}
\altemailmark\orcid{0000-0002-8185-9882} 
Takayuki \textsc{Muto}
\altaffilmark{3}
\altemailmark
Naomi \textsc{Hirano},
\altaffilmark{2}
\altemailmark\orcid{0000-0001-9304-7884}
Ryohei \textsc{Kawabe},
\altaffilmark{4, 5}
\altemailmark\orcid{0000-0002-8049-7525}
Takashi \textsc{Tsukagoshi},
\altaffilmark{6}
\altemailmark\orcid{0000-0002-6034-2892}
and
Masahiro \textsc{N. Machida}
\altaffilmark{7}
\altemailmark\orcid{0000-0002-0963-0872}
}
\altaffiltext{1}{Department of Earth and Planetary Sciences, Graduate School of Science, Kyushu University, 744 Motooka, Nishi-ku, Fukuoka 819-0395, Japan}
\altaffiltext{2}{Academia Sinica Institute of Astronomy and Astrophysics, 11F of ASMA Building, No.1, Sec. 4, Roosevelt Rd, Taipei 106216, Taiwan}
\altaffiltext{3}{Division of Liberal Arts, Kogakuin University, 1-24-2 Nishi-Shinjuku, Shinjuku-ku, Tokyo 163-8677, Japan}
\altaffiltext{4}{National Astronomical Observatory of Japan, 2-21-1 Osawa, Mitaka, Tokyo 181-8588, Japan}
\altaffiltext{5}{Department of Astronomy, School of Science, The Graduate University for Advanced Studies (SOKENDAI), Osawa, Mitaka, Tokyo 181-8588, Japan}
\altaffiltext{6}{Faculty of Engineering, Ashikaga University, Ohmae-cho 268-1, Ashikaga, Tochigi 326-8558, Japan}
\altaffiltext{7}{Department of Earth and Planetary Sciences, Faculty of Science, Kyushu University, 744 Motooka, Nishi-ku, Fukuoka 819-0395, Japan}

\maketitle

\begin{abstract}
This study focuses on Class I, Flat Spectrum (FS), and Class II disks in the Ophiuchus molecular cloud, a nearby active star-forming region with numerous young stellar objects (YSOs), to unveil signs of substructure formation in these disks. 
We employ two-dimensional super-resolution imaging based on Sparse Modeling (SpM) for ALMA archival Band 6 continuum data, achieving images with spatial resolutions comparable to a few au ($0\farcs02$–$0\farcs2$) for 78 dust disks, all of which are spatially resolved. 
In our sample, we confirm that approximately 30–40\% of the disks exhibit substructures, and we identify new substructures in 15 disks (4 Class I, 7 Class FS, and 4 Class II objects).
Compared to the eDisk sample in terms of bolometric temperature, $T_{\rm bol}$, our targets are in a relatively later accretion phase.
By combining our targets with the eDisk sample, we confirm that substructure detection in available data is restricted to objects where $T_{\rm bol}$ exceeds 200–300\,K and the dust disk radius, $R_{\rm dust}$, is larger than $\sim$30\,au.
Moreover, we find that the distribution of inclination angles for Class II disks has a deficit of high values and is not consistent with being random.
Analyzing molecular line emission data around these objects will be crucial to constrain disk evolutionary stages further and understand when and how substructures form.
\end{abstract}

\section{Introduction}\label{sec:intro}
Planet formation is considered to begin in the protoplanetary disks around young stellar objects (YSOs) (e.g., \cite{Hayashi_1981,Hayashi_1985,Shu_1987}), where variations in disk structures and chemical properties are expected to influence the architecture of planetary systems (e.g., \cite{Walter_1988,Skrutskie_1990,Haisch_2001}). 
Over the past decade, the Atacama Large Millimeter/submillimeter Array (ALMA) has provided numerous observations of protoplanetary disks in star-forming regions near the solar system (e.g., \cite{Ansdell_2016,Barenfeld_2017,Cieza_2019,Long_2019,Cazzoletti_2019,Villenave_2021}).
Many observations with higher spatial resolution than $0\farcs1$ have revealed various substructures in protoplanetary disks around Class II (T Tauri) stars, including rings, gaps, and spirals, as well as asymmetries in brightness distribution (e.g., \cite{ALMA_2015,Perez_2016,Andrews_2018_DSHARP,Cazzoletti_2018,Tsukagoshi_2019,Hashimoto_2021,Orihara_2023}).

These substructures are believed to be formed through various mechanisms, including photoevaporation \citep{Hollenbach_1994, Hardy_2015}, gravitational instability \citep{Youdin_2011,Loren-Aguilar_2016,Takahashi_2016}, magnetorotational instability \citep{Flock_2015}, planet-disk interactions (e.g. \cite{Takeuchi_1996,Kley_2012,Baruteau_2014,Zhu_2012}), and chemical processes \citep{Zhang_2015,Okuzumi_2016}, or combinations of these.
Planet-disk interaction is of particular interest to understanding the planet formation process.
Indeed, observations by several telescopes (ALMA, VLT/SPHERE, Subaru, HST, LBTI/ALES) have identified not only ring-shaped protoplanetary disks but also circumplanetary disks within protoplanetary disks, such as those around PDS~70, AB~Aur, HD~169142, and MWC~758 \citep{Keppler_2018,Benisty_2021,Currie_2022,Wagner_2023}.
It is also known that rings and spirals cause localized pressure bumps, which facilitate the capture of relatively large dust particles and promote dust growth and planetesimal formation 
\citep{Youdin_2005,Johansen_2007,Johansen_2009}. 
Substructures in protoplanetary disks are, therefore, important indicators for studying the different stages of planetary formation, as they are by-products of the planet formation process \citep{Dong_2015,Zhang_2018}.

However, it remains uncertain what primary physical processes are responsible for substructure formation and when the substructures emerge.
Previous observational and theoretical studies suggest that the amount of dust in protoplanetary disks (Class II disks) after the accretion phase is insufficient for planet formation.
Thus, planet formation may begin in Class 0/I disks, which contain larger amounts of gas and dust than Class II disks (e.g., \cite{Manara_2018, Tychoniec_2020,Tsukamoto_2017}).

The large ALMA project, eDisk, has conducted detailed observations of 12 Class 0 and 7 Class I disks with a spatial resolution of higher than $0\farcs1$ to search for substructures associated with planet formation during the early accretion phase (e.g., \cite{Ohashi_2023}). 
The project identified relatively few substructures in Class 0/I disks compared to Class II disks. 
\citet{Ohashi_2023} concluded that substructures form rapidly from the Class I to Class II stages. 
Other observations have reported substructures in some Class I disks \citep{Sheehan_2018, Flores_2023, Yamato_2023, Shoshi_2024, Hsieh_2024}; however, the sample size is too small to assess their universality.
Therefore, a larger sample size consisting of disks imaged with $<0\farcs1$ resolution is needed to identify substructures and provide a clearer understanding of their universality, as well as valuable insights into the origin and formation processes of disk substructures.

Recently, super-resolution imaging with Sparse Modeling (SpM) has been proposed as another way to reconstruct interferometric images.
This technique can produce high-fidelity images with spatial resolution improved by a factor of $\sim$2-3, which are comparable to a few au ($0\farcs02$–$0\farcs1$) when applied to protoplanetary disk data observed by ALMA \citep{Yamaguchi_2020,Yamaguchi_2021}. 
Using this technique, \citet{Yamaguchi_2024} analyzed ALMA archival data for 43 objects in the Taurus-Auriga region (at a distance of ~140\,pc). 
In their study, clear gap structures were found in half of their targets. 
The study also discussed planet-disk interactions and examined relationships between the widths and depths of gaps measured by SpM images.

In this study, we apply SpM to ALMA archival Band 6 continuum data of YSOs in the Ophiuchus molecular cloud (at a distance of $\sim140$\,pc; \cite{Ortiz-León_2018,Gaia_2023} and a characteristic age of $\sim 1$\,Myr\footnote[1]{We note that the Ophiuchus molecular cloud has age dispersion, with different ages in different regions \citep{Esplin_2020}.}; \cite{Williams_2019}). 
We obtain higher spatial resolution images ($0\farcs02$-$0\farcs2$), improved by a median factor of 3.8 for 78 disks (15 Class I, 24 Class Flat Spectrum, and 39 Class II) compared to the results after the CLEAN imaging.
We identify characteristic substructures in the high-resolution images obtained by SpM. 
We also present statistical analyses of stellar and disk properties and discuss the relationship between disk evolutionary stages and substructure formation.

This paper is organized as follows. In Section~\ref{sec:obsdata_imaging}, we describe the data reduction and our imaging methods (CLEAN and SpM). 
Section~\ref{sec:spm_image} presents continuum images of 78 disks and mentions their quality. 
In Section~\ref{sec:disk_characterustics}, we discuss methods for measuring basic disk properties and define the categorization of disk substructures. 
Section~\ref{sec:discussion} addresses the differences between the eDisk project and this study, as well as the potential misclassification of disk evolutionary stages. 
Finally, Section~\ref{sec:summary} provides the conclusions.

\section{Sample Selection and Imaging}\label{sec:obsdata_imaging}
\subsection{Observation Data: Targets and Reduction}\label{subsec:obs_details}
We focus on YSOs at different evolutionary stages in the Ophiuchus molecular cloud, as identified in the "Cores to Disks" (c2d) Spitzer Legacy Programme \citep{Evans_2009}. 
YSOs in the c2d program are M-K-type stars with masses below 1\,$M_\odot$, excluding early-type stars like Herbig Ae/Be stars. 
The categorization of evolutionary stages in YSOs follows the spectral slopes, formulated as 
\begin{align}
    \alpha_{\rm IR}=\frac{\log\left(\lambda_1 F_{\lambda_1}\right)-\log\left(\lambda_0 F_{\lambda_0}\right)}{\log\left(\lambda_1\right)-\log\left(\lambda_0\right)},\label{eq:sed}
\end{align}
where $F_\lambda$ is the flux density at $\lambda_0$=2\,$\mu$m and $\lambda_1$=20\,$\mu$m \citep{Greene_1994,Chen_1995}.
Following \citet{Evans_2009} and \citet{Cieza_2019}, we classify the objects into three categories: Class I sources embedded in their envelope ($\alpha_{\rm IR}>0.3$), Class Flat Spectrum (FS) sources with detectable emissions from a reduced envelope ($0.3>\alpha_{\rm IR}>-0.3$), and Class II sources with infrared excess mainly from optically thick disks ($-0.3>\alpha_{\rm IR}>-1.6$).

For this study, we use the same ALMA archival data (Project 2016.1.00545.S, PI Lucas A. Cieza) as the Ophiuchus Disk Survey Employing ALMA (ODISEA; \cite{Cieza_2019}), which includes YSOs selected from the c2d survey based on spectral slopes. 
The samples in this data include Class I and FS sources with $[K]-[24] > 6.75$\,mag and Class II sources brighter than 10\,mag in $K$ band.
The observations were conducted on July 13 and 14, 2017, using 42–45 ALMA antennas in C40-5 configurations with baseline lengths ranging from 17\,m to 2647\,m. 
The data include five spectral windows (SPWs): two SPWs for continuum observations, with central frequencies of 218 and 233\,GHz in Band 6, and three SPWs for line emissions of $^{12}$CO, $^{13}$CO, and C$^{18}$O ($J$=2-1). 
In this study, we use only the continuum SPWs and reconstruct images using two imaging methods: CLEAN and SpM.

The raw data were calibrated using the ALMA pipeline in the Common Astronomy Software Application package (CASA; \cite{CASA_2022}) version 4.7.2. 
The quasars J1517-2422 and J1733-1304 were used as flux calibrators, while J1517-2422 and J125-2527 served as bandpass and phase calibrators, respectively.

\subsection{Imaging with CLEAN and Source Selection}\label{subsec:method_CLEAN}
After the calibration with the ALMA pipeline, the data were imaged using the CASA task \texttt{tclean} in version 6.1.0 of CASA.
We consistently employed multi-frequency synthesis (\texttt{nterm}$=2$; \cite{Rau_2011}) and the Cotton-Schwab algorithm \citep{Schwab_1984} with Briggs weighting with \texttt{robust}=0.5.
We detected a total of 125 systems with emissions above 5$\sigma$ ($\sigma=$0.16\,mJy\,beam$^{-1}$) in the initial CLEAN images. 
Our images have detected five additional sources (2MASS~J16222099-2304025, ISO-Oph~3, ISO-Oph~85, WSB~49, and 2MASS~J16453548-2414226) compared to the 120 systems reported in \citet{Cieza_2019} due to the higher sensitivity achieved by setting Briggs weighting to \texttt{robust}$=0.5$, compared to their images generated with \texttt{robust}$=0.0$.

The self-calibration process was applied to datasets to improve the SNR of the reconstructed images by correcting gain errors, including antenna- and baseline-based errors. 
In this study, the integration time \texttt{solint} and gain calibration settings were determined based on the SNR of the initial CLEAN image. 
For images with an SNR of at least 100, we performed two rounds of phase self-calibration (\texttt{calmode=p}) and one round of amplitude and phase self-calibration (\texttt{calmode=ap}), using integration times of OST, OST/5, and OST in sequence, where OST indicates on-source time. 
For images with 30\,$\leq$\,SNR\,$<$\,100, we applied one round each of phase and amplitude and phase self-calibration with integration times of OST and OST/5, respectively. 
The final CLEAN image with the highest SNR was selected as the final output from the self-calibration process.

In 77 sources with total flux $\gtrsim$ 5.0\,mJy (61\% of the sample), their final CLEAN images show a signal-to-noise ratio (SNR) improvement by a factor of 1.1–2.0 after the self-calibration. 
Faint sources with total flux $<$ 5.0\,mJy, except for 2MASS~J16262404-2424480 (hereafter J16262404), show no improvement with self-calibration, so we used non-self-calibrated data for them.
The field of view ($19\farcs0$ in radius) centered on J16262404 (total flux of 2.6\,mJy) also includes the disk around ISO-Oph~37 (total flux of 23.5\,mJy), which has a relatively strong intensity that enabled phase adjustment between antennas.
This allowed us to correct phase and amplitude differences more efficiently and achieve an SNR improvement, unlike other sources with total flux $<$ 5\,mJy. 

We carefully selected datasets with relatively high SNRs >20 on their CLEAN images, as SpM imaging may produce low-fidelity images for lower-SNR datasets \citep{Yamaguchi_2024}.
After the SpM imaging, we collected 78 systems, including 15 Class I, 24 Class FS, and 39 Class II sources (for details, see \S\ref{sec:spm_image}). 
Our sample represents 26\% of all sources in the c2d survey (30\% of the disks up to Class II), comprising 65 single, 10 binary, and 3 triple star systems. 
Table~\ref{table:stellar_properties} summarizes their stellar properties, including SED classification, distance, spectral type, bolometric luminosity, bolometric temperature, and stellar mass. Distances are taken from Gaia Data Release 2 and 3 (DR2: \cite{Gaia_2018, Williams_2019}; DR3: \cite{Gaia_2023}). 
We referred to \citet{Dunham_2015} for $L_{\rm bol}$ and $T_{\rm bol}$ of all the systems.
We note that most Class I and FS sources, embedded in optically thick envelopes, lack distance references.
Therefore, in our sample, we adopted the average distance of 140.0\,pc, derived from the systems detected by  Gaia DR2 and DR3, for the undetected objects.
In addition, Table~\ref{table:image_info} summarizes the final CLEAN image parameters for detected sources, including CLEAN beam size $\theta_{\rm CLEAN}$, peak intensity, RMS noise level $\sigma_{\rm CLEAN}$ (collected RMS from emission-free areas), and total flux $F_\nu$ (measured from dust emission larger than 5$\sigma_{\rm CLEAN}$).

\subsection{Imaging with Sparse Modeling}\label{subsec:method_SpM}
We conducted SpM imaging with the self-calibrated dataset to reconstruct super-resolution images exceeding the quality of CLEAN images, following the methodology of \citet{Yamaguchi_2024}.
Using the SpM imaging software \texttt{PRIISM} version 0.11.5 \citep{Nakazato_2020} on CASA version 6.1.0, we applied $\ell_1$+TSV imaging with a cross-validation (CV) scheme. \footnote[2]{\texttt{PRIISM} (Python Module for Radio Interferometry Imaging with Sparse Modeling) is the public software for imaging ALMA observations based on the SpM technique, and it is available at $\langle$\url{https://github.com/tnakazato/priism}$\rangle$.}
This imaging technique generates an image by minimizing a cost function composed of a chi-squared error term and two additional convex regularizations, formulated as
\begin{align}
\mathbf{I}_{\rm SpM} &= \underset{\mathbf{I}}{\rm argmin} \left( \|\mathbf{W}(\mathbf{V} - \mathbf{F}\mathbf{I})\|^2_2 + \Lambda_l\sum_i\sum_j |{\rm I}_{i, j}|\right.\notag\\
&\left.+\Lambda_{tsv}\sum_i\sum_j \left(|{\rm I}_{i+1,j} - {\rm I}_{i,j}|^2 + |{\rm I}_{i,j+1} - {\rm I}_{i,j}|^2 \right) \right),\label{eq:spm_algorithms}
\end{align}
where $\mathbf{I}=\{{\rm I}_{i, j}\}$ represents the two-dimensional image being generated, with ${\rm I}_{i, j}$ as the pixel intensity at indices $i$ and $j$. 
Here, $\mathbf{V}$ is the observed visibility (self-calibrated in this study), and $\mathbf{F}$ is the Fourier matrix.
$\mathbf{W}$ is a diagonal weight matrix, with each diagonal element representing the inverse of the squared observational error for each visibility point. 
These elements are normalized by the residual visibility $(\mathbf{V}-\mathbf{F}\mathbf{I})$ in the chi-squared term.
The cost function includes the chi-squared error between the observed visibility and the visibility model, derived from the model image through Fourier transformation, along with two regularization terms: the $\ell_1$-norm and total squared variation (TSV).

The $\ell_1$-norm, the second term in the cost function, adjusts the sparsity of the brightness distribution in the image. 
This term calculates the sum of model image components ${\rm I}_{i, j}$, allowing us to maintain the total flux density in the brightness distribution while controlling low-intensity noise in the emission-free regions (e.g., \cite{Honma_2014}). 
The hyper-parameter $\Lambda_l$ controls the relative weighting of this term and constrains the extent of the image sparsity. 
A larger $\Lambda_l$ results in a more sparse distribution with reduced background noise levels, though the total flux also decreases.

The TSV regularization, the third term in the cost function, controls the smoothness of the brightness distribution. 
It calculates the sum of squared differences between an image component ${\rm I}_{i, j}$ and its neighboring components ${\rm I}_{i+1, j}$ or ${\rm I}_{i, j+1}$ in the vertical and horizontal directions. 
This term helps to reduce artificial, abrupt brightness changes, resulting in high-quality images with a smoother distribution, independent of imaging parameters such as field of view or pixel size \citep{Akiyama_2017a,Kuramochi_2018}. 
The hyper-parameter $\Lambda_{tsv}$ is essential for adjusting the weighting of this term and influences the effective spatial resolution.
As $\Lambda_{tsv}$ increases, the image becomes smoother, but the effective spatial resolution decreases.

We prepared a wide range of combinations for the two hyper-parameters $(\Lambda_l, \Lambda_{tsv})$ and selected the most suitable combination for each image by minimizing the cost function using the 10-fold cross-validation (CV) approach \citep{Yamaguchi_2021}. 
However, images with the optimal hyper-parameters for 11 sources (2MASS~J16313679-2404200, 2MASS~J16271643-2431145, 2MASS~J16230544-2302566, ISO-Oph~52, ISO-Oph~75, ISO-Oph~95, ISO-Oph~105, WSB~63, WSB~19, WSB~12, WSB~67) displayed artificial patchy structures. 
In these cases, we conservatively chose an image generated with the $\Lambda_{tsv}$ value that is one order of magnitude larger than that selected by the CV approach.
The consistency of total fluxes between SpM and CLEAN images was evaluated using the curve-growth method (see \S A.\ref{subsec:curve_growth}). 
We confirmed that the total fluxes of the SpM images were consistent with those of the CLEAN images within 5-10\% error. 
In the CLEAN image, ISO-Oph~147 appears as a single dust disk; however, in the SpM image, it is resolved into two separate disks (hereafter referred to as ISO-Oph~147\,a and b; see \S\ref{subsec:new_substructures} for details). 
The combined flux densities of ISO-Oph~147\,a and b are comparable to the flux observed in the CLEAN image. 
The final values for the regularization hyper-parameters $\left(\Lambda_l, \Lambda_{tsv}\right)$, RMS noise in CLEAN images $\sigma_{\rm RMS}$, peak intensities $I_{\rm peak}$, and total fluxes $F_\nu$ in both SpM and CLEAN images are summarized in Table~\ref{table:image_info}.

The effective spatial resolution of the SpM images is evaluated using the ‘point-source injection’ method, as described in \citet{Yamaguchi_2021}. 
Unlike the CLEAN algorithm, the SpM algorithm does not involve beam convolution to produce an image. 
To test resolution, an artificial point source was first injected into the observed visibility data. 
The point source was positioned in an emission-free region north of the central star, at a distance within the maximum recoverable scale ($\sim$2.1\,arcsec).
The total flux of the point source was set to 5–25\% of the target source’s total flux to ensure its detectability above the continuum sensitivity.
SpM imaging was then carried out using the same hyper-parameters as those applied for the optimal image.
The point source was reconstructed as an elliptical Gaussian feature, and its Full Width at Half Maximum (FWHM) was used to determine the effective spatial resolution, $\theta_{\rm eff}$, of the SpM image.
For reference, we confirmed that the measured spatial resolution varies by only a few percent when the point source is injected to the east, west, north, or south of the central star.
The flux density ratios between targets and point sources (column 7), along with $\theta_{\rm eff}$ of the SpM images, are also summarized in Table~\ref{table:image_info}.

\onecolumn
\begin{longtable}[ht]{llccccccc}
\caption{Host Stellar Properties}
\label{table:stellar_properties}\\
\hline
Source Name & 2MASS & Class & $d$ & SpT & $L_{\rm bol}$ & $T_{\rm bol}$ & $M_\ast$ & Ref. \\
 &  &  & pc &  & $L_\odot$ & K & $M_\odot$ & \\
(1) & (2) & (3) & (4) & (5) & (6) & (7) & (8) & (9) \\
\hline
\hline
\endfirsthead
\multicolumn{9}{c}{(Continued)}\\
\hline
Source Name & 2MASS & Class & $d$ & SpT & $L_{\rm bol}$ & $T_{\rm bol}$ & $M_\ast$ & Ref. \\
 &  &  & pc &  & $L_\odot$ & K & $M_\odot$ & \\
(1) & (2) & (3) & (4) & (5) & (6) & (7) & (8) & (9) \\
\hline
\hline
\endhead
\endfoot
\multicolumn{8}{l}{\hbox to 0pt{\parbox{153mm}{\footnotemark[]\textbf{Notes.}
Column Description:
(1) Source name. 
(2) Name of the host star in Two MicronAll Sky Survey (2MASS; \cite{Skrutskie_2006}).
(3) Classification for the host star mainly referred to \citet{Cieza_2019} and \citet{Williams_2019}.
(4) Distance of the host star adopted mainly from Gaia DR3 and Gaia DR2 parallax \citep{Gaia_2023, Gaia_2018}.
(5) Spectral type of the host star.
(6) Bolometric luminosity of the host star \citep{Dunham_2015} corrected at 140\,pc.
(7) Bolometric temperature of the host star \citep{Dunham_2015}.
(8) Stellar mass.}}}\\
\multicolumn{8}{l}{\hbox to 0pt{\parbox{153mm}{\footnotemark[]\textbf{References.}
a: \citet{Williams_2019},
b: \citet{Cieza_2019},
c: \citet{Cieza_2021},
d: \citet{Testi_2022},
e: \citet{vanderMarel_2021},
f: \citet{vanderMarel_2016},
g: \citet{Gaia_2023},
h: \citet{Manara_2015},
i: \citet{Dunham_2015},
j: \citet{Villenave_2022},
k: \citet{Michel_2021}.
}}}
\endlastfoot
ISO-Oph~54 & J16264046-2427144 & I & 140.0 & M4 & 0.08 & 380 & $\cdots$ & a, c, e, i \\
2MASS~J16214513-2342316 & J16214513-2342316 & I & 140.0 & $\cdots$ & 0.04 & 240 & $\cdots$ & a, i \\
WLY 2-63 & J16313565-2401294 & I & 140.0 & M6 & 2.01 & 270 & $\cdots$ & a, c, e, i \\
ISO-Oph~127 & J16271838-2439146  & I & 140.0 & $\cdots$ & 0.05 & 630 & 0.59 & a, h, i \\
ISO-Oph~99 & J16270524-2436297 & I & 140.0 & $\cdots$ & 0.20 & 97 & $\cdots$ & a, i \\
ISO-Oph~165 & J16273894-2440206  & I & 140.0 & M2.5 & 0.08 & 320 & 0.40 & a, d, i \\
ISO-Oph~21 & J16261722-2423453 & I & 140.0 & $\cdots$ & 0.11 & 490 & $\cdots$ & a, i \\
2MASS~J16313679-2404200 & J16313679-2404200 & I & 140.0 & $\cdots$ & 0.21 & 74 & $\cdots$ & a, i \\
2MASS~J16262548-2423015 & J16262548-2423015 & I & 140.0 & $\cdots$ & 0.13 & 140 & $\cdots$ & a, i \\
ISO-Oph~170 & J16274161-2446447 & I & 140.0 & $\cdots$ & 0.03 & 360 & $\cdots$ & a, h, i \\
2MASS~J16271643-2431145 & J16271643-2431145 & I & 140.0 & $\cdots$ & 0.03 & 620 & $\cdots$ & a, i \\
2MASS~J16230544-2302566 & J16230544-2302566 & I & 140.0 & M3 & 0.01 & 790 & $\cdots$ & a, e, i \\
WL 17 & J16270677-2438149 & I & 140.0 & M3 & 0.60 & 330 & 1.45 & a, b, h, i \\
ISO-Oph~137 & J16272461-2441034 & I & 140.0 & $\cdots$ & 0.39 & 170 & $\cdots$ & a, i \\
ISO-Oph~200 & J16314375-2455245 & I & 150.2 & $\cdots$ & 0.33 & 520 & $\cdots$ & a, g, i \\
\hline
2MASS~J16313124-2426281 & J16313124-2426281 & F & 147.0 & K4 & 0.02 & 750 & 1.20 & a, j, i \\
2MASS~J16254662-2423361 & J16254662-2423361 & F & 140.0 & $\cdots$ & 0.00 & 690 & $\cdots$ & a, i \\
ISO-Oph~37 & J16262357-2424394  & F & 140.0 & K7 & 0.30 & 710 & 0.77 & a, d, i \\
ISO-Oph~94 & J16270359-2420054 & F & 140.0 & M1 & 0.01 & 770 & 0.04 & a, e, h, i \\
2MASS~J16395292-2419314 & J16395292-2419314 & F & 140.0 & $\cdots$ & 0.00 & 980 & $\cdots$ & a, i \\
ISO-Oph~70 & J16264848-2428389 & F & 140.0 & M0 & 0.15 & 440 & $\cdots$ & a, e, h, i \\
ISO-Oph~112 & J16271117-2440466 & F & 140.0 & M4 & 0.26 & 610 & 1.20 & a, e, h, i \\
ISO-Oph~93 & J16270300-2426146 & F & 140.0 & M3 & 0.05 & 380 & 0.31 & a, e, h, i \\
ISO-Oph~51 & J16263682-2415518  & F & 136.6 & M0 & 0.30 & 680 & 0.59 & a, d, g, i \\
ISO-Oph~26 & J16261898-2424142  & F & 140.0 & M6 & 0.04 & 710 & 0.14 & a, d, i \\
ISO-Oph~167 & J16273982-2443150 & F & 140.0 & K6 & 0.90 & 570 & 3.80 & a, e, h, i \\
ISO-Oph~52 & J16263778-2423007 & F & 140.0 & M0 & $\cdots$ & $\cdots$ & 0.17 & a, e, h \\
ISO-Oph~46 & J16263046-2422571 & F & 140.0 & K6 & $\cdots$ & $\cdots$ & 0.66 & a, e, h \\
ISO-Oph~75 & J16265197-2430394 & F & 140.0 & M5 & 0.01 & 680 & 0.13 & a, e, h, i \\
ISO-Oph~129 & J16271921-2428438 & F & 140.0 & K7 & 0.11 & 660 & 0.91 & a, e, h, i \\
ISO-Oph~95 & J16270410-2428299 & F & 140.0 & M4 & 0.10 & 640 & 0.95 & a, e, h, i \\
ISO-Oph~204 & J16315211-2456156 & F & 153.0 & $\cdots$ & 1.51 & 730 & $\cdots$ & a, g, i \\
ISO-Oph~59 & J16264214-2431029 & F & 140.0 & $\cdots$ & 0.04 & 790 & 0.39 & a, h, i \\
ISO-Oph~107 & J16270935-2440224 & F & 140.0 & M4 & 0.05 & 750 & 0.27 & a, e, h, i \\
ISO-Oph~132 & J16272146-2441430 & F & 140.0 & K7 & 1.38 & 610 & 1.74 & a, e, h, i \\
ISO-Oph~212 & J16322105-2430358 & F & 153.6 & $\cdots$ & 0.50 & 820 & $\cdots$ & a, g, i \\
BBRCG 58 & J16273213-2429435  & F & 140.0 & M6 & 0.01 & 850 & 0.15 & a, d, i \\
ISO-Oph~147 & J16273018-2427433  & F & 140.0 & K8 & 1.22 & 500 & 0.60 & a, d, i \\
ISO-Oph~171 & J16274175-2443360 & F & 140.0 & M5 & 0.03 & 720 & 0.17 & a, e, h, i \\
\hline
Elias~27 & J16264502-2423077  & II & 110.1 & K8 & 0.36 & 820 & 0.63 & a, d, g, i \\
DoAr~25 & J16262367-2443138  & II & 138.2 & K6 & 0.60 & 1500 & 0.80 & a, d, g, i \\
Elias~24 & J16262407-2416134  & II & 139.3 & K5.5 & 2.38 & 980 & 1.10 & a, d, g, i \\
WSB~82 & J16394544-2402039 & II & 145.8 & K4 & 0.95 & 1100 & 1.26 & a, g, i, k \\
ISO-Oph~2 & J16253812-2422362 & II & 134.3 & M0 & 0.14 & 1100 & 0.50 & a, d, e, g, i \\
ISO-Oph~196 & J16281650-2436579  & II & 135.0 & M4.5  & 0.18 & 1200 & 0.22 & a, d, g, i \\
DoAr~44 & J16313346-2427372 & II & 146.3 & K3 & 0.94 & 1200 & 1.40 & a, b, e, i  \\
ISO-Oph~17 & J16261033-2420548  & II & 140.0 & K8 & 1.25 & 290 & 0.69 & a, d, i \\
SR~24S & J16265843-2445318  & II & 115.0 & M0 & 1.63 & 840 & 0.86 & a, c, d, i \\
RXJ1633.9-2442 & J16335560-2442049 & II & 143.8 & K7 & 0.21 & 1500 & 0.80 & a, e, g, i \\
Elias~20 & J16261886-2428196  & II & 137.5 & M0 & 0.63 & 990 & 0.88 & a, d, g, i \\
SR~20W & J16282333-2422405  & II & 146.8 & K5 & 0.29 & 1200 & 0.97 & a, d, g, i \\
IRAS16201-2410 & J16230923-2417047 & II & 156.6 & M0 & 0.51 & 1200 & 1.12 & a, e, g, k, i \\
SR~13 & J16284527-2428190  & II & 115.5 & M2 & 0.72 & 1300 & 0.37 & a, d, g, i \\
SR~4 & J16255615-2420481  & II & 134.8 & K6 & 1.12 & 1100 & 0.80 & a, d, g, i \\
DoAr~43 & J16313087-2424399 & II & 135.9 & K2 & 0.74 & 990 & $\cdots$ & a, e, g, i \\
ISO-Oph~105 & J16270910-2434081  & II & 134.6 & K7 & 0.28 & 1000 & 0.67 & a, d, g, i \\
WSB~52 & J16273942-2439155  & II & 135.3 & M0 & 0.46 & 1000 & 0.50 & a, d, g, i \\
DoAr~33 & J16273901-2358187  & II & 141.6 & K5 & 0.38 & 1600 & 0.98 & a, d, g, i \\
WSB~63 & J16285407-2447442  & II & 136.5 & M1.5 & 0.21 & 1400 & 0.39 & a, d, g, i \\
ISO-Oph~117 & J16271382-2443316  & II & 140.7 & M3 & 0.12 & 830 & 0.29 & a, d, g, i \\
WSB~19 & J16250208-2459323  & II & 142.0 & M3 & 0.21 & 1200 & 0.29 & a, d, i \\
WSB~12 & J16221852-2321480 & II & 136.9 & K5 & 0.53 & 1400 & 1.30 & a, e, g, f, i \\
ISO-Oph~83 & J16265677-2413515  & II & 136.5 & K7 & 0.16 & 1300 & 0.72 & a, d, g, i \\
ISO-Oph~72 & J16264897-2438252  & II & 132.5 & M3 & 0.10 & 1000 & 0.29 & a, d, g, i \\
WSB~14 & J16222497-2329553 & II & 137.5 & $\cdots$ & 0.14 & 1500 & $\cdots$ & a, g, i \\
ISO-Oph~163 & J16273832-2436585  & II & 139.5 & K5.5 & 0.45 & 1100 & 0.98 & a, d, g, i \\
WSB~67 & J16302339-2454161 & II & 141.1 & M3 & 0.18 & 1300 & 0.50 & a, f, g, i \\
SR~22 & J16252434-2429442  & II & 131.4 & M4 & 0.16 & 1400 & 0.22 & a, d, g, i \\
ISO-Oph~39 & J16262404-2424480  & II & 139.5 & K5.5 & 1.51 & 900 & 1.12 & a, d, g, i \\
DoAr~32 & J16273832-2357324  & II & 141.1 & K5 & 0.50 & 1400 & 0.98 & a, d, g, i \\
ISO-Oph~155 & J16273311-2441152  & II & 136.6 & K5.5 & 0.59 & 1200 & 1.14 & a, d, g, i \\
ISO-Oph~128 & J16271848-2429059  & II & 140.0 & M1.5  & 0.16 & 930 & 0.47 & a, d, g, i \\
ISO-Oph~62 & J16264285-2420299  & II & 136.7 & K7 & 0.53 & 1300 & 0.87 & a, d, i \\
ISO-Oph~36 & J16262335-2420597  & II & 139.2 & K0 & 2.38 & 1100 & 2.60 & a, d, g, i \\
ISO-Oph~20 & J16261706-2420216  & II & 135.3 & K6 & 0.43 & 1600 & 0.83 & a, d, g, i \\
ISO-Oph~116 & J16271372-2418168  & II & 137.2 & M0 & 0.15 & 1300 & 0.53 & a, d, g, i \\
2MASS~J16314457-2402129 & J16314457-2402129 & II & 132.0 & $\cdots$ & 0.13 & 970 & $\cdots$ & a, g, i \\
ISO-Oph~106 & J16270907-2412007  & II & 143.7 & M2.5 & 0.08 & 1400 & 0.34 & a, d, g, i \\
\hline
\end{longtable}
\twocolumn

\begin{figure*}[t]
    \begin{center}
    \includegraphics[width=0.95\linewidth]{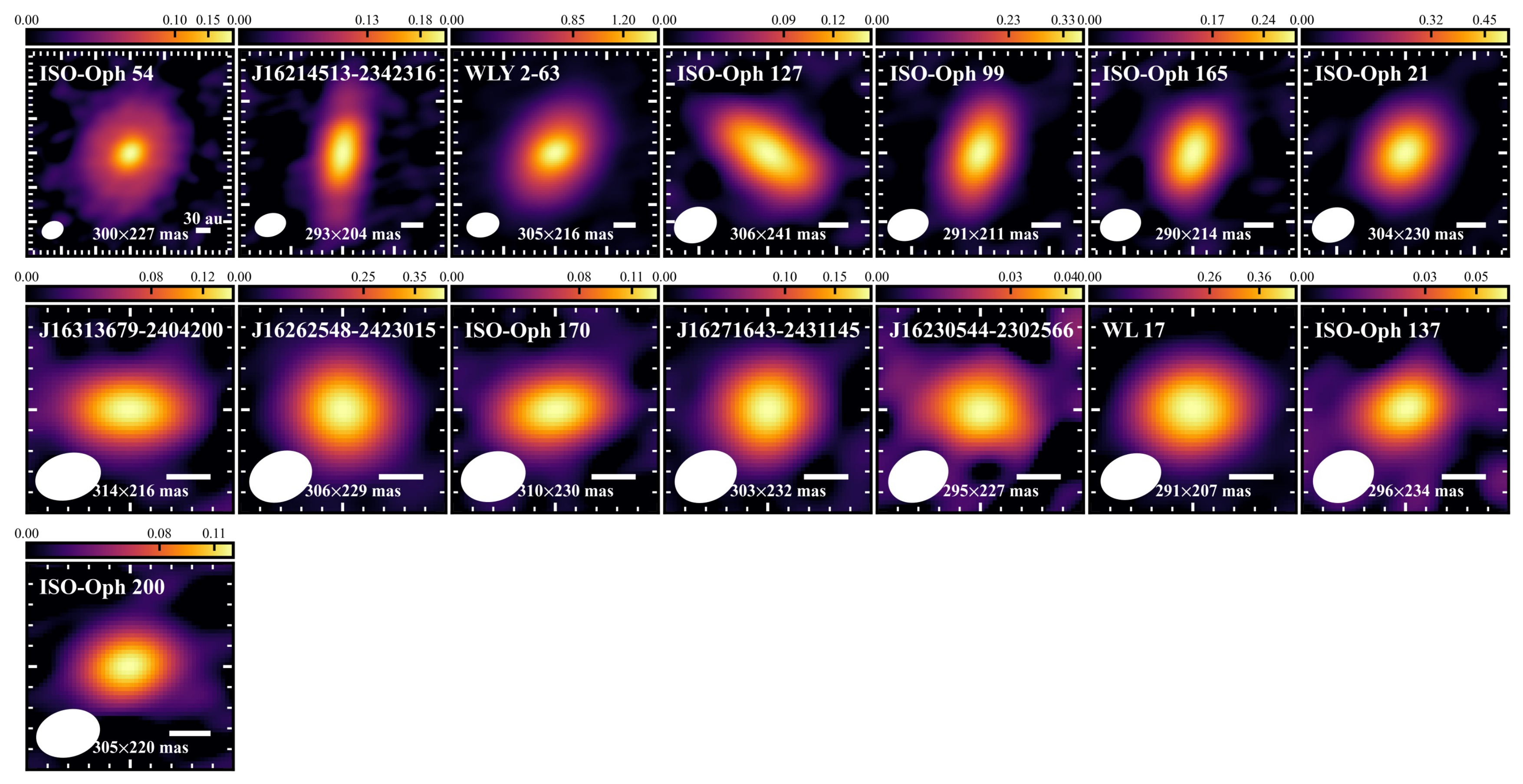}
    \end{center}
    \caption{Gallery of CLEAN images of Class I disks detected through SpM imaging.
    The images are arranged in descending order of disk radius, measured in astronomical units, from upper left to right.
    Each image has a minor axis width of $0\farcs10$, and a color scale following a power law ($\gamma$=0.5) is applied.
    The filled white ellipse in the bottom left corner of each panel represents the synthesized beam $\theta_{\rm CLEAN}$ listed in Table~\ref{table:image_info}.
    All CLEAN images were created using Briggs weighting with a \texttt{robust} parameter of 0.5, and the units were converted from Jy\,beam$^{-1}$ to Jy\,arcsec$^{-2}$.
    }
    \label{fig:clean_classi}
    
    \begin{center}
    \includegraphics[width=0.95\linewidth]{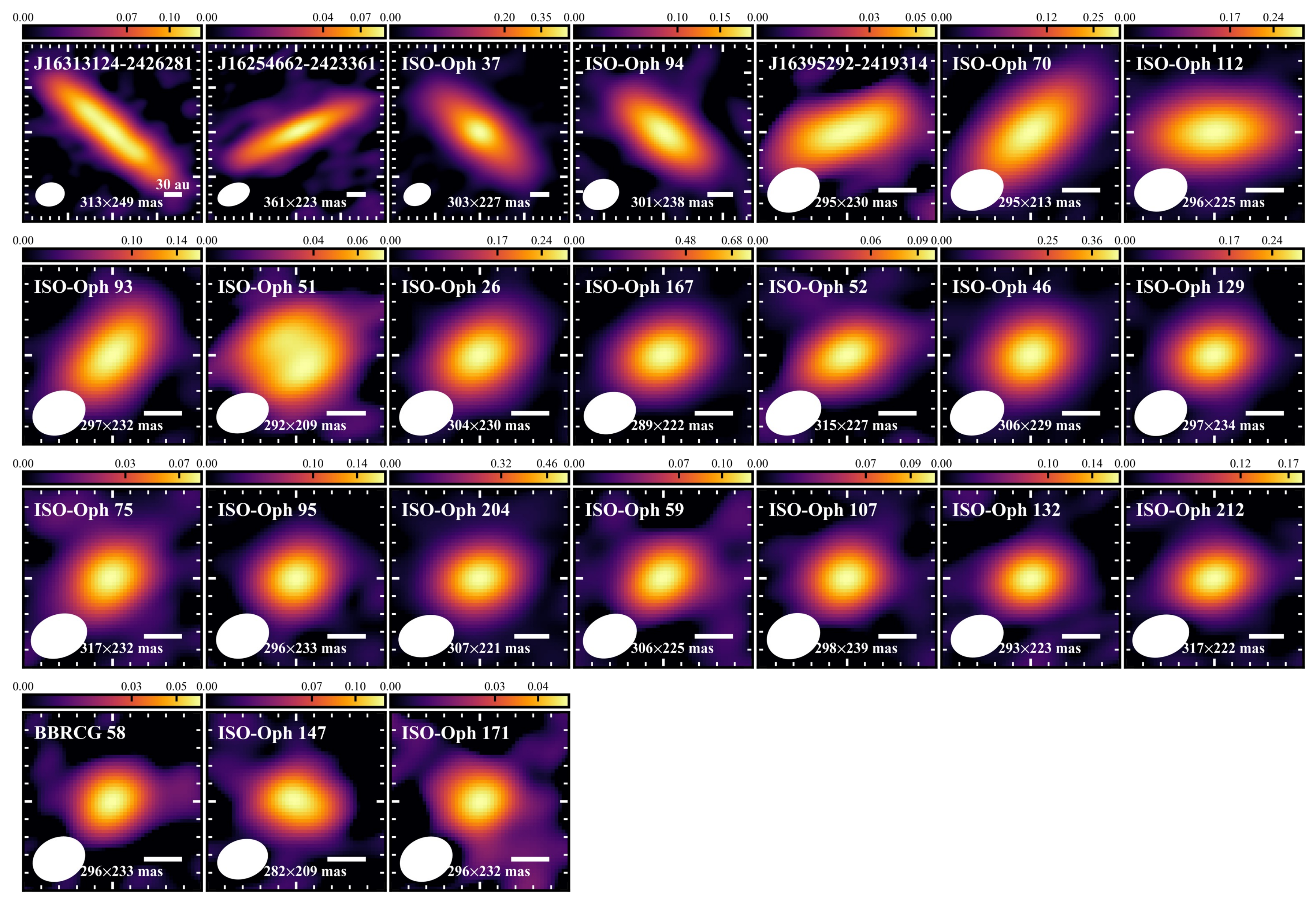}
    \end{center}
    \caption{Same as Figure~\ref{fig:clean_classi} but for Class FS disks.
    Only the brighter component of multiple systems, including ISO-Oph~70, ISO-Oph~167, and ISO-Oph~204, is shown here.
    }
    \label{fig:clean_classflat}
\end{figure*}

\begin{figure*}[t]
    \begin{center}
    \includegraphics[width=\linewidth]{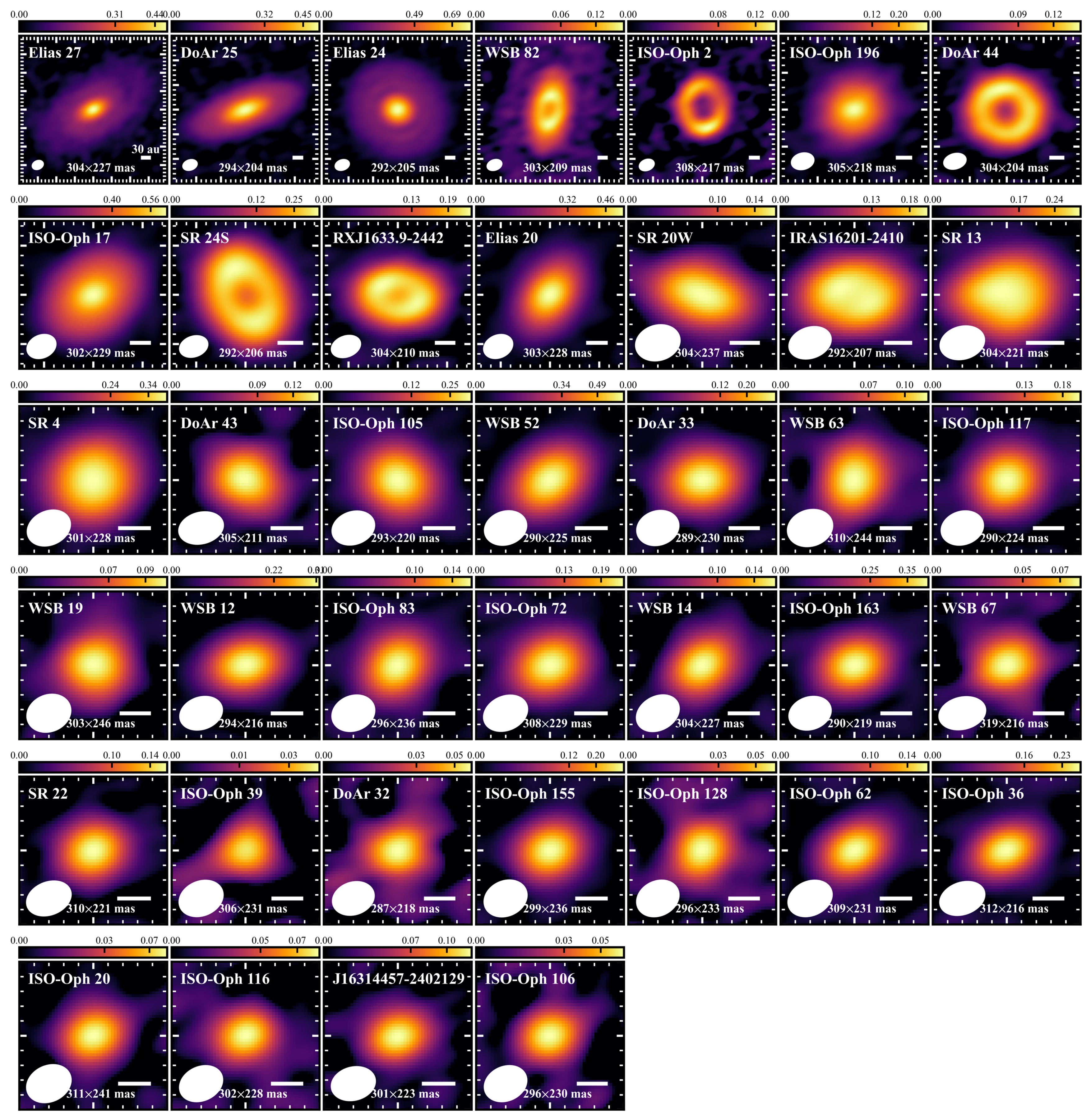}
    \end{center}
    \caption{Same as Figures~\ref{fig:clean_classi}-\ref{fig:clean_classflat} but for Class II disks.
    Only the brighter component of multiple systems, including ISO-Oph~2, SR~24S, DoAr~43, WSB~19, and ISO-Oph~62, is shown here.
    }
    \label{fig:clean_classii}
\end{figure*}

\begin{figure*}[t]
    \begin{center}
    \includegraphics[width=0.95\linewidth]{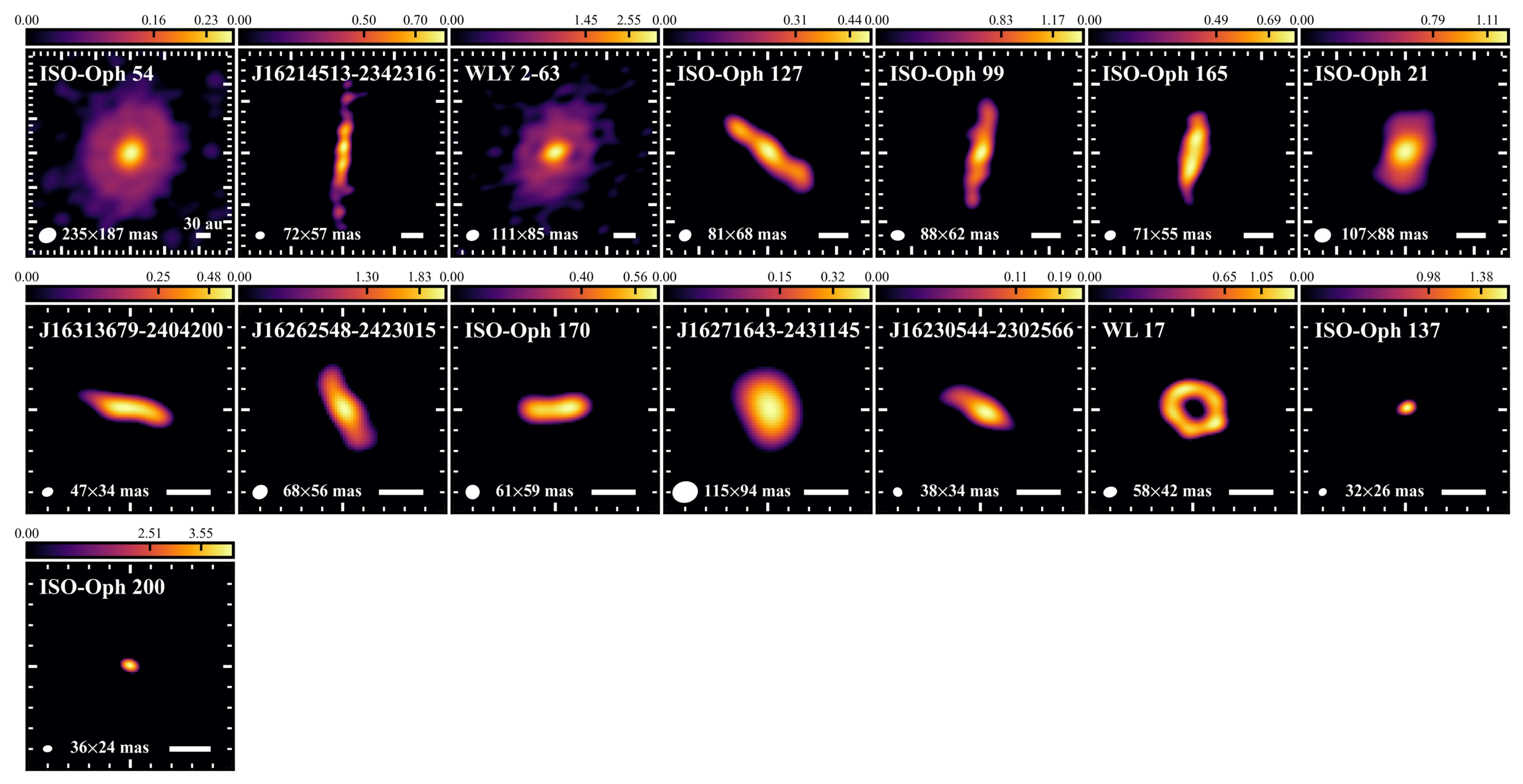}
    \end{center}
    \caption{Gallery of SpM images of Class I.
    The images are arranged in descending order of disk radius, measured in astronomical units, from upper left to right.
    The SpM images are not convolved with the beam, so the unit of the brightness distribution is converted from Jy\,pixel$^{-1}$ to Jy\,arcsec$^{-2}$.
    Each image has a minor axis width of $0\farcs10$, and a color scale following a power law ($\gamma$=0.5) is applied.
    The filled white ellipse in the bottom left corner of each panel represents the effective spatial resolution $\theta_{\rm eff}$ listed in Table~\ref{table:image_info}.
    }
    \label{fig:spm_classi}
    
    \begin{center}
    \includegraphics[width=0.95\linewidth]{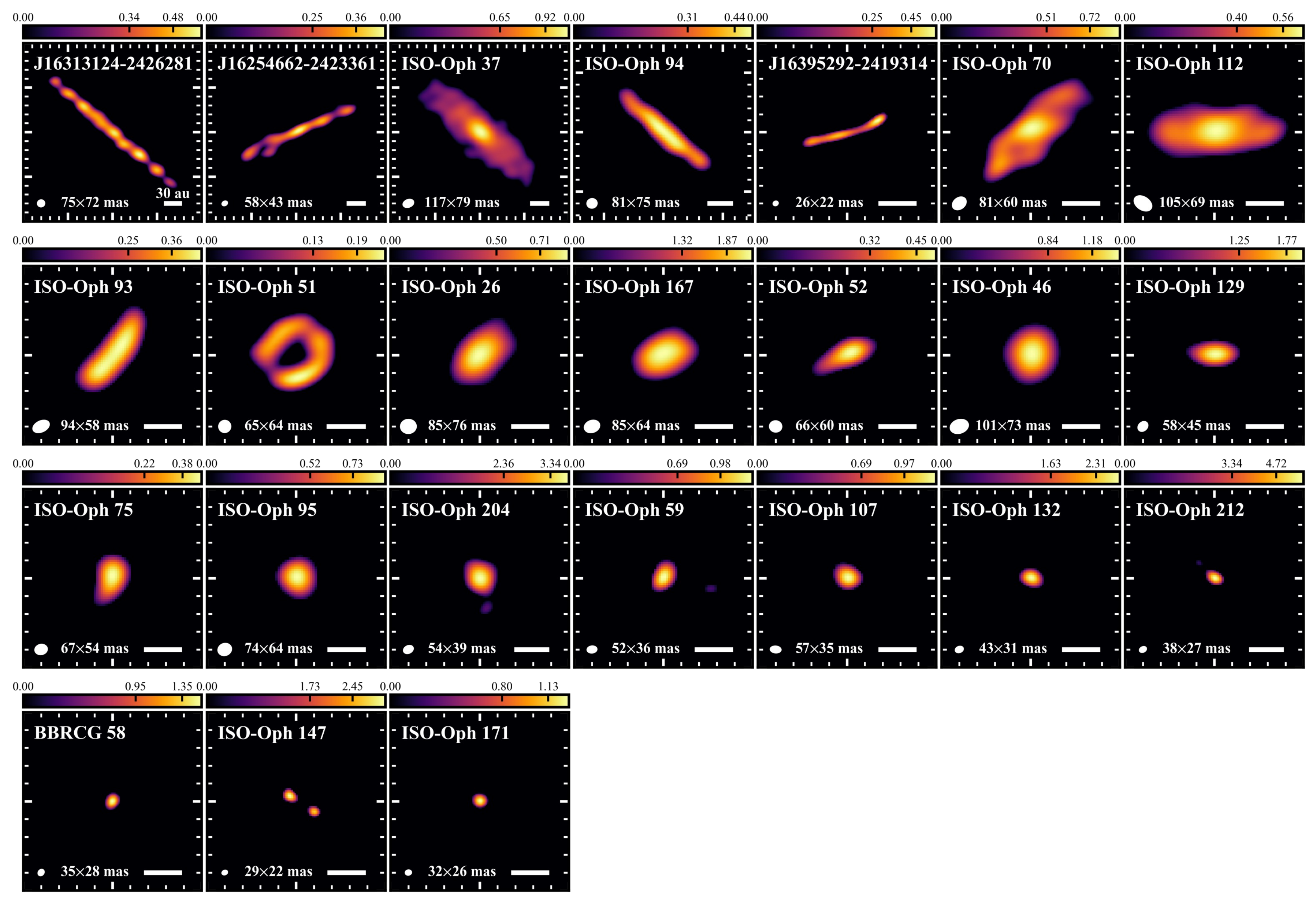}
    \end{center}
    \caption{Same as Figure~\ref{fig:clean_classi} but for Class FS disks.
    Only the brighter component of multiple systems, including ISO-Oph~70, ISO-Oph~167, and ISO-Oph~204, is shown here.
    }
    \label{fig:spm_classflat}
\end{figure*}

\begin{figure*}[t]
    \begin{center}
    \includegraphics[width=\linewidth]{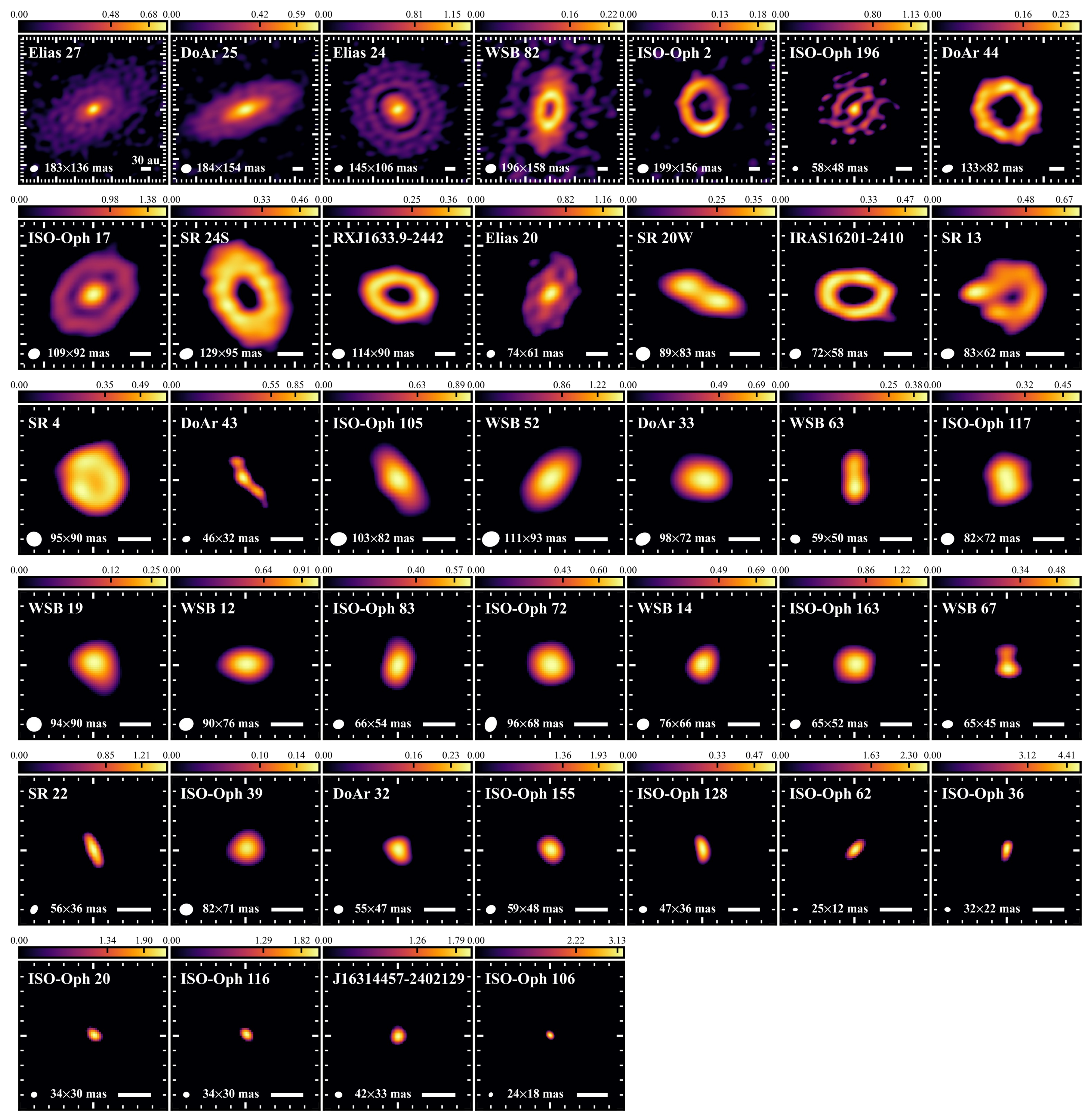}
    \end{center}
    \caption{Same as Figures~\ref{fig:spm_classi}-\ref{fig:spm_classflat} but for Class II disks.
    Only the brighter component of multiple systems, including ISO-Oph~2, SR~24S, DoAr~43, WSB~19, and ISO-Oph~62, is shown here.
    }
    \label{fig:spm_classii}
\end{figure*}

\onecolumn
\begin{landscape}
\begin{longtable}[ht]{lccccccc}
\caption{Information for CLEAN and SpM Images}
\label{table:image_info}\\
\hline 
Source Name & $\theta_{\rm CLEAN}$ (PA) & $\theta_{\rm eff}$ (PA) & $\sigma_{\rm CLEAN}$ & Peak $I_{\rm peak}$ & $F_\nu$ & Point & log($\Lambda_l, \Lambda_{tsv}$)\\
 & {\footnotesize (CLEAN)} & {\footnotesize (SpM)} & {\footnotesize (CLEAN)} & {\footnotesize (CLEAN, SpM)} & {\footnotesize (CLEAN, SpM)} & {\footnotesize (SpM)} & {\footnotesize (SpM)} \\
 & mas (deg) & mas (deg) & mJy\,arcsec$^{-2}$ & mJy\,arcsec$^{-2}$ & mJy & \% &  \\
(1) & (2) & (3) & (4) & (5) & (6) & (7) & (8) \\
\hline
\hline
\endfirsthead
\multicolumn{8}{c}{(Continued)}\\
\hline
Source Name & $\theta_{\rm CLEAN}$ (PA) & $\theta_{\rm eff}$ (PA) & $\sigma_{\rm CLEAN}$ & Peak $I_{\rm peak}$ & $F_\nu$ & Point & log($\Lambda_l, \Lambda_{tsv}$)\\
 & {\footnotesize (CLEAN)} & {\footnotesize (SpM)} & {\footnotesize (CLEAN)} & {\footnotesize (CLEAN, SpM)} & {\footnotesize (CLEAN, SpM)} & {\footnotesize (SpM)} & {\footnotesize (SpM)} \\
 & mas (deg) & mas (deg) & mJy\,arcsec$^{-2}$ & mJy\,arcsec$^{-2}$ & mJy & \% &  \\
(1) & (2) & (3) & (4) & (5) & (6) & (7) & (8) \\
\hline
\hline
\endhead
\endfoot
\multicolumn{8}{l}{\hbox to 0pt{\parbox{192mm}{\footnotemark[]\textbf{Notes.}
Column Description:
(1) Source name.
(2) CLEAN beam $\theta_{\rm CLEAN}$. Briggs \texttt{robust} parameters are set to 0.5 for all images.
(3) SpM beam $\theta_{\rm eff}$, obtained using the point-source injection method (see \S\ref{subsec:method_SpM}).
(4) RMS noise $\sigma_{\rm CLEAN}$ of the CLEAN image in units of mJy\,arcsec$^{-2}$, calculated from an emission-free area.
(5) Peak intensity in CLEAN and SpM images.
(6) Flux density $F_\nu$ of CLEAN and SpM images. CLEAN values are measured from dust brightness distributions exceeding 5$\sigma_{\rm CLEAN}$, while SpM values are derived via the curve-growth method (see \S\ref{subsec:curve_growth}). Uncertainties include a 10\% absolute calibration error for ALMA observations.
(7) Point-source percentage relative to flux density.
(8) Two hyper-parameters for SpM imaging ($\Lambda_l, \Lambda_{tsv}$) in logarithmic scale. The symbol $\ast$ denotes sources for which we conservatively selected images obtained with a value one order of magnitude larger than $\Lambda_{tsv, {\rm min}}$}}.
}\\
\endlastfoot
ISO-Oph~54 & 300$\times$227 (-66.9) & 235$\times$187 (117.2) & 2.0 & 194.0, 263.9 & 90.7, 95.9 & 5 & 4, 11 \\
2MASS~J16214513-2342316 & 293$\times$205 (-72.4) & 72$\times$57 (101.2) & 2.3 & 208.2, 823.7 & 41.4, 38.5 & 5 & 5, 9 \\
WLY 2-63 & 305$\times$216 (-74.5) & 111$\times$85 (111.5) & 2.4 & 1455.3, 3457.8 & 353.2, 353.8 & 5 & 4, 9 \\
ISO-Oph~127 & 306$\times$241 (-66.8) & 81$\times$68 (128.5) & 2.0 & 149.6, 495.0 & 28.3, 27.2 & 5 & 5, 9 \\
ISO-Oph~99 & 292$\times$211 (-72.6) & 88$\times$62 (88.1) & 2.3 & 369.3, 1377.9 & 54.0, 52.5 & 5 & 5, 9 \\
ISO-Oph~165 & 290$\times$214 (-72.3) & 71$\times$55 (115.7) & 2.4 & 281.9, 793.9 & 39.0, 37.4 & 5 & 5, 9 \\
ISO-Oph~21 & 304$\times$230 (-67.2) & 107$\times$88 (99.6) & 2.1 & 512.5, 1239.0 & 75.6, 73.4 & 5 & 5, 9 \\
2MASS~J16313679-2404200 & 314$\times$216 (-75.1) & 47$\times$34 (113.3) & 2.2 & 137.8, 616.2 & 13.9, 13.2 & 5 & 5, 9$^\ast$ \\
2MASS~J16262548-2423015 & 306$\times$229 (-66.3) & 68$\times$56 (126.5) & 2.1 & 415.5, 2097.9 & 47.3, 46.2 & 5 & 5, 8 \\
ISO-Oph~170 & 310$\times$230 (-73.7) & 61$\times$59 (169.0) & 2.1 & 132.0, 633.1 & 12.8, 12.6 & 10 & 5, 9 \\
2MASS~J16271643-2431145 & 303$\times$232 (-64.9) & 115$\times$94 (102.5) & 2.0 & 176.4, 479.5 & 18.3, 18.6 & 5 & 5, 10$^\ast$ \\
2MASS~J16230544-2302566 & 295$\times$227 (-62.1) & 38$\times$34 (38.2) & 2.3 & 47.5, 242.8 & 4.2, 3.8 & 20 & 5, 10$^\ast$ \\
WL 17 & 292$\times$207 (-71.4) & 58$\times$42 (105.9) & 2.4 & 44.7, 1513.7 & 53.2, 52.7 & 5 & 5, 8 \\
ISO-Oph~137 & 296$\times$234 (-65.1) & 32$\times$26 (130.3) & 2.2 & 56.0, 1593.2 & 3.9, 3.8 & 20 & 5, 8 \\
ISO-Oph~200 & 305$\times$220 (-74.8) & 36$\times$24 (96.3) & 2.1 & 121.3, 4209.9 & 8.7, 8.8 & 10 & 5, 7 \\
\hline
2MASS~J16313124-2426281 & 313$\times$249 (-78.7) & 75$\times$72 (107.0) & 2.0 & 118.6, 565.5 & 49.4, 46.4 & 5 & 5, 9 \\
2MASS~J16254662-2423361 & 361$\times$223 (-66.7) & 58$\times$43 (119.0) & 1.8 & 94.3, 427.9 & 23.5, 21.0 & 5 & 5, 9 \\
ISO-Oph~37 & 303$\times$227 (-65.9) & 117$\times$79 (111.0) & 2.2 & 490.7, 1060.4 & 139.3, 133.1 & 5 & 5, 9 \\
ISO-Oph~94 & 301$\times$238 (-67.2) & 81$\times$75 (102.9) & 1.9 & 178.8, 493.4 & 35.8, 34.6 & 5 & 5, 9 \\
2MASS~J16395292-2419314 & 295$\times$231 (-64.2) & 26$\times$22 (118.8) & 2.4 & 51.1, 607.6 & 6.4, 5.3 & 15 & 5, 8 \\
ISO-Oph~70 & 295$\times$213 (-70.5) & 81$\times$60 (127.2) & 2.4 & 344.3, 863.7 & 51.1, 49.2 & 5 & 5, 9 \\
ISO-Oph~112 & 296$\times$225 (-71.4) & 105$\times$69 (56.5) & 2.2 & 289.4, 635.8 & 44.2, 42.1 & 5 & 5, 9 \\
ISO-Oph~93 & 297$\times$232 (-66.8) & 94$\times$58 (115.4) & 2.1 & 161.8, 432.5 & 20.2, 19.6 & 5 & 5, 9 \\
ISO-Oph~51 & 292$\times$209 (-71.1) & 65$\times$64 (154.7) & 2.4 & 69.9, 225.5 & 12.5, 11.1 & 10 & 5, 10 \\
ISO-Oph~26 & 304$\times$230 (-66.8) & 85$\times$76 (92.6) & 2.2 & 279.4, 850.0 & 30.4, 29.2 & 5 & 5, 9 \\
ISO-Oph~167 & 289$\times$222 (-72.3) & 85$\times$64 (109.5) & 2.6 & 770.2, 2184.8 & 75.0, 74.6 & 5 & 5, 9 \\
ISO-Oph~52 & 315$\times$227 (-66.5) & 66$\times$60 (85.4) & 2.0 & 100.0, 505.0 & 9.1, 8.6 & 10 & 5, 9$^\ast$ \\
ISO-Oph~46 & 306$\times$229 (-66.5) & 101$\times$73 (108.4) & 2.1 & 422.4, 1398.9 & 43.9, 43.0 & 5 & 5, 9 \\
ISO-Oph~75 & 317$\times$232 (-68.3) & 67$\times$54 (102.0) & 1.9 & 91.4, 740.3 & 8.5, 8.2 & 10 & 5, 9$^\ast$ \\
ISO-Oph~129 & 297$\times$234 (-63.9) & 58$\times$45 (134.0) & 2.2 & 294.7, 1953.1 & 26.3, 25.6 & 5 & 5, 8 \\
ISO-Oph~95 & 296$\times$233 (-66.8) & 74$\times$64 (110.8) & 2.1 & 170.2, 895.4 & 14.3, 14.5 & 10 & 5, 9$^\ast$ \\
ISO-Oph~204 & 307$\times$221 (-74.5) & 54$\times$39 (116.2) & 2.4 & 521.9, 3724.4 & 45.9, 45.0 & 5 & 5, 8 \\
ISO-Oph~59 & 306$\times$225 (-64.7) & 52$\times$36 (96.5) & 2.0 & 116.6, 1189.6 & 9.4, 9.2 & 10 & 5, 8 \\
ISO-Oph~107 & 298$\times$239 (-66.6) & 57$\times$35 (85.5) & 1.9 & 109.7, 1185.6 & 8.7, 8.8 & 5 & 5, 8 \\
ISO-Oph~132 & 293$\times$223 (-71.5) & 43$\times$31 (110.1) & 2.2 & 168.3, 2661.2 & 12.3, 12.4 & 10 & 5, 8 \\
ISO-Oph~212 & 317$\times$222 (-72.3) & 38$\times$27 (116.6) & 2.3 & 191.4, 5570.8 & 15.0, 14.9 & 25 & 5, 7 \\
BBRCG 58 & 296$\times$233 (-64.6) & 35$\times$28 (142.0) & 2.2 & 54.2, 1512.7 & 3.6, 3.7 & 25 & 5, 8 \\
ISO-Oph~147a & 282$\times$209 (-71.2) & 29$\times$22 (113.9) & 2.3 & 114.6, 2994.4 & 9.6, 6.2 & 10 & 5, 7 \\
ISO-Oph~147b & 282$\times$209 (-71.2) & 29$\times$22 (113.9) & 2.3 & $\cdots$, 2311.0 &  $\cdots$, 3.4 & 10 & 5, 7 \\
ISO-Oph~171 & 296$\times$232 (-65.2) & 32$\times$26 (100.6) & 2.2 & 43.8, 1278.8 & 2.9, 2.9 & 25 & 5, 8 \\
\hline
Elias~27 & 304$\times$228 (-66.2) & 183$\times$136 (113.6) & 2.1 & 475.3, 780.2 & 255.5, 261.3 & 5 & 4, 10 \\
DoAr~25 & 294$\times$204 (-71.6) & 184$\times$154 (105.6) & 2.3 & 505.4, 698.1 & 226.3, 228.4 & 5 & 4, 11 \\
Elias~24 & 292$\times$205 (-71.3) & 145$\times$106 (108.5) & 2.3 & 798.0, 1323.6 & 340.1, 340.9 & 5 & 4, 10 \\
WSB~82 & 303$\times$209 (-71.1) & 196$\times$158 (105.0) & 2.3 & 182.7, 247.6 & 125.2, 120.4 & 5 & 4, 11 \\
ISO-Oph~2 & 308$\times$217 (-69.0) & 199$\times$156 (104.2) & 2.1 & 134.5, 207.5 & 67.0, 69.0 & 5 & 4, 11 \\
ISO-Oph~196 & 305$\times$218 (-76.1) & 58$\times$48 (97.0) & 2.3 & 313.0, 1282.6 & 93.9, 86.3 & 5 & 5, 8 \\
DoAr~44 & 304$\times$204 (-74.9) & 133$\times$82 (111.0) & 2.4 & 149.7, 266.1 & 80.7, 74.3 & 5 & 5, 10 \\
ISO-Oph~17 & 302$\times$229 (-66.1) & 109$\times$92 (121.1) & 2.2 & 634.4, 1588.3 & 173.5, 167.7 & 5 & 5, 9 \\
SR~24S & 292$\times$206 (-71.3) & 129$\times$95 (112.0) & 2.4 & 363.3, 530.9 & 187.8, 181.3 & 5 & 5, 10 \\
RXJ1633.9-2442 & 304$\times$210 (-69.6) & 114$\times$90 (97.2) & 2.3 & 225.7, 420.6 & 76.1, 72.6 & 5 & 5, 10 \\
Elias~20 & 303$\times$228 (-66.7) & 74$\times$61 (118.3) & 2.2 & 520.8, 1355.7 & 96.3, 92.4 & 5 & 5, 8 \\
SR~20W & 304$\times$237 (-73.7) & 89$\times$83 (108.3) & 2.1 & 164.3, 417.9 & 24.0, 23.1 & 5 & 5, 10 \\
IRAS16201-2410 & 292$\times$207 (-71.1) & 72$\times$58 (118.2) & 2.3 & 209.2, 560.3 & 42.7, 41.1 & 5 & 5, 9 \\
SR~13 & 304$\times$221 (-75.1) & 83$\times$62 (102.8) & 2.3 & 293.9, 756.4 & 61.2, 59.1 & 5 & 5, 9 \\
SR~4 & 301$\times$228 (-66.0) & 95$\times$90 (62.9) & 2.2 & 393.6, 595.3 & 66.6, 65.0 & 5 & 5, 9 \\
DoAr~43 & 305$\times$211 (-76.0) & 46$\times$32 (107.9) & 2.3 & 141.2, 1212.4 & 15.0, 14.5 & 10 & 5, 8 \\
ISO-Oph~105 & 293$\times$220 (-71.4) & 103$\times$82 (102.6) & 2.3 & 343.1, 994.2 & 39.8, 39.0 & 5 & 5, 10$^\ast$ \\
WSB~52 & 290$\times$225 (-72.0) & 111$\times$93 (110.6) & 2.3 & 595.5, 1479.7 & 66.6, 65.9 & 5 & 5, 10 \\
DoAr~33 & 289$\times$230 (-71.9) & 98$\times$72 (120.4) & 2.3 & 311.9, 794.5 & 33.4, 32.6 & 5 & 5, 10 \\
WSB~63 & 310$\times$244 (-72.8) & 59$\times$50 (75.7) & 2.0 & 113.2, 466.6 & 12.6, 12.2 & 5 & 5, 9$^\ast$ \\
ISO-Oph~117 & 290$\times$224 (-72.6) & 82$\times$72 (100.2) & 2.3 & 207.6, 514.8 & 21.8, 21.0 & 5 & 5, 10 \\
WSB~19 & 303$\times$246 (-67.4) & 94$\times$90 (83.2) & 2.0 & 109.7, 312.1 & 11.6, 10.8 & 5 & 5, 10$^\ast$ \\
WSB~12 & 294$\times$216 (-70.8) & 90$\times$76 (112.8) & 2.3 & 314.1, 1028.0 & 28.6, 28.7 & 5 & 5, 10$^\ast$ \\
ISO-Oph~83 & 296$\times$236 (-67.9) & 66$\times$54 (110.0) & 2.0 & 158.4, 640.1 & 15.6, 15.0 & 5 & 5, 9 \\
ISO-Oph~72 & 308$\times$229 (-66.1) & 96$\times$68 (158.8) & 2.0 & 225.0, 724.2 & 23.4, 22.8 & 5 & 5, 9 \\
WSB~14 & 304$\times$227 (-61.3) & 76$\times$66 (116.7) & 2.4 & 165.8, 799.2 & 14.8, 14.2 & 10 & 5, 9 \\
ISO-Oph~163 & 290$\times$219 (-71.8) & 65$\times$52 (118.4) & 2.4 & 409.6, 1488.4 & 37.2, 36.4 & 5 & 5, 9 \\
WSB~67 & 319$\times$216 (-76.6) & 65$\times$45 (98.5) & 2.2 & 81.2, 568.2 & 6.9, 6.6 & 15 & 5, 9$^\ast$ \\
SR~22 & 310$\times$221 (-68.6) & 56$\times$36 (149.7) & 2.1 & 153.6, 1459.0 & 13.3, 13.3 & 10 & 5, 8 \\
ISO-Oph~39 & 305$\times$229 (-66.9) & 82$\times$71 (95.1) & 2.1 & 41.5, 153.6 & 2.6, 2.9 & 25 & 5, 10 \\
DoAr~32 & 287$\times$218 (-71.6) & 55$\times$47 (104.3) & 2.5 & 51.3, 270.2 & 3.5, 3.3 & 25 & 5, 10 \\
ISO-Oph~155 & 299$\times$236 (-65.4) & 59$\times$48 (121.2) & 2.1 & 305.1, 2326.6 & 26.3, 26.4 & 5 & 5, 8 \\
ISO-Oph~128 & 296$\times$233 (-64.2) & 47$\times$36 (94.9) & 2.3 & 53.8, 548.5 & 4.0, 3.8 & 20 & 5, 9 \\
ISO-Oph~62 & 309$\times$231 (-66.2) & 25$\times$12 (92.0) & 2.0 & 163.4, 2644.5 & 14.8, 13.2 & 5 & 5, 7 \\
ISO-Oph~36 & 312$\times$216 (-70.6) & 32$\times$22 (83.3) & 2.5 & 262.2, 4856.9 & 20.5, 18.3 & 5 & 5, 7 \\
ISO-Oph~20 & 311$\times$241 (-64.7) & 34$\times$30 (112.0) & 1.9 & 90.8, 2248.2 & 7.2, 7.3 & 10 & 5, 7 \\
ISO-Oph~116 & 302$\times$228 (-65.1) & 34$\times$30 (91.2) & 2.3 & 83.1, 2071.2 & 5.9, 5.8 & 20 & 5, 7 \\
2MASS~J16314457-2402129 & 301$\times$223 (-75.4) & 42$\times$33 (85.0) & 2.1 & 115.7, 1998.3 & 8.6, 8.6 & 10 & 5, 8 \\
ISO-Oph~106 & 296$\times$230 (-66.0) & 24$\times$18 (149.0) & 2.2 & 56.7, 3275.7 & 3.7, 3.7 & 25 & 5, 7 \\
\hline
\end{longtable}
\end{landscape}

\begin{longtable}[ht]{lccccccc}
\caption{Disk Properties estimated from SpM Images}\\
\hline
\label{table:disk_info}
Source Name & PA & $i_{\rm disk}$ & $L_{\rm mm}$ & $R_{68\%}$ & $R_{95\%}$ & $\sigma_{\rm Radius}$ & Category \\
 & deg & deg & mJy & au & au & au &  \\
(1) & (2) & (3) & (4) & (5) & (6) & (7) & (8) \\
\hline
\hline
\endfirsthead
\multicolumn{8}{c}{(Continued)}\\
\hline
Source Name & PA & $i_{\rm disk}$ & $L_{\rm mm}$ & $R_{68\%}$ & $R_{95\%}$ & $\sigma_{\rm Radius}$ & Category \\
 & deg & deg & mJy & au & au & au &  \\
(1) & (2) & (3) & (4) & (5) & (6) & (7) & (8) \\
\hline
\hline
\endhead
\endfoot
\multicolumn{8}{l}{\hbox to 0pt{\parbox{168mm}{\footnotemark[]\textbf{Notes.}
Column Description:
(1) Source name.
(2) Position angle (PA) of the disk.
(3) Inclination angle $i_{\rm disk}$.
(4) Millimeter luminosity $L_{\rm mm}$, calculated as the total flux corrected for a distance of 140\,pc.
(5) Disk radius containing 68\% of the flux density, determined via the curve-growth method.
(6) Disk radius containing 95\% of the flux density, also determined via the curve-growth method. A $\dagger$ symbol indicates that the radius is treated as a `very compact disk', which is defined as one that requires additional long-baseline data to resolve its size (see \S A.\ref{subsec:fidelity_disk}).
(7) Uncertainty in disk radius, approximately equal to the square root of the effective spatial resolution $\theta_{\rm eff}$.
(8) Disk categorization (see \S\ref{subsec:categorization}). An $\ast$ symbol denotes newly detected substructures identified in this study (see \S\ref{subsec:new_substructures}).
}}}\\
\endlastfoot
ISO-Oph~54 & 151.2$^{+0.41}_{-0.46}$ & 31.6$^{+0.24}_{-0.21}$ & 95.8 & 81.9 & 123.7 & 12.5 & Single/Inflection \\
2MASS~J16214513-2342316 & 174.1$^{+0.02}_{-0.02}$ & 80.2$^{+0.02}_{-0.02}$ & 38.5 & 46.1 & 85.9 & 3.8 & Single/Ring$^\ast$ \\
WLY~2-63 & 146.6$^{+0.04}_{-0.04}$ & 46.9$^{+0.03}_{-0.03}$ & 353.5 & 49.7 & 79.0 & 5.8 & Single/Inflection \\
ISO-Oph~127 & 51.3$^{+0.02}_{-0.03}$ & 77.5$^{+0.02}_{-0.02}$ & 27.2 & 43.7 & 60.0 & 4.4 & Single/Ring$^\ast$ \\
ISO-Oph~99 & 168.2$^{+0.02}_{-0.02}$ & 75.8$^{+0.02}_{-0.02}$ & 52.5 & 34.9 & 51.1 & 4.4 & Single/Ring$^\ast$ \\
ISO-Oph~165 & 168.2$^{+0.03}_{-0.03}$ & 73.6$^{+0.03}_{-0.03}$ & 37.4 & 30.6 & 43.3 & 3.7 & Single/Ring$^\ast$ \\
ISO-Oph~21 & 161.9$^{+0.05}_{-0.05}$ & 49.0$^{+0.04}_{-0.04}$ & 73.3 & 27.3 & 40.0 & 5.8 & Single/Smooth \\
2MASS~J16313679-2404200 & 78.4$^{+0.05}_{-0.05}$ & 78.5$^{+0.05}_{-0.05}$ & 13.2 & 21.7 & 30.6 & 2.4 & Single/Smooth \\
2MASS~J16262548-2423015 & 28.8$^{+0.02}_{-0.02}$ & 67.5$^{+0.02}_{-0.02}$ & 46.2 & 20.0 & 29.1 & 3.7 & Single/Smooth \\
ISO-Oph~170 & 94.2$^{+0.07}_{-0.06}$ & 71.4$^{+0.06}_{-0.05}$ & 12.6 & 19.1 & 26.7 & 3.6 & Single/Ring or Binary/Smooth? \\
2MASS~J16271643-2431145 & 21.2$^{+0.20}_{-0.19}$ & 55.7$^{+0.13}_{-0.12}$ & 18.6 & 18.2 & 25.5 & 6.2 & Single/Smooth \\
2MASS~J16230544-2302566 & 59.4$^{+0.40}_{-0.39}$ & 69.7$^{+0.39}_{-0.39}$ & 3.8 & 15.5 & 24.1 & 2.1 & Single/Smooth \\
WL~17 & 63.0$^{+2.30}_{-2.30}$ & 34.8$^{+0.80}_{-0.80}$ & 52.7 & 18.7 & 22.4 & 2.9 & Single/Ring \\
ISO-Oph~137 & 110.2$^{+0.25}_{-0.29}$ & 42.3$^{+0.37}_{-0.45}$ & 3.8 & 4.8$^\dagger$ & 6.8$^\dagger$ & 1.7 & Single/Smooth \\
ISO-Oph~200 & 72.3$^{+0.12}_{-0.12}$ & 46.0$^{+0.10}_{-0.09}$ & 10.1 & 5.5$^\dagger$ & 6.8$^\dagger$ & 1.9 & Single/Smooth \\
\hline
2MASS~J16313124-2426281 & 48.9$^{+0.01}_{-0.01}$ & 86.0$^{+0.01}_{-0.01}$ & 51.2 & 103.0 & 144.5 & 4.6 & Single/Ring \\
2MASS~J16254662-2423361 & 113.2$^{+0.02}_{-0.02}$ & 85.1$^{+0.02}_{-0.02}$ & 21.0 & 79.0 & 126.1 & 3.0 & Single/Ring$^\ast$ \\
ISO-Oph~37 & 48.4$^{+0.02}_{-0.02}$ & 71.1$^{+0.01}_{-0.01}$ & 133.0 & 63.7 & 96.4 & 5.7 & Single/Ring \\
ISO-Oph~94 & 49.0$^{+0.02}_{-0.02}$ & 79.2$^{+0.02}_{-0.02}$ & 34.6 & 47.3 & 69.1 & 4.6 & Single/Ring$^\ast$ \\
2MASS~J16395292-2419314 & 107.4$^{+0.05}_{-0.05}$ & 85.4$^{+0.04}_{-0.04}$ & 5.3 & 33.8 & 54.1 & 1.4 & Single/Ring$^\ast$ \\
ISO-Oph~70 & 133.9$^{+0.03}_{-0.04}$ & 70.0$^{+0.03}_{-0.04}$ & 49.2 & 37.4 & 51.1 & 4.1 & Multiple/Ring$^\ast$ \\
ISO-Oph~112 & 92.0$^{+0.03}_{-0.03}$ & 69.0$^{+0.03}_{-0.03}$ & 42.1 & 36.4 & 49.1 & 5.1 & Single/Inflection$^\ast$ \\
ISO-Oph~93 & 142.0$^{+0.04}_{-0.04}$ & 71.9$^{+0.04}_{-0.04}$ & 19.6 & 29.1 & 41.8 & 4.4 & Single/Ring$^\ast$ \\
ISO-Oph~51 & 132.7$^{+4.40}_{4.40}$ & 24.5$^{+0.70}_{-0.70}$ & 10.6 & 23.6 & 29.2 & 3.7 & Single/Ring$^\ast$ \\
ISO-Oph~26 & 135.0$^{+0.08}_{-0.08}$ & 50.4$^{+0.06}_{-0.06}$ & 29.2 & 18.2 & 27.3 & 4.8 & Single/Smooth \\
ISO-Oph~167 & 113.8$^{+0.05}_{-0.05}$ & 48.0$^{+0.03}_{-0.03}$ & 74.5 & 16.6 & 24.2 & 4.4 & Multiple/Smooth \\
ISO-Oph~52 & 112.2$^{+0.10}_{-0.10}$ & 65.7$^{+0.10}_{-0.10}$ & 8.6 & 14.6 & 23.7 & 3.7 & Single/Smooth \\
ISO-Oph~46 & 167.0$^{+0.18}_{-0.19}$ & 27.2$^{+0.09}_{-0.08}$ & 43.0 & 14.6 & 20.0 & 5.1 & Single/Smooth \\
ISO-Oph~75 & 165.6$^{+0.16}_{-0.16}$ & 47.7$^{+0.25}_{-0.22}$ & 8.2 & 12.7 & 18.2 & 3.6 & Single/Smooth \\
ISO-Oph~129 & 90.0$^{+0.04}_{-0.04}$ & 57.7$^{+0.03}_{-0.03}$ & 25.6 & 12.7 & 18.2 & 3.0 & Single/Smooth \\
ISO-Oph~95 & 27.9$^{+1.44}_{-1.06}$ & 38.0$^{+0.34}_{-0.39}$ & 14.5 & 10.9 & 16.4 & 4.1 & Single/Smooth \\
ISO-Oph~204 & 26.6$^{+0.15}_{-0.14}$ & 29.5$^{+0.07}_{-0.07}$ & 53.7 & 9.7 & 15.3 & 3.0 & Multiple/Smooth \\
ISO-Oph~59 & 151.0$^{+0.14}_{-0.14}$ & 45.9$^{+0.10}_{-0.10}$ & 9.2 & 9.1 & 12.7 & 2.6 & Single/Smooth \\
ISO-Oph~107 & 63.6$^{+0.15}_{-0.17}$ & 32.2$^{+0.35}_{-0.30}$ & 8.8 & 7.3 & 10.9 & 2.5 & Single/Smooth \\
ISO-Oph~132 & 70.9$^{+0.19}_{-0.22}$ & 40.3$^{+0.13}_{-0.12}$ & 12.4 & 6.4 & 8.9 & 2.2 & Single/Smooth \\
ISO-Oph~212 & 61.5$^{+0.07}_{-0.07}$ & 45.8$^{+0.05}_{-0.05}$ & 17.9 & 5.6 & 8.4 & 2.1 & Single/Smooth \\
BBRCG 58 & 149.8$^{+0.36}_{-0.35}$ & 37.1$^{+0.60}_{-0.56}$ & 3.7 & 3.8$^\dagger$ & 6.7$^\dagger$ & 1.9 & Single/Smooth \\
ISO-Oph~147 & 54.9$^{+0.28}_{-0.26}$ & 40.2$^{+0.18}_{-0.16}$ & 6.2 & 3.8 & 6.4 & 1.5 & Multiple/Smooth \\
ISO-Oph~171 & 76.3$^{+0.79}_{-0.82}$ & 22.9$^{+2.08}_{-2.23}$ & 2.9 & 3.9$^\dagger$ & 5.8$^\dagger$ & 1.7 & Single/Smooth \\
\hline
Elias~27 & 118.1$^{+0.04}_{-0.04}$ & 56.2$^{+0.03}_{-0.03}$ & 161.5 & 105.9 & 178.9 & 7.4 & Single/Spiral \\
DoAr~25 & 110.3$^{+0.03}_{-0.03}$ & 65.5$^{+0.03}_{-0.03}$ & 222.4 & 102.1 & 152.5 & 9.9 & Single/Inflection \\
Elias~24 & 55.6$^{+2.00}_{-2.00}$ & 22.7$^{+0.80}_{-0.80}$ & 337.3 & 86.8 & 132.6 & 7.3 & Single/Ring \\
WSB~82 & 170.5$^{+1.10}_{-1.10}$ & 50.9$^{+0.30}_{-0.30}$ & 130.5 & 76.9 & 118.1 & 10.9 & Single/Ring \\
ISO-Oph~2 & 5.8$^{+0.40}_{-0.40}$ & 36.5$^{+1.00}_{-1.00}$ & 63.4 & 68.4 & 80.6 & 10.0 & Multiple/Ring \\
ISO-Oph~196 & 131.4$^{+5.00}_{-5.00}$ & 36.1$^{+1.30}_{-1.30}$ & 80.2 & 43.0 & 70.0 & 3.0 & Single/Ring \\
DoAr~44 & 70.9$^{+4.70}_{-4.70}$ & 23.3$^{+0.70}_{-0.70}$ & 81.2 & 57.3 & 67.9 & 6.5 & Single/Ring \\
ISO-Oph~17 & 128.3$^{+1.30}_{-1.30}$ & 40.5$^{+0.50}_{-0.50}$ & 167.6 & 50.9 & 63.7 & 6.0 & Single/Ring \\
SR~24S & 26.4$^{+0.06}_{-0.06}$ & 47.8$^{+0.05}_{-0.05}$ & 122.3 & 49.1 & 61.4 & 5.4 & Multiple/Ring \\
RXJ1633.9-2442 & 82.2$^{+1.10}_{-1.10}$ & 47.3$^{+0.30}_{-0.30}$ & 76.6 & 44.5 & 55.0 & 6.2 & Single/Ring \\
Elias~20 & 153.5$^{+0.05}_{-0.05}$ & 51.5$^{+0.04}_{-0.04}$ & 89.2 & 35.8 & 51.8 & 3.9 & Single/Inflection \\
SR~20W & 67.3$^{+0.06}_{-6.00}$ & 70.5$^{+0.05}_{-0.05}$ & 25.4 & 32.1 & 45.4 & 5.4 & Single/Ring$^\ast$ \\
IRAS16201-2410 & 87.8$^{+3.70}_{-3.70}$ & 49.9$^{+0.80}_{-0.80}$ & 51.4 & 34.8 & 41.8 & 4.3 & Single/Ring$^\ast$ \\
SR~13 & 143.3$^{+5.00}_{-5.00}$ & 31.0$^{+1.30}_{-1.30}$ & 40.2 & 24.2 & 32.6 & 3.5 & Multiple/Ring$^\ast$ \\
SR~4 & 21.5$^{+0.30}_{-0.31}$ & 23.9$^{+0.11}_{-0.13}$ & 60.2 & 24.5 & 31.5 & 5.3 & Single/Ring \\
DoAr~43 & 37.3$^{+0.04}_{-0.05}$ & 74.7$^{+0.04}_{-0.04}$ & 13.7 & 19.8 & 30.9 & 2.2 & Multiple/Ring$^\ast$ \\
ISO-Oph~105 & 33.8$^{+0.07}_{-0.07}$ & 60.3$^{+0.06}_{-0.06}$ & 36.0 & 20.4 & 29.9 & 5.3 & Single/Smooth \\
WSB~52 & 139.1$^{+0.05}_{-0.05}$ & 52.1$^{+0.04}_{-0.04}$ & 61.5 & 20.9 & 29.5 & 5.8 & Single/Smooth \\
DoAr~33 & 82.8$^{+0.15}_{-0.15}$ & 43.4$^{+0.11}_{-0.11}$ & 33.3 & 18.0 & 27.1 & 5.0 & Single/Smooth \\
WSB~63 & 1.5$^{+0.08}_{-0.08}$ & 73.4$^{+0.07}_{-0.06}$ & 11.6 & 19.5 & 26.6 & 3.1 & Single/Ring or Binary/Smooth? \\
ISO-Oph~117 & 6.8$^{+0.32}_{-0.32}$ & 37.4$^{+0.18}_{-0.19}$ & 21.2 & 16.6 & 24.3 & 4.6 & Single/Smooth \\
WSB~19 & 50.8$^{+0.64}_{-0.65}$ & 41.5$^{+0.33}_{-0.34}$ & 11.1 & 16.6 & 24.0 & 5.5 & Multiple/Smooth \\
WSB~12 & 85.9$^{+0.14}_{-0.12}$ & 54.9$^{+0.10}_{-0.09}$ & 27.4 & 15.8 & 21.9 & 4.8 & Multiple/Smooth \\
ISO-Oph~83 & 166.8$^{+0.14}_{-0.14}$ & 49.1$^{+0.10}_{-0.10}$ & 14.3 & 14.2 & 21.3 & 3.5 & Single/Smooth \\
ISO-Oph~72 & 63.0$^{+0.52}_{-0.47}$ & 24.5$^{+0.24}_{-0.21}$ & 20.4 & 13.8 & 18.9 & 4.5 & Single/Smooth \\
WSB~14 & 150.5$^{+0.31}_{-0.25}$ & 36.5$^{+0.18}_{-0.15}$ & 13.7 & 12.5 & 17.9 & 4.1 & Single/Smooth \\
ISO-Oph~163 & 92.3$^{+0.36}_{-0.37}$ & 24.1$^{+0.14}_{-0.14}$ & 36.1 & 12.7 & 17.8 & 3.4 & Single/Smooth \\
WSB~67 & 60.1$^{+0.37}_{-0.32}$ & 43.0$^{+0.28}_{-0.28}$ & 6.7 & 12.2 & 17.6 & 3.2 & Single/Ring or Binary/Smooth? \\
SR~22 & 22.7$^{+0.07}_{-0.05}$ & 69.0$^{+0.05}_{-0.05}$ & 11.7 & 12.0 & 16.7 & 2.5 & Single/Smooth \\
ISO-Oph~39 & 105.5$^{+3.94}_{-4.47}$ & 30.0$^{+1.68}_{-1.54}$ & 2.9 & 10.9 & 16.3 & 5.1 & Single/Smooth \\
DoAr~32 & 29.4$^{+0.93}_{-1.17}$ & 33.8$^{+2.02}_{-1.85}$ & 3.4 & 9.7 & 13.6 & 3.0 & Single/Smooth \\
ISO-Oph~155 & 36.0$^{+0.11}_{-0.11}$ & 35.5$^{+0.06}_{-0.06}$ & 25.1 & 9.1 & 12.7 & 3.2 & Single/Smooth \\
ISO-Oph~128 & 11.6$^{+0.25}_{-0.24}$ & 60.1$^{+0.30}_{-0.29}$ & 3.8 & 8.7 & 12.6 & 2.4 & Single/Smooth \\
ISO-Oph~62 & 137.8$^{+0.04}_{-0.04}$ & 59.1$^{+0.03}_{-0.04}$ & 12.6 & 7.1 & 10.7 & 1.0 & Multiple/Smooth \\
ISO-Oph~36 & 166.8$^{+0.04}_{-0.04}$ & 59.8$^{+0.03}_{-0.03}$ & 18.1 & 6.2 & 9.9 & 1.6 & Multiple/Smooth \\
ISO-Oph~20 & 51.4$^{+0.09}_{-0.09}$ & 44.1$^{+0.13}_{-0.13}$ & 6.8 & 5.5 & 9.1 & 1.9 & Single/Smooth \\
ISO-Oph~116 & 45.5$^{+0.11}_{-0.10}$ & 45.5$^{+0.18}_{-0.14}$ & 5.6 & 5.4 & 8.9 & 1.9 & Single/Smooth \\
2MASS~J16314457-2402129 & 162.8$^{+0.47}_{-0.30}$ & 25.4$^{+0.67}_{-1.41}$ & 7.6 & 4.8$^\dagger$ & 8.4$^\dagger$ & 2.1 & Single/Smooth \\
ISO-Oph~106 & 46.2$^{+0.30}_{-0.37}$ & 38.9$^{+0.76}_{-0.44}$ & 3.9 & 3.0$^\dagger$ & 5.0$^\dagger$ & 1.3 & Single/Smooth \\
\hline
\end{longtable}
\twocolumn

\begin{figure*}
    \begin{center}
    \includegraphics[width=\linewidth]{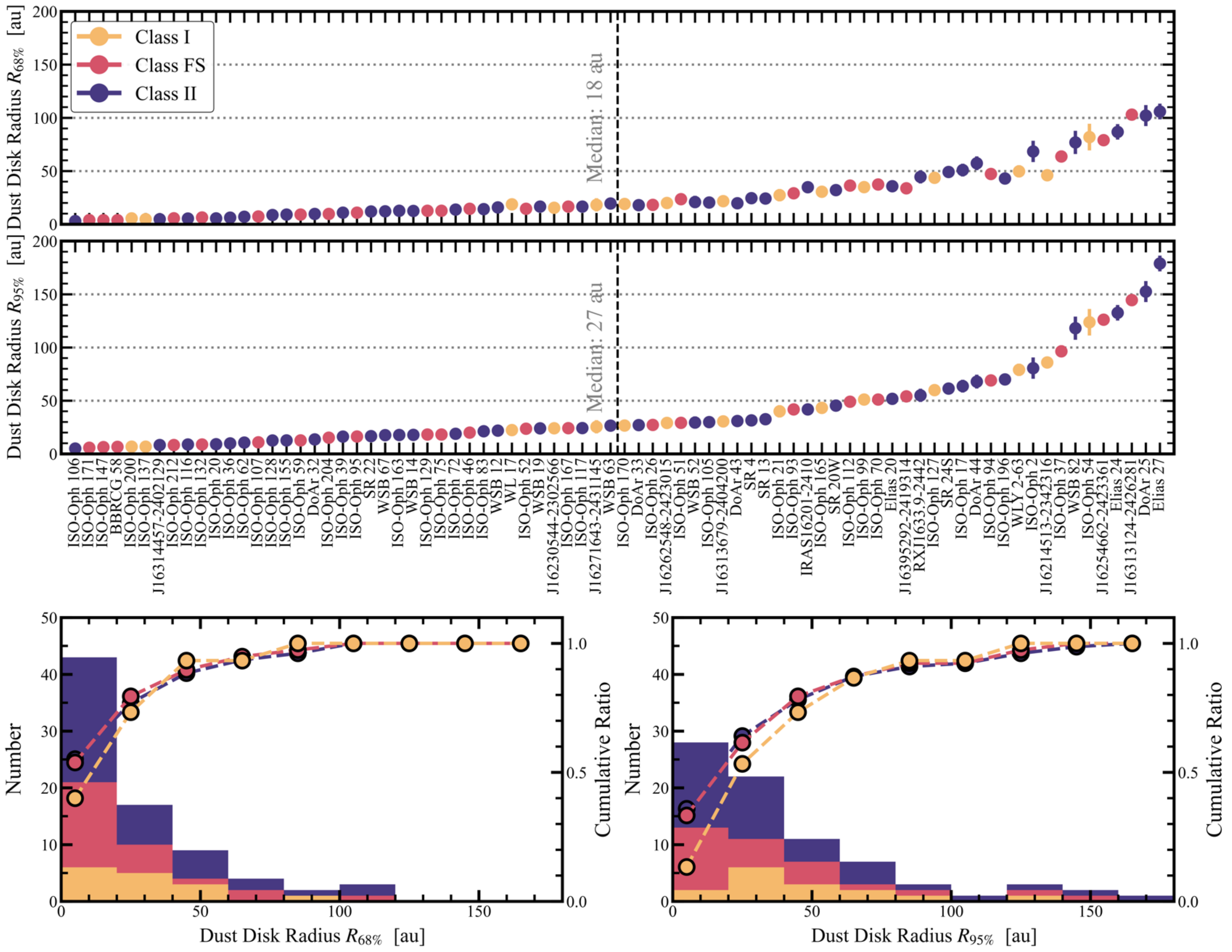}
    \end{center}
    \caption{
    (Top and middle) Dust disk radius ($R_{68\%}$ and $R_{95\%}$) for each source, shown in ascending order. The disk radii range from 3 to 106\,au with a median of 18\,au for $R_{68\%}$, and from 5 to 179\,au with a median of 27\,au for $R_{95\%}$.
    (Bottom) Histograms and cumulative density distribution of dust disk radii, showing that disks with radii less than 30\,au make up the majority of our sample. Disks with $R_{95\%} >$100\,au account for 10\% (7/78) of our targets.
    In all panels, circles, bars, and dashed lines in yellow, red, and violet represent Class I, FS, and II disks, respectively.
    }
    \label{fig:disk_radius}
\end{figure*}

\begin{figure*}
    \begin{center}
    \includegraphics[width=0.95\linewidth]{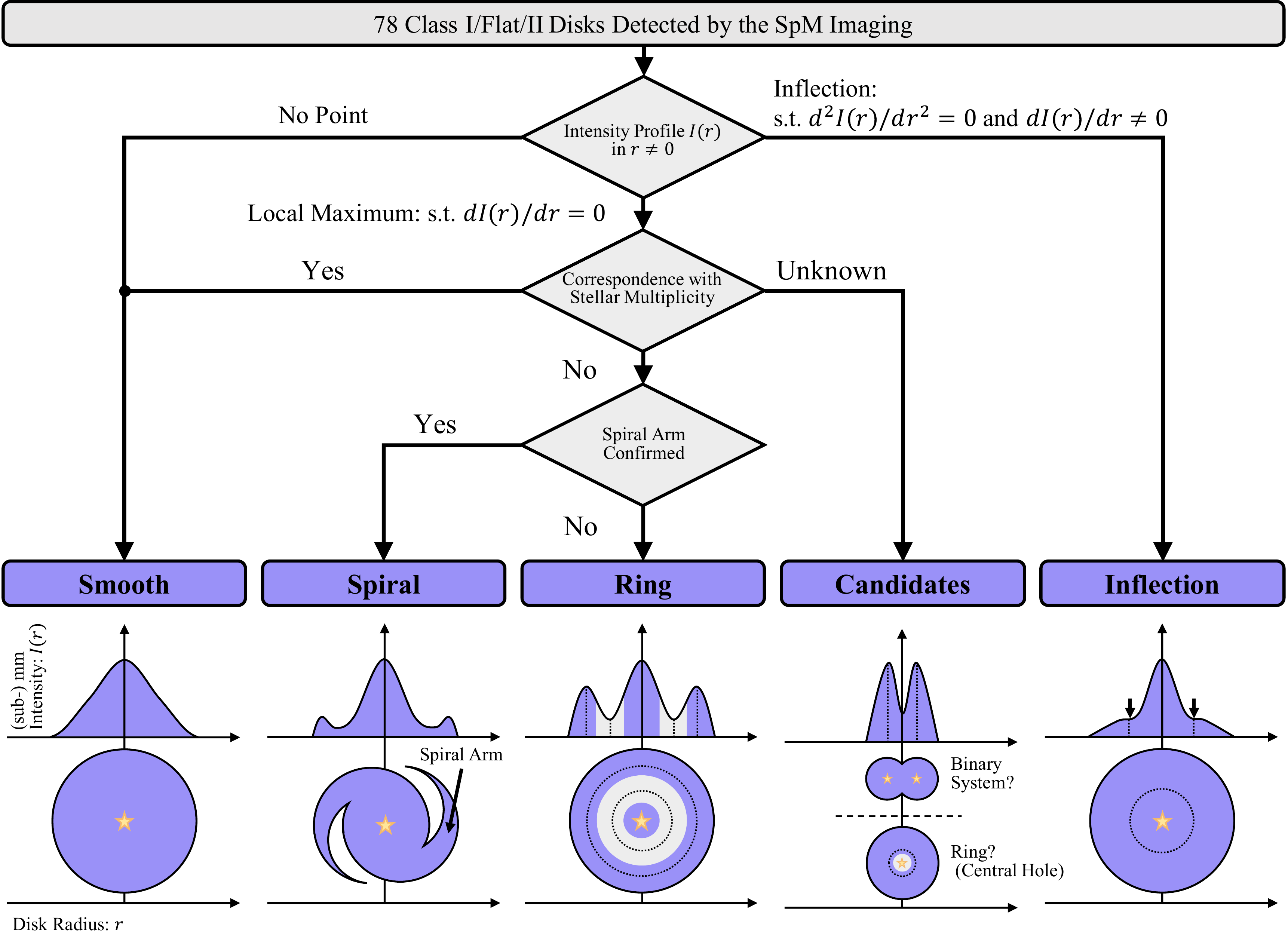}
    \end{center}
    \caption{
    Flow chart for determining (sub-)mm dust disk substructures. 
    We define five categories: `Spiral', `Ring', `Inflection', `Smooth', and `Candidates'. The fifth category, `Candidates', includes circumstellar disks around binary systems or disks with a ring (central hole) around a single star. 
    $I_{\rm radial}(r)$ and $I_{\rm major}$ represent the radial intensity profile and the profile along the major axis of the disk, respectively. 
    We classify the disks as follows: 43 `Smooth', 26 `Ring', one `Spiral', 5 `Inflection’, and 3 candidates for circumstellar disks around a binary system or a nearly edge-on disk with `Ring'.
    }
    \label{fig:how_to_categorize}
\end{figure*}

\begin{figure}[t]
    \begin{center}
    \includegraphics[width=\linewidth]{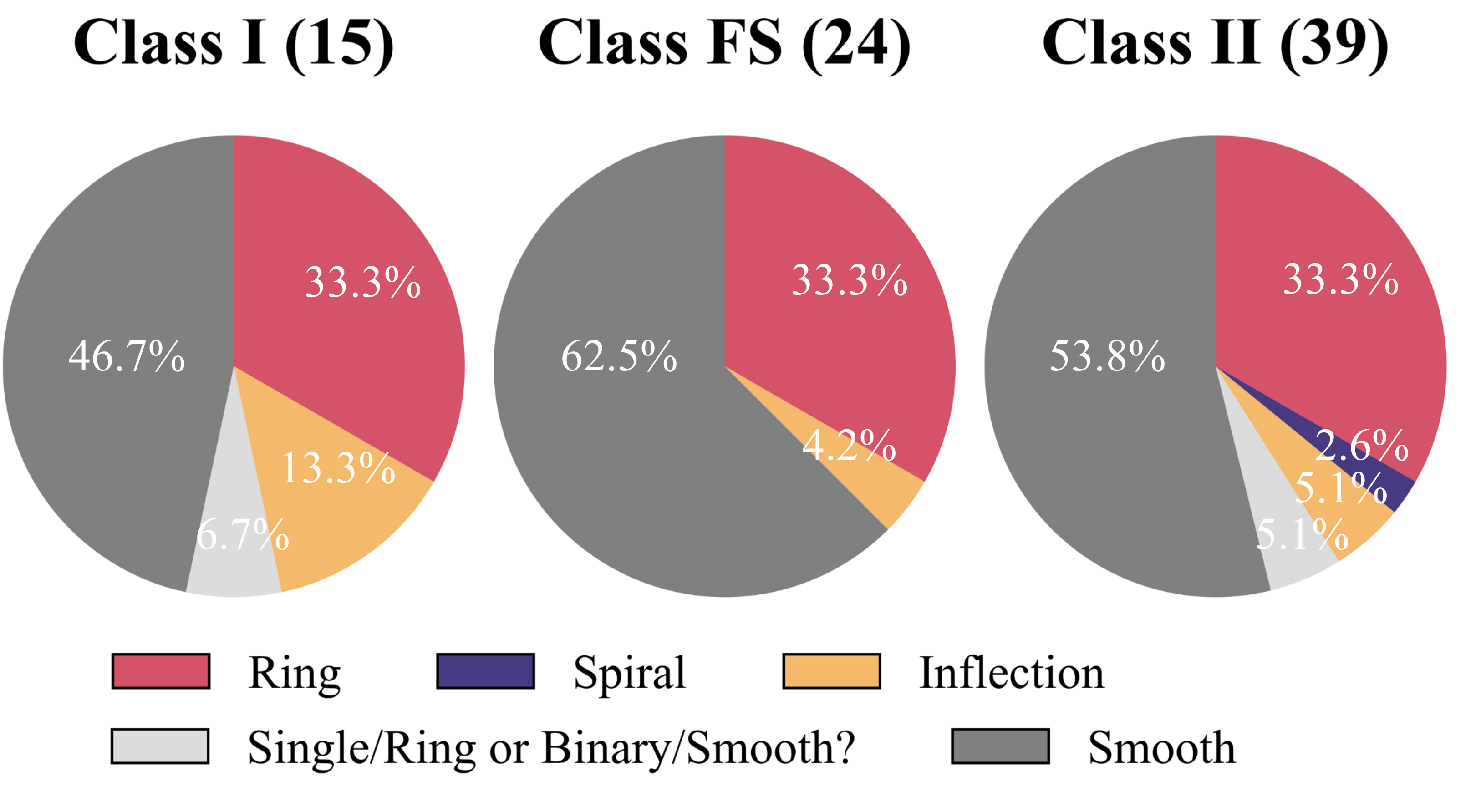}
    \end{center}
    \caption{Percentage distributions for Class I (left), FS (center), and Class II (right) disks. 
             The red, violet, and yellow regions in each panel correspond to disks identified as `Ring', `Spiral', and `Inflection', respectively. 
             The light gray sections represent disks categorized as the candidates for nearly edge-on disks with `Ring' features or circumstellar disks around binary systems. 
             The dark gray areas indicate disks with `Smooth' brightness distributions.
             }
    \label{fig:pie_subst}
\end{figure}

\begin{figure*}[ht]
    \centering
    \includegraphics[width=\linewidth]{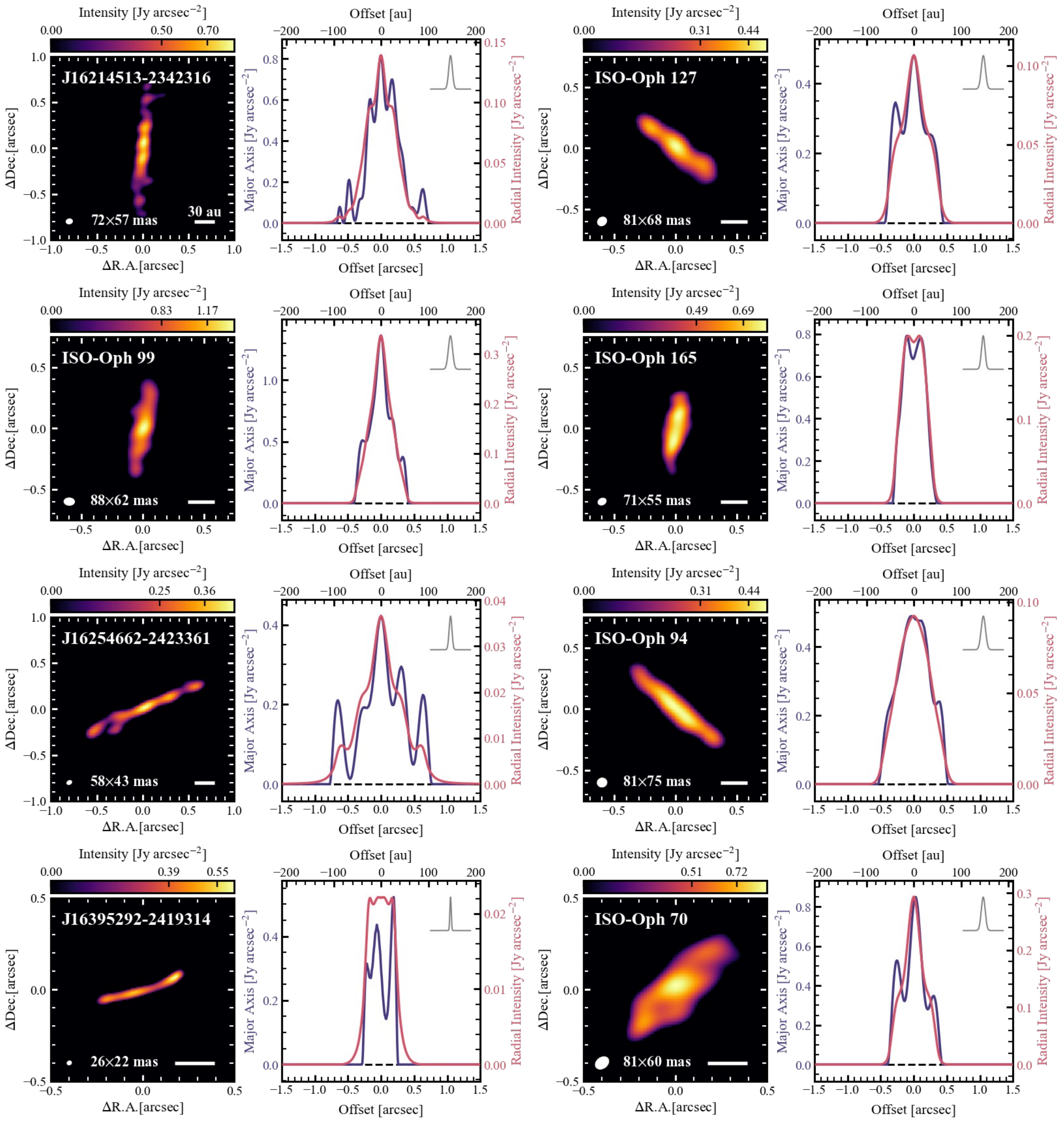}
    \caption{Gallery of SpM images and intensity profiles of 4 Class I and 4 Class FS disks with newly detected substructures.
    The continuum images are identical to those in Figures~\ref{fig:spm_classi}-\ref{fig:spm_classii}.
    The panel to the right of the continuum image displays the intensity profile: violet curves represent profiles along the major axis (aligned with the PA direction), and red curves show radial profiles averaged over all azimuthal angles. 
    Negative components of the red curves are linearly symmetrical to the positive ones. 
    Gray curves in the upper right indicate the effective spatial resolution $\theta_{\rm eff}$ of the SpM images.
    }
    \label{fig:new_vol1}
\end{figure*}

\begin{figure*}[ht]
    \centering
    \includegraphics[width=\linewidth]{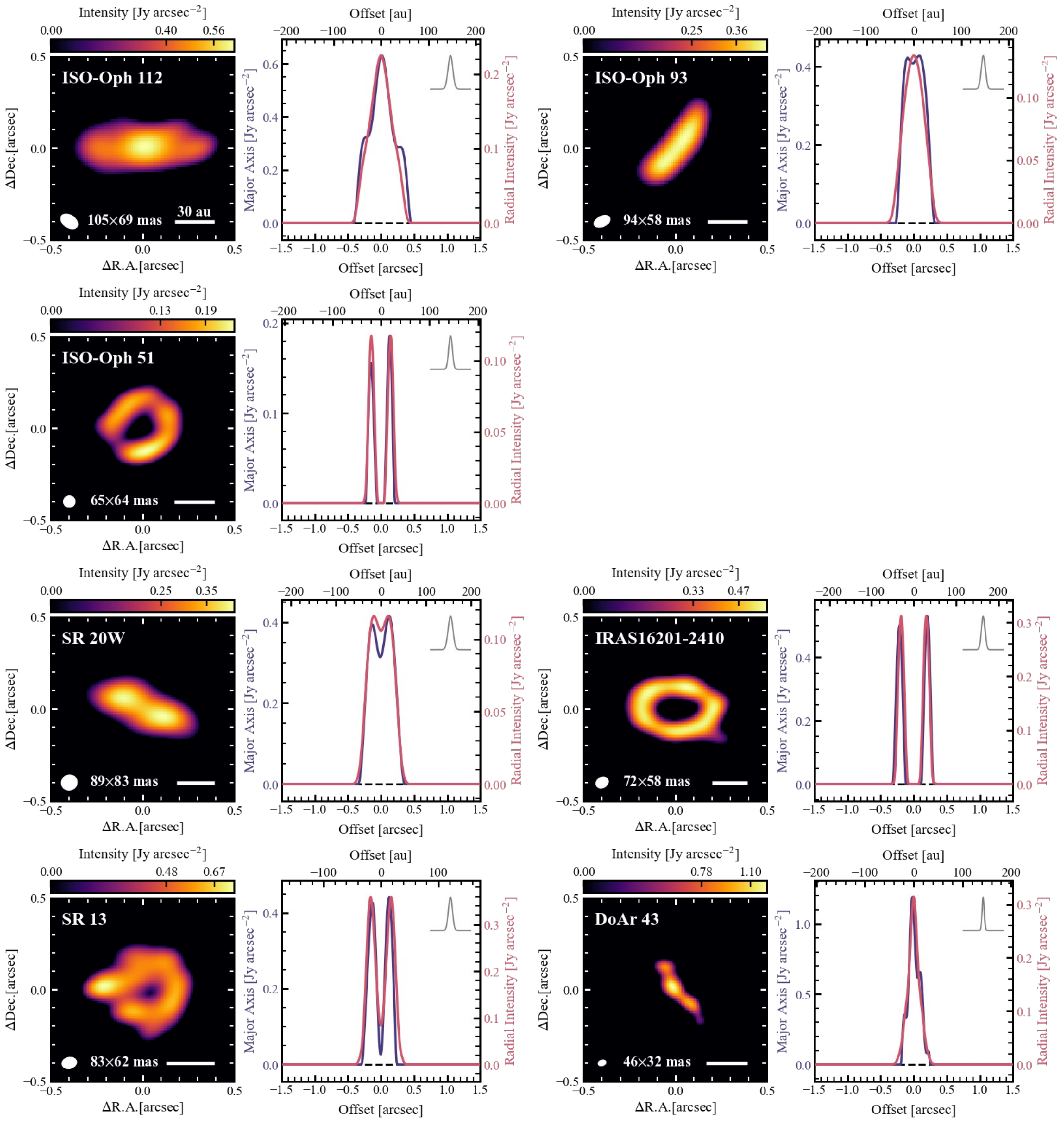}
    \caption{Same as Figures~\ref{fig:new_vol1} but for different 3 Class FS and 4 Class II disks with newly detected substructures.
    }
    \label{fig:new_vol2}
\end{figure*}

\section{Reconstructed SpM Images}\label{sec:spm_image}
Figures~\ref{fig:clean_classi}-\ref{fig:spm_classii} show galleries of the 78 disks imaged by CLEAN (Figs.\ref{fig:clean_classi}-\ref{fig:clean_classii}) and SpM (Figs.\ref{fig:spm_classi}-\ref{fig:spm_classii}).
Our sample consists of 70 single stars and the primary stars\footnote[3]{Note that a primary star in a binary system is regarded as brighter in optical observation or a more massive star in general.}, which are defined in this paper as ones with brighter intensities at (sub-)millimeter wavelengths, of 8 multiple systems (see Table~\ref{table:disk_info} and  \S\ref{subsec:categorization}).
In the SpM algorithm, no post-processing Gaussian convolution was applied, and the intensity units are maintained as Jy\,pixel$^{-1}$, differing from the units of CLEAN images, which are in Jy\,beam$^{-1}$.
For consistency, we use Jy\,arcsec$^{-2}$ as the intensity unit to match the units between the CLEAN and SpM images.
The spatial resolution in CLEAN images is about $0\farcs30\times0\farcs22$ (42$\times$31\,au).
The SpM images achieve an effective spatial resolution that is a median of 3.8 times higher than that of the CLEAN images.
We also confirm that the resolution ratios ($\theta_{\rm eff}/\theta_{\rm CLEAN}$) correlate positively with the disk sizes in the SpM images and the SNRs in 
the CLEAN images.
\citet{Yamaguchi_2024} shows a negative correlation between SNR and the resolution ratio, which differs from our result. 
The difference in the distribution of observational visibility causes the distinction in the correlation (for details, see \S A.\ref{subsec:imp_res}).

The SpM algorithm can also constrain the disk radii (see \S\ref{subsec:disk_properities}) of most detected disks by evaluating observation visibilities (for details, see \S A.\ref{subsec:fidelity_disk}).
For six disks (ISO-Oph~200, ISO-Oph~137, BBRCG~58, ISO-Oph~171, 2MASS~J16314457-2402129, and ISO-Oph~106) with radii less than 10\,au, the maximum baseline lengths of the visibilities are not sufficient to constrain the disk radii.
Considering their inclination angles and total fluxes and comparing the model visibilities from SpM images with the observed visibilities, we treated them as `very compact disks' that require additional long-baseline data to resolve their detailed characteristics (see Table~\ref{table:disk_info} and \S A.\ref{subsec:fidelity_disk}).
The disks around seven systems (Elias~27, DoAr~25, Elias~24, Elias~20, SR~4, WSB~52, and DoAr~33) were observed in the DSHARP project \citep{Andrews_2018_DSHARP} with a long baseline exceeding 10\,km.
Then, we compared the disk radii and brightness distributions between our study and those in the DSHARP project,  matching the 5\,au spatial resolution of the DSHARP project to the effective spatial resolution in the SpM images (see \S A.\ref{subsec:comp_dsharp}).
We confirmed almost the same brightness distributions and substructures in the SpM images as those of the high-resolution observations, which are generated from data with an observation time of less than one minute.
Thus, the SpM images shown in Figures~\ref{fig:spm_classi}-\ref{fig:spm_classii} data achieve high spatial resolution and fidelity in disk radius and substructure, which are used to extract disk characteristics in the following section.

\section{Identified Disk Characteristics}\label{sec:disk_characterustics}
In this section, we describe the disk characteristics based on their high spatial resolution images obtained with SpM. 
First, we estimate basic disk properties, such as position angle (PA), inclination angle, and disk radius, with a primary focus on the distribution of dust disk radius in \S\ref{subsec:disk_properities} (for inclination angle distribution, see \S\ref{subsec:bias_inc}).
Next, we describe the categorization of disk substructure types and summarize the statistics of each type across Class I to II in \S\ref{subsec:categorization}.
Finally, in \S\ref{subsec:new_substructures}, we provide details on the disks with substructures newly identified in this study.

\subsection{Measurements and Statistical Analysis of Disk Sizes}\label{subsec:disk_properities}
We first measured the position angle PA and inclination angle $i_{\rm disk}$.
We applied the Gaussian fitting with the Markov Chain Monte Carlo (MCMC) method (for details, see \S A.\ref{subsec:gauss_mcmc}) to all the SpM images, which are unaffected by the beam convolution and also fitted 13 disks with visually distinguishable ring structures (WL~17, ISO-Oph~51, Elias~24, WSB~82, ISO-Oph~2, ISO-Oph~196, DoAr~44, ISO-Oph~17, SR~24S, RXJ1633.9-2442, IRAS16201-2410, SR~13, and SR~4) with ellipses (see \cite{Yamaguchi_2021}).
These results of the ellipse fitting are consistent with those obtained from Gaussian fitting with MCMC, within a 10\% error.
We used measurements from Gaussian fitting with MCMC and ellipse fitting based on the deprojections with PA and $i_{\rm disk}$ of the brightness distributions.
The PA and $i_{\rm disk}$ values are summarized in Table~\ref{table:disk_info}.

Then, we measured disk radii $R_{68\%}$ and $R_{95\%}$, corresponding to the radii enclosing 68\% and 95\% of the flux density, using the curve-growth method (for details, see \S A.\ref{subsec:curve_growth}).
We estimated the uncertainty of disk radius as $\sigma_{\rm Radius}=\left<\theta_{\rm eff}\right>/2\sqrt{2\ln2}$, where $\left<\theta_{\rm eff}\right>$ is the geometric mean of the major and minor diameters in the effective spatial resolution, $\theta_{\rm eff}$.
Table~\ref{table:disk_info} lists the millimeter luminosity, $L_{\rm mm}=F_\nu\times\left(d/140\,{\rm pc}\right)^2$, which is the flux density corrected for a distance 140\,pc, along with $R_{68\%}$, $R_{95\%}$ and $\sigma_{\rm Radius}$.
Note that, for six disks categorized as compact disks, the radii $R_{68\%}$ and $R_{95\%}$ are used as the reference values because they require more long-baseline data to constrain their sizes (for details, see \S A.\ref{subsec:fidelity_disk}).
Figures~\ref{fig:spm_classi}-\ref{fig:spm_classii} show disks at each evolutionary stage, arranged by disk radii $R_{95\%}$.

Figure~\ref{fig:disk_radius} shows the disk radii and corresponding histograms for $R_{68\%}$ and $R_{95\%}$.
The $R_{68\%}$ and $R_{95\%}$ values range from 3 to 106\,au and from 5 to 179\,au, with medians of 18 and 27\,au, respectively. 
The histogram in Figure~\ref{fig:disk_radius} shows that 64\% of our targets have radii of $R_{95\%}\leq$40\,au, while only a few disks (9\%) have radii of $R_{95\%}\geq$100\,au.
Other observations also show the same trend that smaller disks are more frequent (e.g., \cite{Tobin_2020,Hsieh_2024}) and align with the theoretical prediction described by \citet{Tsukamoto_2020} and \citet{Yen_2024}.

Moreover, we can see that the cumulative density distributions are nearly identical across all evolutionary stages.
Other observations \citep{Tobin_2020,Hsieh_2024} also show similar radius distributions between Class I and FS disks.
We applied the Kolmogorov-Smirnov test (KS test; \cite{Wall_2012}) to all stage combinations, resulting in $p$-values greater than 0.10, which support the similarity in distributions.
Note that the observation time per source was less than one minute, which was not enough to detect weak emissions from the envelope surrounding the disk.
The radii of gas disks in the early evolutionary stage could be larger than measured, and we use dust disk radii in this study.
Considering this, our results suggest that disk radii may either decrease or remain unchanged as the disk evolves.
\citet{Dasgupta_2025}  shows similar results from ALMA Band 8 observations of Ophiuchus disks.

However, at face value, these observational findings seem to contradict theoretical studies based on various models, which indicate that the disk radius generally expands over time (e.g., the ballistic approximation model; \cite{Ulrich_1976,Cassen_1981}, $\alpha$-disk model; \cite{Shakura_1973} and non-ideal MHD models; e.g., \cite{Machida_2014,Tsukamoto_2015}). 
On the other hand, this trend toward matching disk radii is in line with the disk evolution model that incorporates MHD disk winds or radial dust drift. 
However, we note that this does not necessarily reject such theoretical predictions (e.g., \cite{Bai_2016,Birnstiel_2010}).
Their observational indication for dust continuum emission should be discussed carefully, including the growth and radial drift of dust particles.

\subsection{Categorization of Substructures}\label{subsec:categorization}
We identified characteristic substructures by analyzing both the intensity profile and the brightness distribution. 
Here, substructures refer to variations in brightness that either increase or decrease in intensity in a non-uniform manner. 
Figure~\ref{fig:how_to_categorize} provides an overview of our method for categorizing disk substructures in this study, which largely follows the categorization in \citet{Yamaguchi_2024}.
We employed two types of profiles: a radial profile and a profile along the major axis of the disk. 
For disks with low to moderate inclination, azimuthally averaged profiles are useful for detecting substructures since the averaging suppresses the noise level.
For disks with high inclination angles, the intrinsic asymmetry between major and minor axes may introduce some artifacts on the averaged profiles. 
Hence, it is necessary to use radial profiles in the major axes.
Using $i_{\rm disk}$ and PA derived in \S\ref{subsec:disk_properities}, we deprojected the brightness distribution to a face-on view, averaging it over all azimuthal angles to obtain the radial intensity profile $I_{\rm radial}(r)$, where $r$ is the disk radius.
In addition, by setting only the PA of the disk, we determined the intensity profile along its major axis, $I_{\rm major}(r)$, using the CASA viewer.

We summarize the radial intensity profiles $I_{\rm radial}(r)$, represented by red curves, and the profiles along the major-axis direction $I_{\rm major}(r)$, represented by violet curves, in \S A.\ref{sec:gallery_profile}.
Based on these two intensity profiles, $I_{\rm radial}(r)$ and $I_{\rm major}(r)$, we categorized the detected disks as either having a `Smooth' distribution or one of three distinct substructure types: `Ring', `Inflection', and `Spiral'.

We first explain the characteristics of the substructures.
A `Ring' is defined as a disk with multiple local maxima and minima, excluding $r=0$, in either $I_{\rm radial}(r)$ or $I_{\rm major}(r)$.
In the SpM images, clear rings can be seen in face-on views.
In addition, several symmetric components are visible in edge-on views, such as 2MASS~J16214513-234316, ISO-Oph~127, and 2MASS~J16254662-2423361, likely corresponding to ring edges where the intensity appears to overlap due to high inclination angles (for details, see \S\ref{subsec:new_substructures}).
The two peaks in the brightness distribution may represent the ring edge of a nearly edge-on disk with a ring structure or correspond to circumstellar disks around a close binary system (like `Candidates' shown in Figure~\ref{fig:how_to_categorize}).

To determine if it is a single star or not, we revist observations provided by SHARP I at the Max-Planck-Institute for Extraterrestrial Physics \citep{Ratzka_2005}, Keck NIRC2 \citep{Cheetham_2015, Ruíz-Rodríguez_2016}, VLT \& VLBA \citep{Loinard_2009}, and ALMA \citep{Cox_2017, Cieza_2019}.
We also refer to Gaia DR3 observations, which can detect the close pairs with a separation of $0\farcs18$–$0\farcs40$.
In the case of a disk where the Gaia DR3 observations show a single star and the separation between the two peaks in $I_{\rm major}(r)$ is greater than $0\farcs18$, we treat the disk as a nearly edge-on one with a `Ring' structure.
At the same time, a disk unsatisfied with this criterion is treated as candidates for either a ring-structured nearly edge-on disk or circumstellar disks around binary systems.

We also identify distinct distribution patterns that deviate from typical `Ring' structures. 
For instance, Elias~27 shows several asymmetric local peaks in the major intensity profile $I_{\rm major}(r)$, suggesting a spiraling structure toward the center rather than a clear ring in the SpM image. 
In this case, we classify Elias~27 as a `Spiral', consistent with observations at a spatial resolution of 5\,au \citep{Andrews_2018_DSHARP}.
For other objects, we observe disks with inflection points ($d^2I(r)/dr^2=0$ and $dI(r)/dr\neq0$) rather than local maxima in $I_{\rm radial}(r)$ or $I_{\rm major}(r)$. 
While these could represent `Ring' structures, we conservatively categorize them as `Inflection', based on their brightness distributions in the SpM images.

We finally focus on a disk with a single local peak at $r=0$ in $I_{\rm radial}(r)$ and $I_{\rm major}(r)$, which we define as a `Smooth' distribution. 
Among these `Smooth' disks, eight sources (2MASS~J16313679-2404200, 2MASS~J16271643-2431145, 2MASS~J16230544-2302566, ISO-Oph~52, ISO-Oph~75, WSB~19, WSB~12, and ISO-Oph~128) show slightly distorted distortion in the 2D SpM images, indicating the existence of some asymmetric structures.
The lack of observational data over long baselines may cause these distributions, potentially showing only a part of an asymmetric structure (for details, see \S A.\ref{subsec:comp_dsharp}). 

In other compact disks with $R_{95\%}<$15\,au, the maximum baseline length is insufficient to confirm the presence of a null point in the visibility profile, making it difficult to determine if ring-like substructures exist (see \S A.\ref{subsec:fidelity_disk}).
In this paper, we focus on clearly defined substructures based on $I_{\rm radial}(r)$ and $I_{\rm major}(r)$; hence, we note that `Smooth' disks may also contain hidden substructures.

Table~\ref{table:disk_info} describes the categorizations of host stars and disks.
We classify the disks as follows: 43 `Smooth', 26 `Ring', one `Spiral', and 5 `Inflection'.
In addition, there are 3 candidates for near edge-on disks with `Ring' or circumstellar disks around a binary system.
Figure~\ref{fig:pie_subst} shows the fraction of substructures identified in our analysis at each evolutionary stage.
The figure indicates that approximately 30-40\% of the disks display substructures across all stages.
Among these, 13 disks with ring structures have inclination angles greater than 60\,degrees, with Class I and FS disks comprising 85\% of them.
We suspect this may reflect a selection bias in the disk evolutionary stage, which is discussed further in \S\ref{subsec:bias_inc}.

With Figure~\ref{fig:pie_subst}, we describe the fraction of disks with rings and inner cavities.
We assume that candidates for disks with rings around single protostars or circumstellar disks in binary systems should be included in the 'Ring' category and that they have inner cavities.
The probabilities of Class I, Class FS, and Class II disks having ring structures are 40\%, 33\%, and 38\%, respectively, showing close agreement.
This suggests that the formation of ring structures may begin as early as the Class I stage.
In \S\ref{subsec:comp_edisk}, we discuss the comparison with the eDisk project and describe when the formation of  substructures, including rings, begins.

It should also be noted that the samples are limited to those resolved by SpM imaging, so the number of disks at each evolutionary stage is still limited. 
The numbers shown in Figure~\ref{fig:pie_subst} may still have considerable uncertainties.
Other spatially unresolved disks may alter these fractions.
The detection rates of substructures could increase when the systems evolve from Class I to Class II stages.

\subsection{New Substructure Candidates}\label{subsec:new_substructures}
In our categorization described in \S\ref{subsec:categorization}, we detected 32 disks with distinct substructures (`Ring', `Spiral', and `Inflection'), accounting for 41\% of all detected disks.
Excluding targets previously confirmed in high spatial resolution observations, such as DSHARP and ODISEA (e.g., \cite{Andrews_2018_DSHARP, Cieza_2021, Villenave_2022}), we identified 15 disks as new substructure candidates, including Class I (2MASS~J16214513-2342316, ISO-Oph~127, ISO-Oph~99, and ISO-Oph~165), Class FS (2MASS~J16254662-2423361, ISO-Oph~94, 2MASS~J16395292-2419314, ISO-Oph~70, ISO-Oph~112, ISO-Oph~93, and ISO-Oph~51), and Class II disks (SR~20W, IRAS16201-2410, SR~13, and DoAr~43), as shown in Figures~\ref{fig:new_vol1} and \ref{fig:new_vol2}.
In the following, we categorize these new detections and describe each in detail.

Most of the new detections (2MASS~J16214513-2342316, ISO-Oph~127, ISO-Oph~99, ISO-Oph~165, 2MASS~J16254662-2423361, ISO-Oph~94, 2MASS~J16395292-2419314, ISO-Oph~70, ISO-Oph~112, ISO-Oph~93, SR~20W, and DoAr~43) are nearly edge-on disks with bumpy brightness distributions.
Their intensity profiles $I_{\rm major}(r)$ along the major axes exhibit local maxima, excluding $r=0$.
Similar distributions have been confirmed in previous ALMA observations \citep{Alves_2020, Sai_2023}.
In particular, \citet{Alves_2020} showed that the 1.3\,mm dust continuum emission for [BHB2007]~1 revealed an inner disk surrounding the central star and a distinct ring-gap structure outside this inner disk, with two peaks corresponding to the ring edges.
For nearly edge-on disks with ring structures, material (or dust) at the ring edges overlaps along the line of sight, increasing the optical depth at the edges and enhancing the (dust thermal) emission.
Therefore, we classify disks with multiple peaks as belonging to the `Ring' category, as described above, and include them in the new detections.

We identified ISO-Oph~51 and IRAS16201-2410 as new candidates for transition disks in the ‘Ring` categorization, which is defined as disks with a clear ring structure and a cavity in the SpM images \citep{Yamaguchi_2024}.
The disk of ISO-Oph~51 has a cavity with a radius of 20\,au and shows an asymmetry with peak intensity on the south side.
The cavity radii of transition disks around Herbig Ae/Be and T Tauri stars (CIDA9, CS Cha, GM Aur, MHO2, PDS99, RXJ1852.3-3700, and Sz91) with a single ring structure, as categorized in \citet{Francis_2020}, range from 28 to 86\,au. 
In comparison, it is unusual for ISO-Oph~51, a young Class FS protostar, to be surrounded by a disk with both the smallest cavity and a ring structure, suggesting that substructure formation may begin during the accretion phase.
Meanwhile, the cavity radius of IRAS16201-2410 is 30\,au, which is comparable to those of CIDA9 and MHO2.
The brightness distribution of its disk is very similar to RXJ1633.9-2442, as seen in \citet{Cieza_2021} and this study.
Note that other observations with higher spatial resolution than that of the SpM images could detect inner disks within their cavities.
Previous ALMA observations with a spatial resolution of 5\,au revealed the inner disks around SR~4 and WL~17 \citep{Cieza_2021,Shoshi_2024}.
However, we could not confirm their inner disks in the SpM images due to the lack of long-baseline data used in this study (for details, see \S A.\ref{subsec:comp_dsharp}).

Furthermore, we confirmed the characteristic distribution of 1.3\,mm dust emission around SR~13. 
This source is a triple system in which a protostar (B) orbits central binary stars (Aa \& Ab) with an orbital period of $\sim400$\,years. 
The orbital positions and flux ratios were measured by \citet{Schaefer_2018} using adaptive optics imaging at the Keck Observatory.
In this study, we identified, for the first time, both the circumbinary disk with a ring-gap structure and the disk around SR~13B.
Note that the single star and its disk are located at the peak intensity on the northeast side relative to the two central stars.
\citet{Kennedy_2019} presented a similar case in the quadruple system HD~98800, where one binary orbits the other with an orbital period of $251$\,years. However, they could detect only the central circumbinary disk with a ring structure.
Thus, this is the first case confirming both a ring-shaped disk and a circumbinary disk orbiting the two central stars, providing valuable insights into disk-disk interactions.

Additionally, we found a candidate binary system from the SpM images. 
The ALMA observation of ISO-Oph~147, with a spatial resolution of 14-18\,au, showed an unresolved disk with two peaks, as reported in \citet{Hsieh_2024}. 
However, we resolved the disk around ISO-Oph~147 into two distinct dust emissions, separated by $0\farcs17$ (ISO-Oph~147a and ISO-Oph~147b).
This result is consistent with the near-infrared observation at the VLT through the L' filter \citep{Dunchêne_2007}, and we consider ISO-Oph~147 a binary system. 
Note that the weak emission from these disks may not have been fully resolved, as the observation time was less than one minute.

\begin{figure*}[t]
    \begin{center}
    \includegraphics[width=0.90\linewidth]{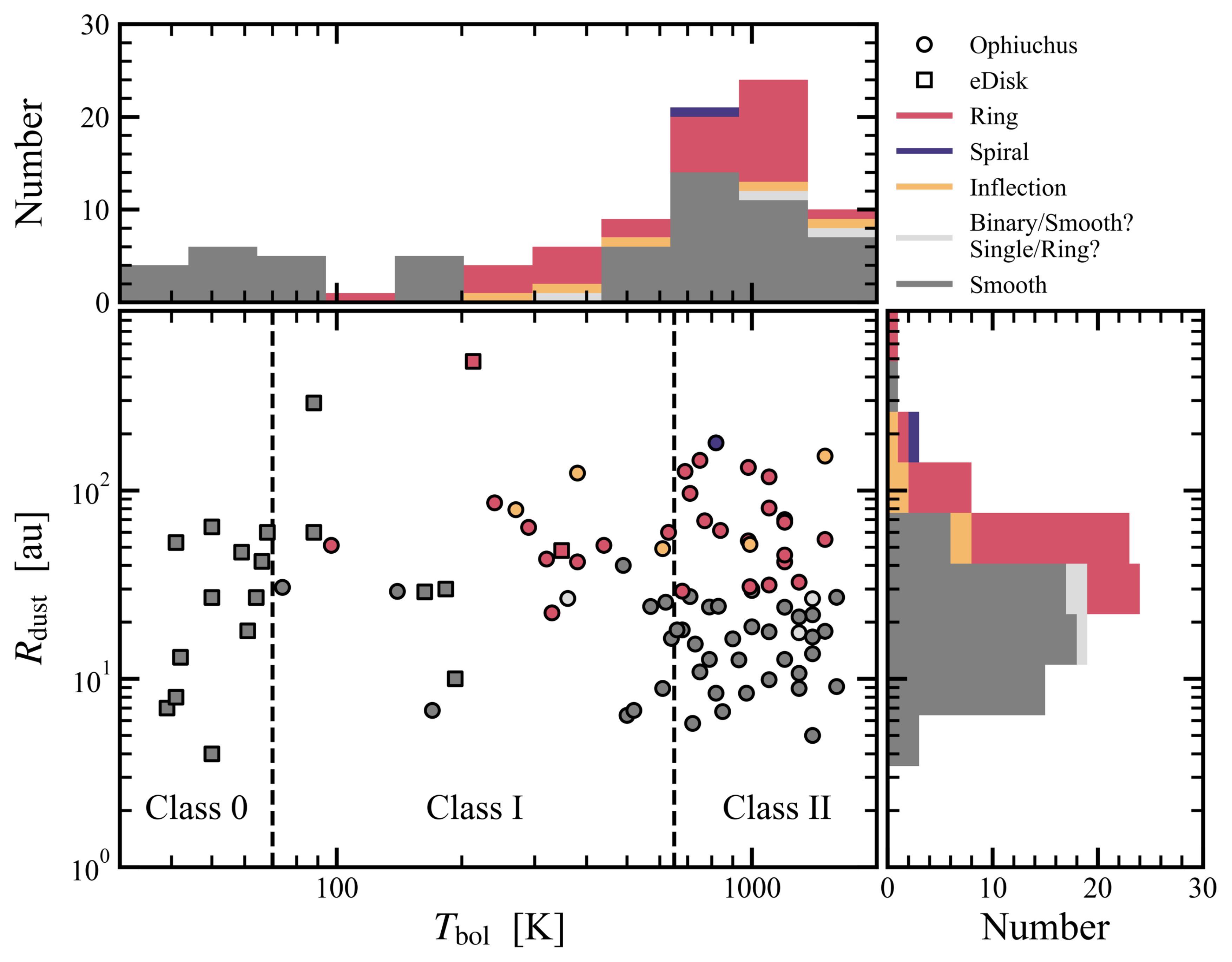}
    \end{center}
    \caption{
    Relationship between bolometric temperature and dust disk radius, including $R_{95\%}$ of 76 Ophiuchus disks (circles) from this study and $R_{\rm disk}$ of 19 eDisk samples (squares) as described in \citet{Yen_2024_edisk}.
    The red, violet, yellow, light gray, and dark gray colors represent disks categorized as `Ring', `Spiral’, `Inflection’ and the candidates for nearly edge-on disks with `Ring' features or circumstellar disks around binary systems, and `Smooth' brightness distributions, respectively.
    The top and right panels show the histograms of $T_{\rm bol}$ and $R_{\rm dust}$ with eleven bins spanning their minimum and maximum values.
    }
    \label{fig:Tbol_Rdust}
\end{figure*}

\begin{figure*}[t]
    \begin{center}
    \includegraphics[width=0.77\linewidth]{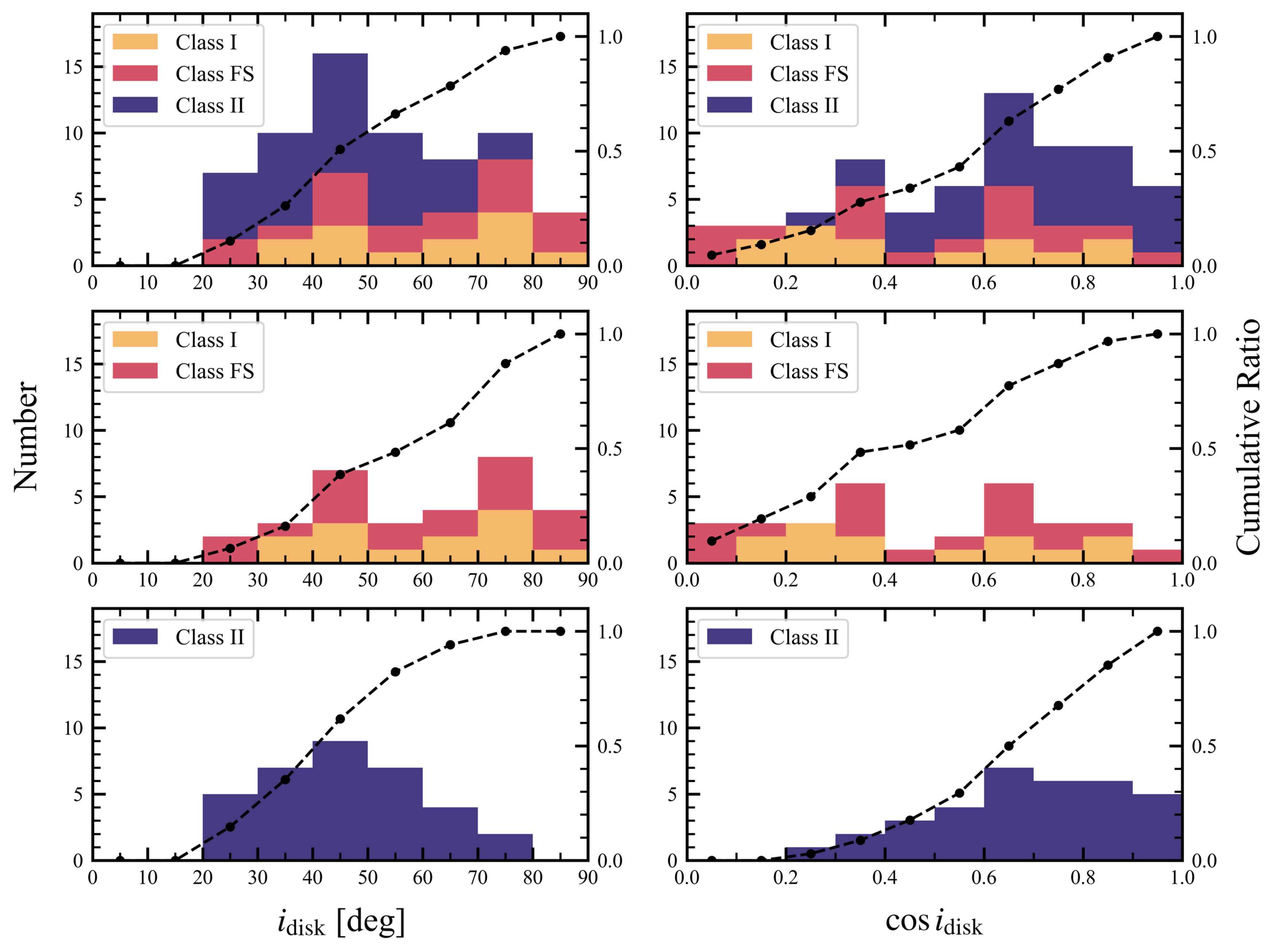}
    \end{center}
    \caption{
    Histograms of inclination angles $i_{\rm disk}$ (left) and $\cos\,i_{\rm disk}$ (right).
    All disks (top), Class I/FS disks (middle), and Class II disks only (bottom) are plotted.
    The black dashed line in each panel represents the cumulative density distribution.
    }
    \label{fig:hist_inc_cos}
    \begin{center}
    \includegraphics[width=0.77\linewidth]{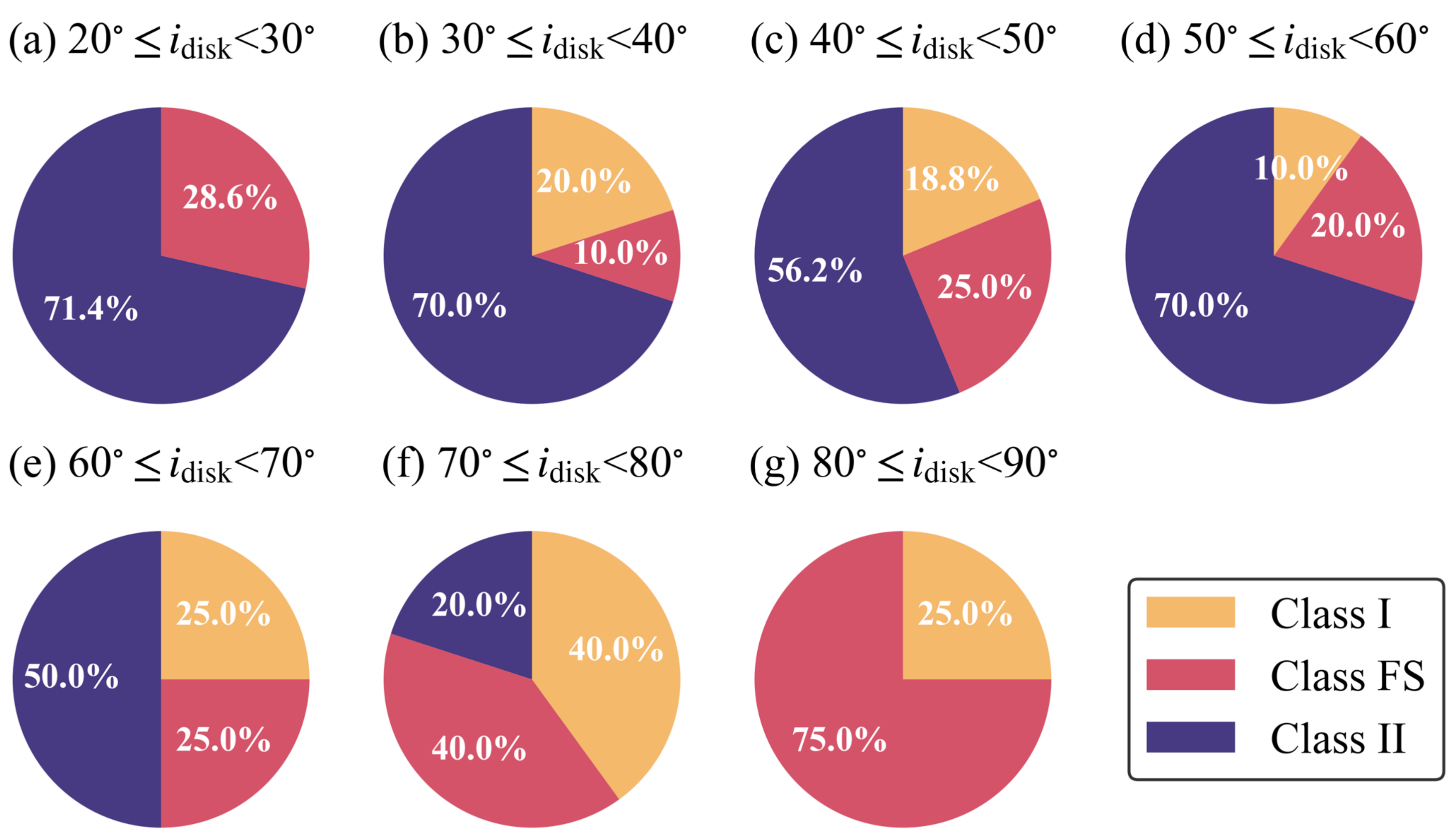}
    \end{center}
    \caption{
     Pie charts of disk classification by inclination angles $i_{\rm disk}$, ranging from 20 to 90\,degrees in 10\,degree increments.
     The number of Class I, FS, and Class II disks in each chart is referenced in Figure~\ref{fig:hist_inc_cos}.
    }
    \label{fig:pie_inc}
\end{figure*}

\section{Discussion}\label{sec:discussion}
We have applied SpM imaging to ALMA archival data for the Ophiuchus disks, achieving images with spatial resolutions 1.2-15 times higher than those obtained by the conventional CLEAN method for 78 disks.
As a result, we identified substructures categorized as `Ring', `Spiral', and `Inflection' in 32 disks, 15 of which are newly identified in this study.
We also confirmed 14 Class I and FS disks with ring structures, indicating the ubiquity of substructures regardless of disk evolutionary stage.
These findings could be related to dust growth and planet formation in the early evolutionary stages, which is crucial for identifying when planet formation begins.
To deepen our understanding of the results, we compare the findings of this study with those of the eDisk project in \S\ref{subsec:comp_edisk} and discuss the possibility of misclassification in disk evolutionary stages in \S\ref{subsec:bias_inc}.
In \S\ref{subsec:indentify_youth}, we describe how to identify disks in truly early evolutionary stages from SED misclassification.

\subsection{Comparison with Targets in eDisk Project}\label{subsec:comp_edisk}
The eDisk is an ALMA large project observing disks around 12 Class 0 and 7 Class I protostars in nearby star-forming regions, with high sensitivity and a spatial resolution of 7\,au (e.g., \cite{Ohashi_2023}).
The targets were selected from Class 0/I disks classified by bolometric temperature $T_{\rm bol}$.
They are located within 200 pc from the Sun and have relatively high luminosities ($L_{\rm bol}>0.1\,L_\odot$).
Thus, these targets are suitable for achieving sufficient sensitivity for continuum and molecular line emission observations within $\sim100$ hours of observation time (for details, see Subsection~3.1 in \cite{Ohashi_2023}).
One of the project aims is to identify disk substructures thought to be formed by planets and to determine when and how planet formation begins.
They showed that while the continuum emission for L1489~IRAS and Oph~IRS63 revealed ring-gap and shoulder structures \citep{Yamato_2023,Flores_2023}, no clear substructures were observed in other sources.
Note that \citet{Segura-Cox_2020} presented a disk with multiple ring gaps around Oph~IRS63.
Other papers from the same project (e.g., \cite{Kido_2023}) indicated that most samples without substructures exhibit asymmetric structures along the major or minor axes.
The eDisk project concluded that disks around Class 0/I protostars exhibit relatively few characteristic substructures compared to Class II disks and that substructures may rapidly develop as the system evolves from Class I to Class II stages \citep{Ohashi_2023}.

As in the eDisk project, we focused on Class I and FS disks in the accretion phase, located in the nearby Ophiuchus molecular cloud.
The detection rates of substructures differ significantly between the eDisk project and this study.
We confirmed substructures in 30-35\% of the Class I and FS disks.
We consider that the difference between these studies is primarily caused by sample selection.
The eDisk project applied a classification using $T_{\rm bol}$ \citep{Evans_2009} because Class 0 protostars, embedded in dust and gas, are optically thick, which prevents stellar blackbody radiation from reaching observers.
On the other hand, we used the classification based on the spectral slope between near- and mid-infrared to select our samples.
To compare the disks selected in both the eDisk project and this study, we reclassified our targets based on $T_{\rm bol}$.

Figure~\ref{fig:Tbol_Rdust} shows the relationship between bolometric temperature $T_{\rm bol}$ and dust disk radius $R_{\rm dust}$ for the samples from the eDisk project and this study.
We selected 19 disks from the eDisk sample. 
They are either disks around single stars or the brightest ones of multiple disks. 
Note that the eDisk samples are located in various star-forming regions, including Ophiuchus (for details, see \cite{Ohashi_2023}).
The bolometric temperatures $T_{\rm bol}$ and dust disk radii $R_{\rm disk}$ for the eDisk samples are taken from \citet{Yen_2024_edisk}.
$R_{\rm dust}$ in Figure~\ref{fig:Tbol_Rdust} consists of the dust disk radii $R_{\rm disk}$ from \citet{Yen_2024_edisk} and $R_{95\%}$ of 76 Ophiuchus disks presented in this study, excluding ISO-Oph~52 and ISO-Oph~46.
Using bolometric temperature $T_{\rm bol}$, we classified our samples into 22 Class I (70~K$\leq T_{\rm bol}\leq$650~K) and 54 Class II (650$< T_{\rm bol}\leq$2800\,K) disks.
The increase of the number of Class II objects largely comes from those originally identified as Class FS falling into Class II category in the classification based on $T_{\rm bol}$.
\citet{Evans_2009,Evans_2009_disk} suggest to a range of of $T_{\rm bol}$ from 350\,K to 950\,K.
For our sample, this range of $T_{\rm bol}$ corresponds to all Class FS objects except for 2MASS~J16395292-2419314 (see \S A.~\ref{sec:IRslope} for the comparison between classifications based on spectral slope and bolometric temperature).
If we consider this Class FS category based on $T_{\rm bol}$, the number of disks in each evolutionary stage categorized based on the infrared spectral slope still holds, at least qualitatively.
In Figure~\ref{fig:Tbol_Rdust}, the eDisk samples are distributed in the lower $T_{\rm bol}$ range, while our samples are more concentrated in the higher range.
This suggests that the eDisk samples are in an early accretion phase, whereas our study focuses on disks in a later accretion phase.

We focus on the eDisk and our sample with substructures (shown in red in Figure~\ref{fig:Tbol_Rdust}). 
Interestingly, many disks with substructures, except for ISO-Oph~99, appear in the region where $T_{\rm bol}$ exceeds 200-300\,K, and their disk radius in this study is larger than the median of 27\,au.
This trend arises due to the identification of numerous substructures in the SpM images, which achieve spatial resolution nearly equivalent to that of the eDisk project.
However, we need to consider the following two caveats.
The first caveat involves the limitations of the maximum baseline length, uncertainties in radius estimation, and challenges in detecting substructures in some compact disks with radii smaller than 15-20\,au (see \S A.\ref{subsec:fidelity_disk}).
The second caveat is the insufficient number of samples with $T_{\rm bol}$ values between 100 and 300\,K, which link our sample to the eDisk sample. 
Taking these factors into account, the number of samples may reduce the threshold values of $T_{\rm bol}$ and $R_{\rm dust}$.
At the very least, combining our samples with the eDisk samples suggests that substructures are more easily detected in disks with relatively larger dust disk radii $R_{\rm dust}$ and higher bolometric temperatures $T_{\rm bol}$.
This could be a contributing factor to the differing detection rates of substructures between our study and the eDisk project.

\subsection{Selection Bias of Disk Evolutionary Stages}\label{subsec:bias_inc}
In the previous subsection, we focused on the differences in disk substructure detection rates between the eDisk project and this study. 
In this subsection, however, we describe the similarities in disk shape or inclination angle in the Class I and FS stages; both samples tend to show elongated disks.

Figure~\ref{fig:hist_inc_cos} represents the histograms of inclination angles $i_{\rm disk}$ and $\cos i_{\rm disk}$ measured in \S\ref{subsec:disk_properities}.
We note that the inclination angles were measured from the SpM images that were unaffected by beam convolution.
We used 65 disks (13 Class I, 18 Class FS, and 34 Class II) whose diameters, considered as twice the disk radii $2\times R_{95\%}$, are at least three times larger than the major axes of the effective spatial resolution $\theta_{\rm eff}$.
The histogram of $i_{\rm disk}$ for Class I and FS disks, shown in the middle left panel of Figure~\ref{fig:hist_inc_cos}, has two peaks at 40-50 and 70-80\,degrees.
On the other hand, we can see the low abundance of Class II disks with $i_{\rm disk}\geq 70$\,degrees.
Figure~\ref{fig:pie_inc} shows pie charts of the classification percentage for different ranges of inclination angles.
In the range of $i_{\rm disk}\geq 60$\,degrees, Class I and FS disks account for more than 50\%.
We performed a KS test \citep{Wall_2012} on the $i_{\rm disk}$ distributions of Class I and FS compared to Class II disks.
The $p$-value is 3.2$\times10^{-3}$, indicating that the distribution of Class I and FS disks differs from that of Class II disks.

If the disk inclination angles are randomly distributed, the distribution of $\cos i_{\rm disk}$ should exhibit a flat histogram or a linear cumulative distribution.
To test this, we prepared sample data with equal numbers of Class I/FS and Class II disks.
In this sample, $\cos i_{\rm disk}$ was randomly distributed between 0 and 1.
We then compared it with $\cos i_{\rm disk}$ for Class I/FS and Class II disks using the KS test.
We adopted the Monte Carlo routine to repeat this analysis ten thousand times to avoid uncertainty caused by the selection of the sample data.
After that, we also applied a Gaussian fitting to the histogram of $p$-values for the Class I and FS disks.
We found that the mean value of the $p$-values is 0.76, indicating that we cannot reject the possibility that $\cos i_{\rm disk}$ of Class I and FS disks are randomly distributed.
For Class II disks, the $p$-value is below the significance level of 0.05 for our seven thousand iterations out of the ten thousand iterations.
This means that the distribution of Class II disks is not consistent with a random distribution.
We can see the lack of low values of $\cos i_{\rm disk}$ in the bottom left panel of Figure~\ref{fig:hist_inc_cos}.

Some previous studies, including the eDisk project, show a similar distribution in inclination angles for Class I and FS disks.
The eDisk project observed 9 Class 0/I disks with inclination angles greater than 70\,degrees out of 22 samples, including binary system companions.
The VANDAM survey observed Class 0/I disks in the Orion star-forming region with a spatial resolution of 32-40\,au and showed a histogram of inclination angles peaking at 60\,degrees (see Figure~13 in \cite{Tobin_2020}).
Similarly, the CAMPOS project \citep{Hsieh_2024} observed Class 0/I (FS) disks in seven nearby star-forming regions with a spatial resolution of 14-18\,au, showing a histogram of inclination angles peaking at 64-68\,degrees (see Figure~11 in \cite{Hsieh_2024}).
Despite inclination angles measured from CLEAN images with beam convolution, their distributions are consistent with our results for Class I and FS disks from SpM images without beam convolution.

We discuss the contamination due to the misclassification of disk evolutionary stages as a possible explanation for the inclination bias in Class II distribution.
In \citet{Furlan_2016}, a simple protostellar model was adopted, including a disk within an infalling envelope with outflow cavities, to generate thirty thousand model SEDs and determine the best-fit parameters for each protostar.
They showed that, in some models, a low inclination angle results in a flatter SED overall in the near- and mid-infrared wavelength regions.
In contrast, an increased inclination angle enhances the silicate absorption feature at 10\,$\mu$m and a steeper slope beyond this point.
This occurs because a high inclination angle increases the column density, raising the optical depth along the line of sight and absorbing near-infrared emission.

Envelope extinction does not affect emission at wavelengths from the far-infrared to millimeter range, so the inclination angle does not impact the SED in these ranges.
In cases of decreased near-infrared emission, simple classification using two infrared wavelengths could cause a protostar with a nearly edge-on disk to be classified as younger than it actually is.
\citet{Masunaga_2000} also pointed out that the contamination of evolved edge-on sources into earlier evolutionary stages is non-negligible.
The detected nearly edge-on disks in this work may correspond to the misclassification and may have evolved to at least later stages than their classification suggests.
Thus, the misclassification of disks with high inclination angles can result in contamination from different evolutionary stages.

The effect described in \citet{Furlan_2016} could influence the distribution of inclination angles for disks in all evolutionary stages.
However, we see that the distribution of the inclination angles for Class I and FS disks is not inconsistent with the random distribution (Figure~\ref{fig:hist_inc_cos}).  
Thus, the contamination from highly inclined Class II disks does not seem to cause any excess in the inclination angle distribution of Class I and FS disks.
The dust emission from truly young, nearly edge-on disks in Class I and FS stages could be too weak and embedded in the surrounding gas, making them unobservable.
As a result, despite including the contamination from the Class II stage, we can see the nearly unbiased distribution of Class I and FS disks.
The lack of high inclination systems for Class II objects suggests that there is no contamination from high inclination Class III objects to Class II disks.
Our Class II sources are limited  to relatively bright ones in the K band (for details, see \S\ref{subsec:obs_details} and \cite{Cieza_2019}).
Therefore, unselected Class II objects may be affected by contamination from highly inclined Class III objects.
We note that verifying our hypothesis requires additional studies of Class I and FS disks with large inclination angles, which cannot be conclusively determined from the current dust continuum emission data alone.
We describe the detailed approach in \S~\ref{subsec:indentify_youth}.

\subsection{Approaches for Identifying Disks in Truly Young Evolutionary Stages}\label{subsec:indentify_youth}
In previous subsections, we discussed characteristic substructures and evolutionary stages of disks, and we concluded that nearly edge-on disks in the accretion phase are related to the misclassification in SED.
\citet{Villenave_2022} showed that the Class FS protostar 2MASS~J16313124-2426281 (one of our targets) has a ring-shaped, nearly edge-on disk ($i_{\rm disk}$=84 degrees).
They also confirmed the presence of $^{12}$CO $J$=2-1 molecular line emission, confirming the existence of a gas disk with Keplerian rotation.
For this object,  they could not detect any protostellar outflow, suggesting that ejection and accretion do not occur significantly.
In other words, the disk is in a later evolutionary stage, and the accretion stage is already complete.
However, a nearly edge-on Class I disk with both a substructure and accretion phenomenon exists, although its properties are influenced by the inclination angle.
The Class I protostar L1489~IRS, one of the eDisk samples located in the Taurus molecular cloud, has a nearly edge-on disk ($i_{\rm disk}$=71\,degrees) with a ring-gap structure \citep{Sai_2020,Yamato_2023}.
Past observations of CO molecular line emission have identified an envelope around the disk and an outflow driven by the disk \citep{Yamato_2023, Tamura_1991, Hogerheijde_1998}.
According to the difference in the observation results of the two sources, it is possible to determine whether the object is in the accretion phase in more detail by observing the molecular emission lines than the categorization based on $T_{\rm bol}$ or the spectral slope.

Moreover, disks with substructures in the accretion stage are also important for constraining the timescale of substructure formation. \citet{Shoshi_2024} showed that the blue- and red-shifted outflows, observed in the $^{12}$CO $J$=2-1 line emission, are associated with the ring-shaped disk around the Class I protostar WL~17 (one of our targets). 
Based on the disk mass and the dynamical timescale of the outflows, they derived the mass accretion rate onto the disk and suggested rapid substructure formation within $\sim 10^4$ years.
It is crucial to analyze molecular line emissions around nearly edge-on disks with substructures. 
This analysis helps identify tracers of star-formation phenomena or objects in the accretion phase, such as protostellar outflows or infalling envelopes, and confirm that they are indeed in the accretion stage.
The disk mass and substructure growth timescale would also provide more detailed information on the evolutionary stage than the spectral slope or bolometric temperature measured from the SED. 
This analysis is similarly applicable to face-on disks with clear ring structures, such as ISO-Oph~51.

\section{Summary and Future Works}\label{sec:summary} 
We use ALMA archival Band 6 continuum data to present 2D super-resolution imaging of 78 Ophiuchus disks, comprising 15 Class I, 24 FS, and 39 Class II disks. 
We employ a 2D super-resolution imaging technique based on Sparse Modeling (SpM), which produces images with high fidelity to observed visibility and enhanced spatial resolution. 

Our main findings are summarized as follows:

\begin{enumerate}
\item 
All dust disks show an improvement in spatial resolution by a median factor of 3.8 compared to the conventional CLEAN method. 
Except for six sources (ISO-Oph200, ISO-Oph137, BBRCG58, ISO-Oph171, 2MASSJ16314457-2402129, and ISO-Oph106), they are successfully spatially resolved, allowing us to constrain their dust disk radii. 
The radii range from 5 to 179\,au, with a median radius of 27\,au.
\item 
Based on the intensity profiles, we classified 43 disks as ‘Smooth,’ 26 as ‘Ring,’ one as ‘Spiral,’ and 5 as ‘Inflection,’ in addition to 3 candidates of nearly edge-on disks with ‘Ring’ or circumstellar disks around a binary system.
We confirmed substructures in approximately 30-40\% of disks across all evolutionary stages. 
This trend could only be confirmed for disks detected by SpM imaging, as spatially unresolved disks could alter these percentages.
\item 
Except for objects from previous high spatial resolution observations, we identified 15 disks with new substructures: Class I 
(2MASS~J16214513-2342316, ISO-Oph~127, ISO-Oph~99, and ISO-Oph~165), Class FS (2MASS~J16254662-2423361, ISO-Oph~94, 2MASS~J16395292-2419314, ISO-Oph~70, ISO-Oph~112, ISO-Oph~93 and ISO-Oph~51) and Class II disks (SR~20W, IRAS16201-2410, SR~13, and DoAr~43).
\item 
Many new detections are nearly edge-on disks with rings, displaying bumpy brightness distributions in their intensity profiles along the major axes. 
ISO-Oph~51 and IRAS16201-2410 are also new candidates for transition disks with a clear ring structure and a cavity. 
In the case of the triple system SR~13, we have, for the first time, confirmed both a ring-shaped circumbinary disk and an additional disk orbiting a companion star that is distant from the central two stars.
\item 
Compared to the eDisk samples, our targets are in relatively late accretion stages. 
In addition, the combination of eDisk samples and ours indicate that substructures may emerge when $T_{\rm bol}\gtrsim$200-300\,K and $R_{\rm dust}\gtrsim$27\,au.
This finding could explain the difference in substructure detection rates between our study and the eDisk project.
\item 
We confirmed different distributions between Class I/FS and Class II disks, which show that the Class II distribution lacks inclination angles larger than 70\,degrees.
The low existence could be related to the misclassification of SED.
We need more detailed observations of molecular lines around Class I and FS disks with large inclination angles.
\end{enumerate}

Utilizing stellar and disk properties measured in \S\ref{subsec:disk_properities} is valuable for gaining a deeper understanding of the disk formation process.
The SpM images in Figures~\ref{fig:spm_classi}-\ref{fig:spm_classii}, achieve higher spatial resolutions than the conventional CLEAN method and are free from beam convolution.
This enables us to resolve small disks and obtain detailed information about many disks. 
For example, we can measure the width and depth of rings, which are related to planet mass \citep{Yamaguchi_2024}. By combining these disk properties with protostellar quantities identified in previous studies (e.g., \cite{Evans_2009}), we aim to discuss the disk evolutionary process, including stellar-disk and planet-disk interactions. 
This project is progressing, and we will summarize the results in our subsequent paper.

\begin{ack}
The authors thank the anonymous referee for all of the comments and advice that helped improve the manuscript and the contents of this study.
This work is part of the ASIAA Summer Student Program 2023, and I appreciate the support of the Academia Sinica Institute of Astronomy and Astrophysics.
The authors thank Dr. Takeshi Nakazato and Dr. Shiro Ikeda for their technical support, and Dr. Munetake Momose for giving valuable comments.
This work was supported by a NAOJ ALMA Scientific Research grant (No. 2022-22B; MNM) and by JSPS KAKENHI 20K04017 (TT), 24K07097 (TT) and 23K03463 (TM).
M.Y. acknowledge support from the National Science and Technology Council (NSTC) of Taiwan with grant NSTC 112-2124-M-001-014 and NSTC 113-2124-M-001-008.
This study uses the following ALMA data: ADS/JAO.ALMA \#2016.1.00545.S and \#2016.1.00484.L. 
ALMA is a partnership of ESO (representing its member states), NSF (USA) and NINS (Japan), together with NRC (Canada), MOST and ASIAA (Taiwan), and KASI (Republic of Korea), in cooperation with the Republic of Chile. 
The Joint ALMA Observatory is operated by ESO, AUI/NRAO and NAOJ.
This work presents results from the European Space Agency (ESA) space mission Gaia. Gaia data are being processed by the Gaia Data Processing and Analysis Consortium (DPAC). Funding for the DPAC is provided by national institutions, in particular, the institutions participating in the Gaia MultiLateral Agreement (MLA). The Gaia mission website is $\langle$\url{https://www.cosmos.esa.int/gaia}$\rangle$. The Gaia archive website is $\langle$\url{https://archives.esac.esa.int/gaia}$\rangle$.
This research has made use of the VizieR catalogue access tool, CDS, Strasbourg, France (DOI : 10.26093/cds/vizier).
The original description of the VizieR service was published in 2000, A\&AS 143, 23.
Data analysis was carried out on the Multi-wavelength Data Analysis System operated by the Astronomy Data Center (ADC), National Astronomical Observatory of Japan.
This paper made use of the following software: AnalysisUtilities $\langle$\url{https://casaguides.nrao.edu/index.php?title=Analysis_Utilities}$\rangle$, Astropy \citep{astropy_2022}, CASA \citep{CASA_2022}, Linmix \citep{Kelly_2007}, matplotlib \citep{Hunter_2007}, PRIISM \citep{Nakazato_2020}, SciPy \citep{Virtanen_2020}, and NumPy \citep{Harris_2020}.
\end{ack}

\appendix
\section{Measurements of Disk Properties}\label{sec:measurements_disk}
\subsection{Gaussian Fitting with Markov Chain Monte Carlo Methods}\label{subsec:gauss_mcmc}
In this Appendix, we explain Gaussian fitting with MCMC for measuring position angles and inclination angles. 
Generally, it is appropriate to use the CASA task \texttt{imfit} for image domains and \texttt{uvmodelfit} for observational visibility when measuring PA and $i_{\rm disk}$.
\citet{Mart_2014} emphasized that \texttt{uvmodelfit} is much more effective than analyzing deconvolved images alone, as it extracts maximum information when observations approach their sensitivity and resolution limits.
However, \texttt{uvmodelfit} is most effective for visibility data from single components (note that the Nordic ALMA Regional Center Node offers a versatile \texttt{uvmultifit} package for multiple components). 
Additionally, \texttt{imfit} calculates uncertainties based on CLEAN beam sizes, which is unsuitable for deriving uncertainties in SpM images without beam convolution. 
Considering these factors, determining disk properties directly in the SpM image domain is more appropriate.

At first, using \texttt{modeling.Gaussian2D} from the Python module \texttt{astropy} (e.g., \cite{astropy_2022}), we prepare a brightness distribution model, $\mathbf{I}_{\rm model}$, represented as:
\begin{align}
\mathbf{I}_{\rm model}=I_{\rm peak}\exp\left[-\frac{1}{2}\left(\frac{x-{\rm dRA}}{\sigma_{\rm disk}}\right)^2-\frac{1}{2}\left(\frac{y-{\rm dDec}}{\sigma_{\rm disk}\cos i_{\rm disk}}\right)^2\right]\label{eq:gauss_func},
\end{align}
where $(x, y)$ are the horizontal and vertical coordinates of the model image, $I_{\rm peak}$ is the peak intensity, (dRA, dDec) are center coordinates, $\sigma_{\rm disk}$ is the major axis length of the disk, and $i_{\rm disk}$ is the inclination angle.
We then rotated the model image by the position angle of the disk PA using \texttt{modeling.Rotation2D} from the same module.
In summary, we use six parameters for the model image: $I_{\rm peak}$, (dRA, dDec), $\sigma_{\rm disk}$, $i_{\rm disk}$, and PA.

By adjusting these parameters and creating numerous model images, we sought the minimum residual between the model and the SpM image. 
In this process, we used the MCMC approach with the Python module \texttt{emcee} \citep{Foreman_2013} to determine the best-fit model image from the posterior probability distribution. 
We set the number of iterations to 2000 and the number of walkers to 100 and adopted $i_{\rm disk}$ and PA as the disk parameters of the appropriate model image.

We applied Gaussian fitting with MCMC to the SpM images for all detected disks. For single disks with smooth distributions, we used \texttt{imfit} and \texttt{uvmodelfit} to verify that the PA and $i_{\rm disk}$ values were consistent with these methods within a 10\% error. 
Table~\ref{table:disk_info} summarizes the measured values and uncertainties of PA and $i_{\rm disk}$ obtained using this method.

\subsection{Curve Growth Method for Disk Radius}\label{subsec:curve_growth}
We used the curve-growth method (e.g., \cite{Ansdell_2016,Yamaguchi_2024}) to determine disk radii. 
First, we converted the brightness distribution in the SpM image into a face-on view by deprojecting it using the measured PA and $i_{\rm disk}$ values listed in Table~\ref{table:disk_info}.
Next, we prepared a radial intensity profile, $I_{\rm radial}(r)$, averaged over the full azimuthal angles (see \S\ref{sec:gallery_profile} and the red curves shown in Figures~\ref{fig:profile_vol1}-\ref{fig:profile_vol4} in Appendix~\ref{sec:gallery_profile}).
Using these profiles, we then calculated the incremental flux density, $F_\nu(r)$, represented as: 
\begin{align}
F_\nu\left(r\right)=2\pi\int^r_0I_{\rm radial}\left(r^\prime\right)r^\prime dr^\prime,\label{eq:curve_growth_method}
\end{align}
where $r$ is the disk radius, and the total flux density is the limiting value of $F_\nu(r)$ as $r\rightarrow\infty$. 
In practice, we calculated $F_\nu(r)$ by gradually increasing $r$ and attempted to identify the constant value of $F_\nu$. 
We consider the value to be constant when the difference in $F_\nu(r)$ between two radii, $r_1$ and $r_2$ $(r_1<r_2)$, is less than 0.01\% of $F_\nu(r_2)$.

We determined two disk radii, $R_{68\%}$ and $R_{95\%}$, where $F_\nu(r)$ equals 0.68$F_\nu$ and 0.95$F_\nu$, respectively. 
The uncertainty $\sigma_{\rm Radius}$ was estimated from the effective spatial resolution $\theta_{\rm eff}$. 
We assumed $\sigma_{\rm Radius}=\left<\theta_{\rm eff}\right>/2\sqrt{2\ln2}$, where $\left<\theta_{\rm eff}\right>$ represents the geometric mean of the spatial resolution.
Table~\ref{table:image_info} summarizes the total flux $F_\nu$, while Table~\ref{table:disk_info} lists the millimeter flux $L_{\rm mm}$ ($F_\nu$ corrected for a distance of 140\,pc), disk radii $R_{68\%}$ and $R_{95\%}$, and the uncertainty in disk radius $\sigma_{\rm Radius}$.

\begin{figure*}[t]
    \begin{center}
    \includegraphics[width=\linewidth]{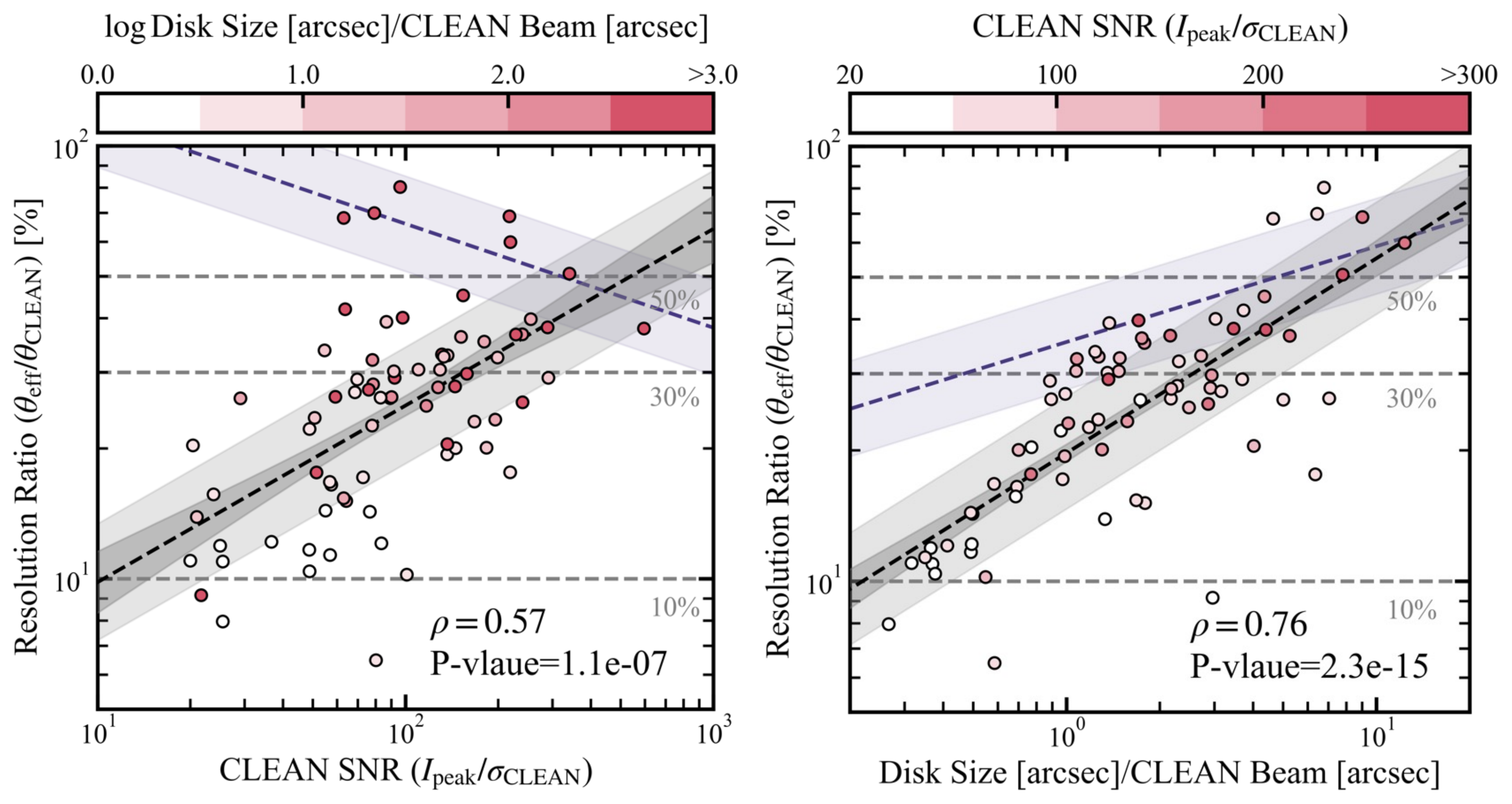}
    \end{center}
    \caption{
    (Left) Relationship between the resolution ratio of SpM to CLEAN ($\theta_{\rm eff}/\theta_{\rm CLEAN}$) and CLEAN SNR (peak intensity $I_{\rm peak}$ to RMS noise $\sigma_{\rm CLEAN}$) in logarithmic scale.
    The samples are color-coded by the disk size $(=2R_{95\%})$ normalized by CLEAN beam size $\theta_{\rm CLEAN}$.
    (Right) Relationship between the resolution ratio $\theta_{\rm eff}/\theta_{\rm CLEAN}$ and the disk size normalized by CLEAN beam size.
    The samples are color-coded by the CLEAN SNR.
    The dashed line represents the median scaling relation from the Bayesian linear regression, with the dark gray area indicating the 68\% confidence interval for the regression.
    The light gray area corresponds to the inferred scatter.
    Pearson's correlation coefficient and the p-value calculated from the sample distribution are shown in the bottom right of each panel.
    The violet dashed lines show the relationships of Equation (1) and (2) in \citet{Yamaguchi_2024}, and the light violet area corresponds to their inferred scatter.
             }
    \label{fig:imp_spm}
\end{figure*}

\section{Performance and Fidelity of SpM Images}\label{sec:spm_perfo}
\subsection{Improvements of Spatial Resolution}\label{subsec:imp_res}
In this appendix, we describe the ratio of the effective spatial resolution $\theta_{\rm eff}$ in SpM images to the CLEAN beam size $\theta_{\rm CLEAN}$, which quantifies the improvement in spatial resolution.
\citet{Yamaguchi_2024} reported that 42\% of their targets showed a two- to three-fold improvement in the spatial resolution of SpM images compared to the conventional CLEAN method. 
They also confirmed a negative correlation (Pearson correlation coefficient $\rho = -0.65$, $p$-value $= 4 \times 10^{-6}$) between the resolution ratio and the SNR in CLEAN images, as well as a positive correlation ($\rho = 0.78$, $p$-value $= 2 \times 10^{-9}$) between the resolution ratio and the disk size normalized by the CLEAN beam. 
In this subsection, we investigated whether a similar relationship exists between the resolution ratio, SNR, and disk size of CLEAN images, as reported by \citet{Yamaguchi_2024}.

For the discussion on spatial resolution ratios, we used 75 disks, excluding those categorized as candidates for `Ring' structures or circumstellar disks around binary systems (see, \S\ref{subsec:categorization}). 
These excluded disks prevent us from determining the positions of protostars (or T Tauri stars) as disk centers, making it difficult to derive accurate disk radii.
The effective spatial resolutions in SpM images, $\theta_{\rm eff}$, and the CLEAN beams, $\theta_{\rm CLEAN}$, with their geometric mean values used in this evaluation, are listed in Table~\ref{table:image_info}.
The SNRs in CLEAN images were calculated from the peak intensities $I_{\rm peak}$ and RMS noise levels $\sigma_{\rm CLEAN}$ described in Table~\ref{table:image_info}.
We also used disk sizes $(=2R_{95\%})$ normalized by the CLEAN beam size $\theta_{\rm CLEAN}$.

Figure~\ref{fig:imp_spm} shows the improvement ratio in spatial resolution, $\theta_{\rm eff}/\theta_{\rm CLEAN}$. 
We observed a more than three-fold improvement in spatial resolution for 76\% of the disks. 
These results indicate greater resolution ratios compared to previous studies using SpM imaging (e.g., \cite{Yamaguchi_2020, Yamaguchi_2024}). 
We investigated the factors contributing to these advancements in spatial resolution, focusing on SNR ($I_{\rm peak}/\sigma_{\rm CLEAN}$) in CLEAN images and disk sizes normalized by the CLEAN beams $(2R_{95\%}/\theta_{\rm CLEAN})$.

The left panel of Figure~\ref{fig:imp_spm} shows the relationship between the resolution ratio $\theta_{\rm eff}/\theta_{\rm CLEAN}$ and the SNR ($I_{\rm peak}/\sigma_{\rm CLEAN}$) for CLEAN images.
A clear trend is observed, in which the resolution ratio decreases as SNR decreases (Pearson correlation coefficient $\rho = 0.57$, $p$-value $= 1.1 \times 10^{-7}$).
We performed Bayesian linear regression in logarithmic space using the Python module \texttt{Linmix} \citep{Kelly_2007}, and this trend can be described as:
\begin{align}
\log\left(\frac{\theta_{\rm eff}/\theta_{\rm CLEAN}}{\%}\right) = \left(0.58 \pm 0.14\right) + \left(0.41 \pm 0.07\right) \log ({\rm SNR}), \label{eq:imp_snr}
\end{align}
with Gaussian scatter perpendicular to the regression line, having a standard deviation of 0.13 dex and an error of 2.8$\times 10^{-5}$ dex. 
This trend differs from that described in \citet{Yamaguchi_2024}, which is shown by the violet dashed line.
Notably, our results indicate that the resolution can be improved by a factor of 3 or more even when SNR drops below 50 ($\sim 1.7$ dex).

The right panel of Figure~\ref{fig:imp_spm} is a scatter plot showing the relationship between the resolution ratio $\theta_{\rm eff}/\theta_{\rm CLEAN}$ and the disk size normalized by the CLEAN beam $(2R_{95\%}/\theta_{\rm CLEAN})$.
The plot reveals a positive correlation, indicating that smaller disk sizes are associated with greater resolution ratios (Pearson correlation coefficient $\rho = 0.76$, $p$-value $=2.3 \times 10^{-15}$).
Following Bayesian linear regression in logarithmic space, this trend can be expressed as: 
\begin{align}
\log\left(\frac{\theta_{\rm eff}/\theta_{\rm CLEAN}}{\%}\right)=\left(1.29\pm0.02\right)+\left(0.45\pm0.05\right)\log\left(\frac{2R_{95\%}}{\theta_{\rm CLEAN}}\right),\label{eq:imp_disk}
\end{align}
with Gaussian scatter of 0.13 dex and an error of $3.6 \times 10^{-5}$ dex. 
This trend aligns with the findings of \citet{Yamaguchi_2024}, where effective spatial resolutions in SpM images can be improved by a factor of more than two when the disk sizes $2R_{95\%}$ are 2-10 times larger than the CLEAN beam size.

In summary, we confirmed positive correlation trends between resolution ratio $\theta_{\rm eff}/\theta_{\rm CLEAN}$ and both disk size and SNR ($I_{\rm peak}/\sigma_{\rm CLEAN}$) in CLEAN images. 
The trend with disk size is consistent with \citet{Yamaguchi_2024}, though it differs in showing a negative correlation with SNR in CLEAN images.
We suggest that the distribution of observational visibility is the primary factor accounting for the difference between \citet{Yamaguchi_2024} and this study.

Most disks in \citet{Yamaguchi_2024} already display substructures in their CLEAN images, which are expected to show multiple nulls in the real part of the observational visibilities. 
In contrast, the majority of our targets appear compact in the SpM images (shown in Figures~\ref{fig:spm_classi}-\ref{fig:spm_classii}), with fewer nulls in the visibilities up to the maximum baseline length ($\sim$1 M$\lambda$ in the deprojected plane) and a smaller decrease in visibility amplitudes (see \S A\ref{subsec:fidelity_disk}).
The null points of visibility are hard to measure because they are easily affected by noise.
The SpM image may also be affected if the disk visibility contains several nulls.
This suggests that the trend in effective spatial resolution achieved by SpM imaging depends on the presence of null points in the observational visibility. 
The negative correlation between SNR in CLEAN images and spatial resolution ratio observed by \citet{Yamaguchi_2024} likely reflects a pattern visible in data with nulls. 
In the left panel of Figure~\ref{fig:imp_spm}, the seven disks (ISO-Oph54, WLY2-63, Elias27, DoAr25, Elias24, WSB82, and ISO-Oph2) with large radii and substructures are in the violet region, which corresponds to the negative correlation.
In our analysis, the positive correlation with the disk size normalized by the CLEAN beam is more pronounced, likely because fewer visibilities contain null points.

\begin{figure}[t]
    \begin{center}
    \includegraphics[width=\linewidth]{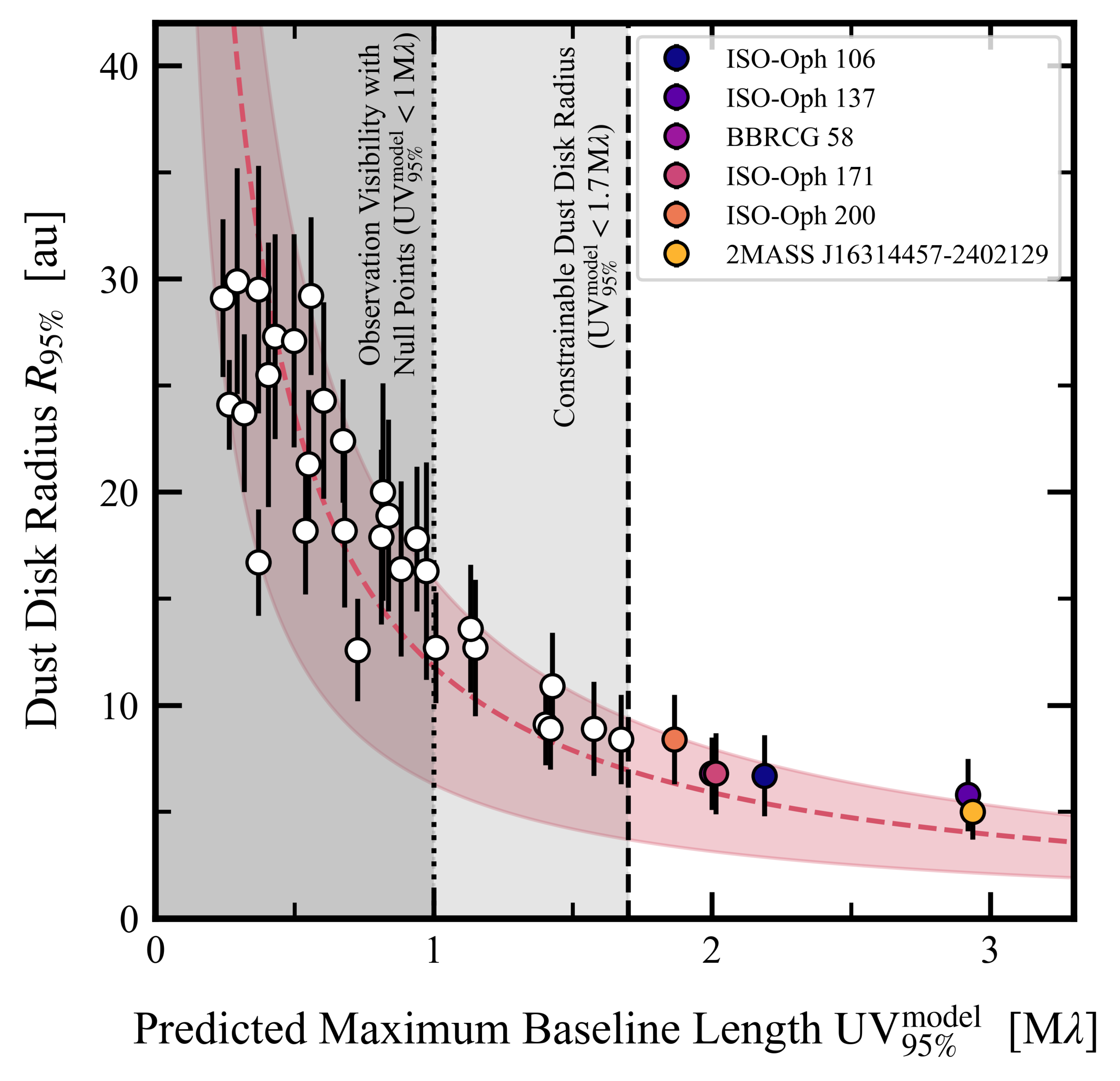}
    \end{center}
    \caption{
    Scatter diagram between the predicted 95th percentile maximum baseline length and disk radius for disks around 36 single systems.
    The dark and light gray areas represent observation visibilities with high fidelity in both disk substructure and radius, and only disk radius, respectively.
    The former threshold is indicated by the dotted line at 1\,M$\lambda$, which is the maximum baseline length in polar coordinates, while the latter is shown by the dashed line at 1.7\,M$\lambda$.
    The colored markers indicate single stars with low fidelity in both measures.
    The red area represents the estimated values derived from Equation~(\ref{eq:disk_evaluation}) adopting $d=140$\,pc and a wide range of inclination angles $i_{\rm disk}$ (30-70\,degrees), where the red dashed line is the case of $i_{\rm disk}=$50\,degrees.
             }
    \label{fig:baseline_comp}
\end{figure}

\begin{figure*}[t]
    \begin{center}
    \includegraphics[width=0.73\linewidth]{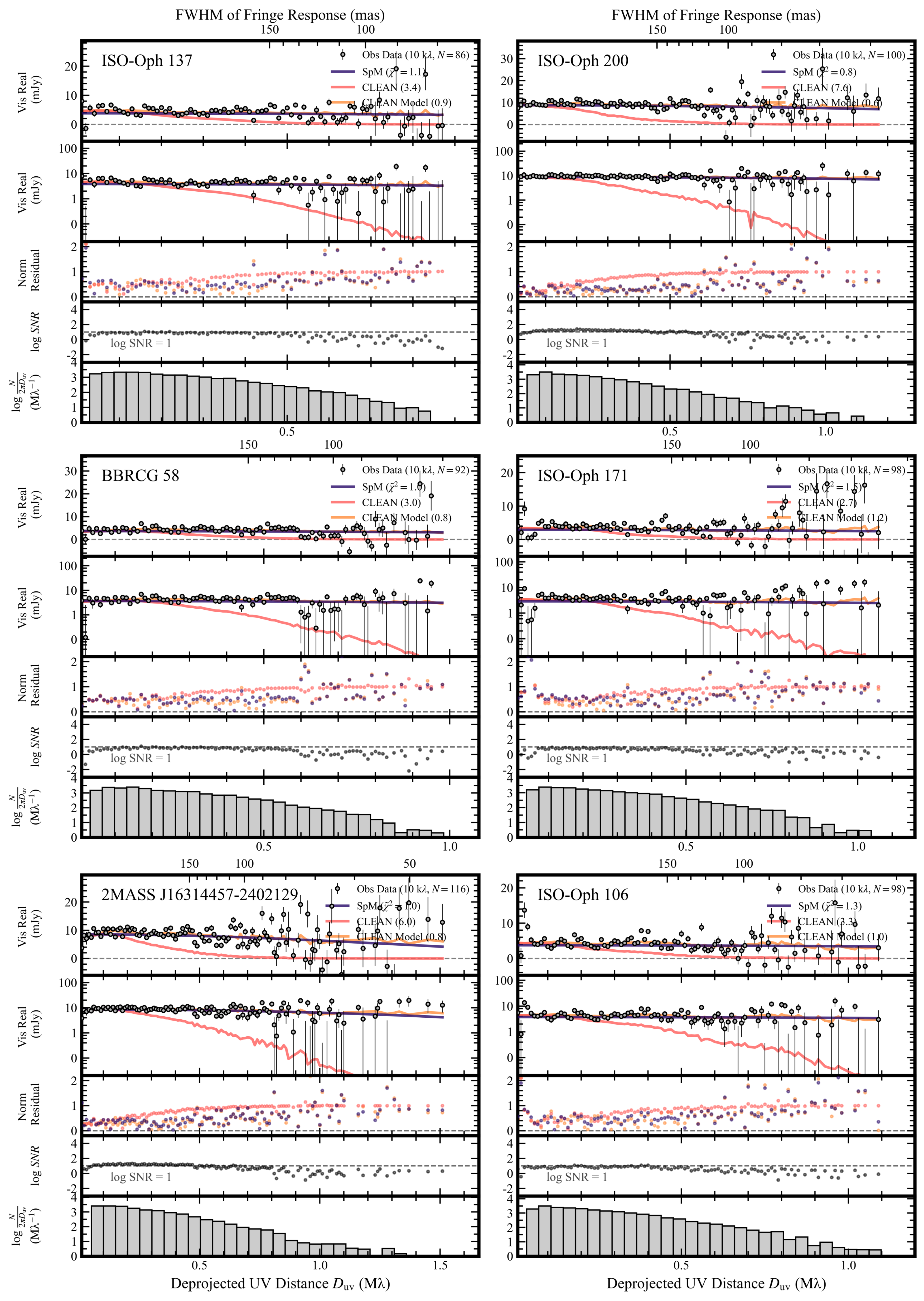}
    \end{center}
    \caption{
    Radial visibility profiles of the six compact disks with $R_{95\%}$ less than 10\,au (top: ISO-Oph~137; ISO-Oph~200, middle: BBRCG~58; ISO-Oph~171, bottom: 2MASS~J16314457-2402129; ISO-Oph~106).
    In each panel, the observed visibility data are shown as dots, and the visibility models for SpM, CLEAN, and CLEAN model are represented by purple, red, and orange lines, respectively.
    The data are binned every 10\,k$\lambda$.
    The reduced-$\chi^2$ values calculated from the observed data and models are shown in the labels of the top panels.
    For each target, the panels display, from top to bottom, the amplitude of the real part of the visibility, its logarithmic scale, the normalized residual between the observation and model, the SNR of visibility within each bin, and the data density of each bin in $uv$-space.
    The SNR is calculated as the ratio of the real part amplitude to its noise.
    Those detailed derivations are described in Appendix~4 in \citet{Yamaguchi_2024}.
            }
    \label{fig:compact_visprofile}
\end{figure*}

\begin{figure*}[t]
    \begin{center}
    \includegraphics[width=\linewidth]{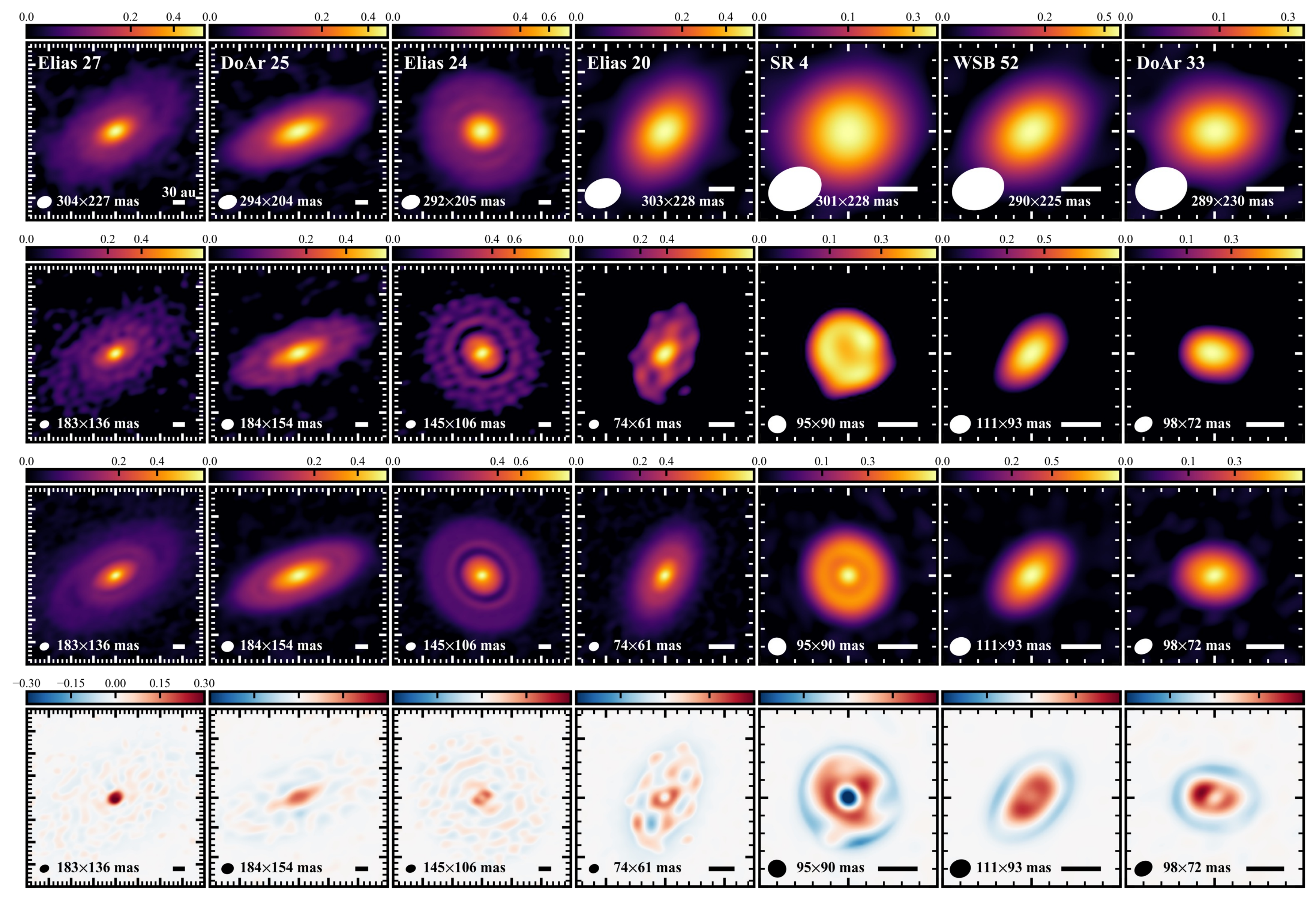}
    \end{center}
    \caption{
    Image Gallery of DSHARP targets (Elias~27, DoAr~25, Elias~24, Elias~20, SR~4, WSB~52, DoAr~33). 
    (Top) CLEAN images from the short-baseline data. 
    (Upper middle) SpM images from the short-baseline data, $\mathbf{I}_{\rm SpM}(x, y)$. 
    (Lower middle) CLEAN reference images, $\mathbf{I}_{\rm Ref}(x, y)$, derived from the long-baseline data in the DSHARP project \citep{Andrews_2018_DSHARP}. 
    (Bottom) Residual maps, $\mathbf{I}_{\rm SpM}(x, y)-\mathbf{I}_{\rm Ref}(x, y)$, normalized by the peak intensity of the CLEAN reference images, ${\rm Peak}(\mathbf{I}_{\rm Ref})$. 
   The filled white and black ellipses in the lower left corner of each panel denote the spatial resolutions of the top and middle images. 
   To minimize the effects of flux calibration, the total flux of the SpM and CLEAN reference images is scaled to match that of the CLEAN reference image.
             }
    \label{fig:comp_dsharp}
\end{figure*}

\begin{figure*}[t]
    \begin{center}
    \includegraphics[width=\linewidth]{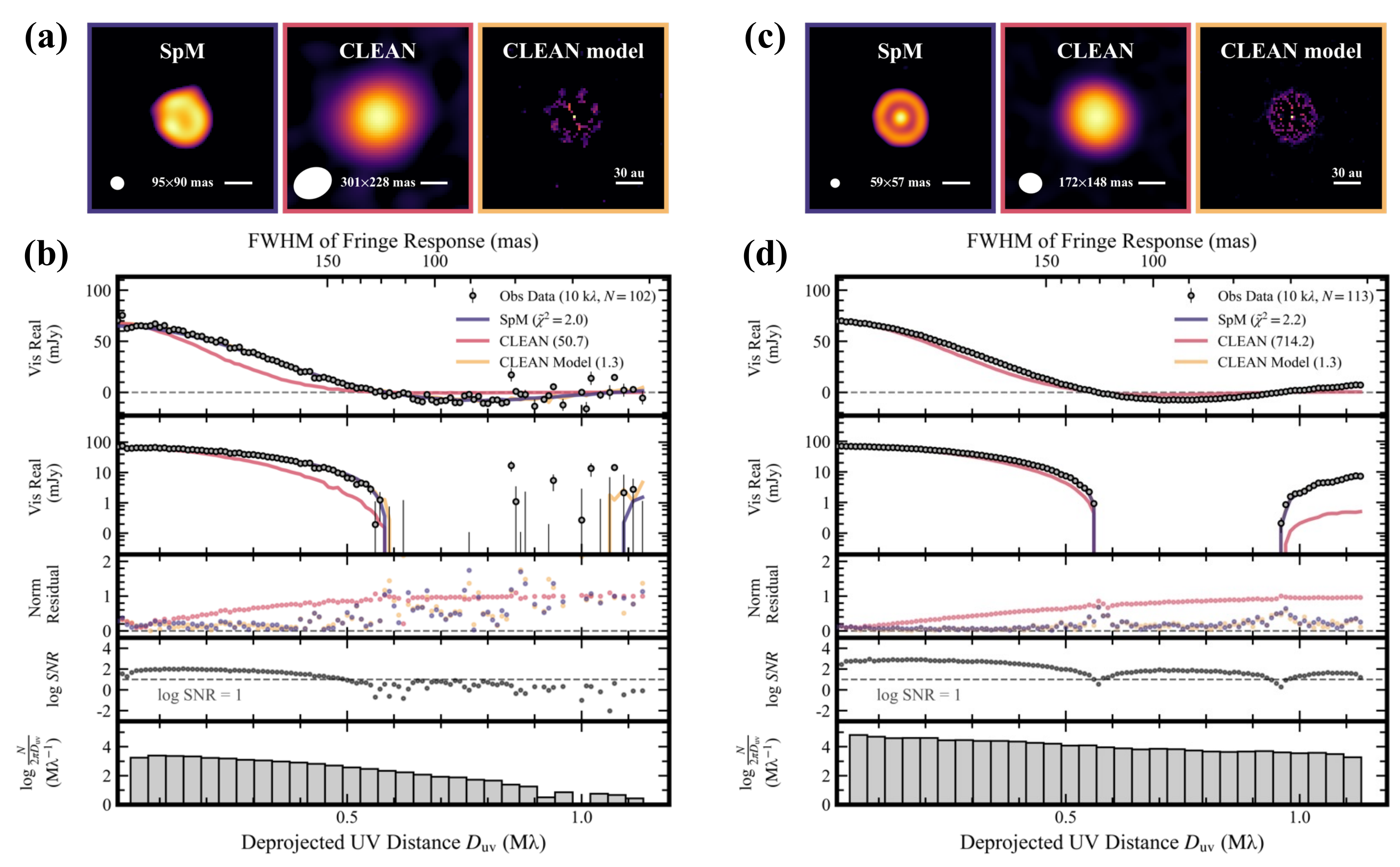}
    \end{center}
    \caption{
    Continuum gallery and azimuthal radial visibility profiles of the compact disk around SR~4 created using data from this study (left) and the DSHARP dataset (right).
    The images in the top panels for SR~4 are, from left to right, the SpM image, beam-convolved CLEAN image, and CLEAN model.
    The radial visibility profiles are plotted with the same settings as described in Figure~\ref{fig:compact_visprofile}.
             }
    \label{fig:comp_sr4}
\end{figure*}

\subsection{Fidelity for Brightness Distribution of Compact Disks}\label{subsec:fidelity_disk}
As mentioned in the previous subsection, the visibility amplitude of a compact disk with a smooth brightness distribution decreases only slightly. 
SpM imaging enables us to reconstruct a super-resolution image by predicting the visibility over longer baselines using the fitting to the observed $uv$-coverage. 
However, it remains uncertain whether applying SpM imaging to limited visibility coverage could introduce biases in estimating disk radii and brightness distributions. 
To evaluate the accuracy of reconstructed disk radii and brightness distributions, we calculated the expected $uv$ distances for 36 disks around single stars with $R_{95\%}$ less than 30\,au in SpM images.
Assuming that disk intensity follows an axisymmetric Gaussian distribution when viewed face-on, we used the following canonical relationship for the Fourier transform: 
\begin{align}
    \left(\frac{{\rm UV}^{\rm model}_{\rm FWHM}}{100\,\rm k \lambda}\right)\left(\frac{\Theta_{\rm FWHM}}{1\,{\rm arcsec}}\right)&=\frac{8\ln 2}{2\pi}\times\frac{360\times3600}{2\pi\times10^5}\times\cos{i_{\rm disk}}\notag\\
    &=1.8204\times\cos{i_{\rm disk}},\label{eq:disk_evaluation}
\end{align}
where ${\rm UV}^{\rm model}_{\rm FWHM}$ is the $uv$ distance at which the visibility amplitude reaches half of its peak, and $\Theta_{\rm FWHM}$ is the disk FWHM. 
The latter was estimated by assuming that twice the standard deviation corresponds to the measured disk radius $R_{95\%}$ in the SpM image as expressed by $\Theta_{\rm FWHM} \sim \sqrt{2 \ln 2} R_{95\%}$.
Based on the derivation for $\Theta_{\rm FWHM}$, we calculated the predicted 95th-percentile maximum baseline length ${\rm UV}^{\rm model}_{95\%}$ as ${\rm UV}^{\rm model}_{95\%} = {\rm UV}^{\rm model}_{\rm FWHM}/\sqrt{2\ln2}$.
Figure~\ref{fig:baseline_comp} compares $R_{95\%}$ with ${\rm UV}^{\rm model}_{95\%}$ for smooth, compact disks identified in the SpM images.

To constrain the disk radius, ${\rm UV}^{\rm model}_{\rm FWHM}$ should be smaller than the maximum baseline length of the observational visibility.
Assuming ${\rm UV}^{\rm model}_{\rm FWHM}/2 = 1$\,M$\lambda$, which corresponds to the maximum baseline length of the visibility used in this study, we estimated the 95th percentile maximum baseline, ${\rm UV}^{\rm model}_{95\%}$, as ${\rm UV}^{\rm model}_{95\%} = {\rm UV}^{\rm model}_{\rm FWHM}/\sqrt{2\ln2} \approx 1.7$\,M$\lambda$, to ensure high-fidelity information on disk radius.
When the predicted maximum baseline length ${\rm UV}^{\rm model}_{95\%}$ exceeds 1.7\,M$\lambda$, fluctuations in visibility amplitude may be insufficient to determine disk radii accurately. 
We define such objects as `very compact disks', which require additional long-baseline data to resolve their detailed characteristics, such as size and structure.
Six disks (ISO-Oph~200, ISO-Oph~137, BBRCG~58, ISO-Oph~171, 2MASS~J16314457-2402129, and ISO-Oph~106) with $R_{95\%}$ of less than 10\,au fall into this category.
However, based on the following three points, we believe it is unlikely that the disk radii are significantly misestimated.
First, the inclination angles calculated from the SpM image aspect ratios match those derived using CASA's \texttt{uvmodelfit} task within a 10\% error (see \S A.\ref{subsec:gauss_mcmc}).
Second, the radial visibility profiles of these six disks, shown in Figure~\ref{fig:compact_visprofile}, indicate that the Fourier-transformed model visibilities from SpM images align well with the observational visibilities (reaching reduced - $\chi^{2}\sim1$).
Finally, the total fluxes of these disks are less than 10\,mJy, suggesting that their radii would not exceed 20\,au based on comparisons with other disks. 
Therefore, we include these disk radii in Figures~\ref{fig:disk_radius}, \ref{fig:Tbol_Rdust}, \ref{fig:hist_inc_cos}, and \ref{fig:pie_inc} and note that these values are provided for reference.

We also performed the same validation for the disk substructures. 
To detect substructures with SpM imaging, the observational visibility must contain some null points up to the maximum baseline length. 
In this study, when the predicted ${\rm UV}^{\rm model}_{95\%}$ is greater than 1\,M$\lambda$, this suggests that the visibility has no nulls, making it difficult to represent substructures with SpM imaging.
The 14 compact disks (ISO-Oph~137, ISO-Oph~200, ISO-Oph~59, ISO-Oph~107, ISO-Oph~132, ISO-Oph~212, BBRCG~58, ISO-Oph~171, DoAr~32, ISO-Oph~155, ISO-Oph~20, ISO-Oph~116, 2MASS~J16314457-2402129, and ISO-Oph~106) fall into this category. 
In this study, we categorized them as having `Smooth' distributions based on our analysis, but we note that we cannot entirely rule out the presence of substructures for these objects.

\subsection{Comparison with the Targets in the DSHARP Project}\label{subsec:comp_dsharp}
In this section, we discuss the fidelity of substructures in the SpM images (Figures~\ref{fig:spm_classi}-\ref{fig:spm_classii}) shown in this study.
We compare the SpM image generated from data with a maximum baseline length of 2.6\,km (referred to as the short-baseline data) with the CLEAN image from the DSHARP project (e.g., \cite{Andrews_2018_DSHARP}), which uses data with a baseline length exceeding 10\,km (referred to as the long-baseline data).

The top and upper middle panels of Figure~\ref{fig:comp_dsharp} show the CLEAN and SpM images, $\mathbf{I}_{\rm SpM}(x, y)$, derived from the short-baseline data of the DSHARP targets (Elias~27, DoAr~25, Elias~24, Elias~20, SR~4, WSB~52, and DoAr~33). 
The total flux of the CLEAN and SpM images of the short-baseline data is scaled to match that of the CLEAN images of the long-baseline data (lower middle panels of Figure~\ref{fig:comp_dsharp}) to minimize the effects of errors in flux calibration. 
For the DSHARP targets, the effective spatial resolutions $\theta_{\rm eff}$ in the SpM images are 1.5 to 4 times higher than those of the CLEAN images (for details, see Table~\ref{table:image_info}). 
In \S\ref{subsec:categorization}, we categorized Elias~24 and SR~4 as `Ring', Elias~27 as `Spiral', DoAr~25 and Elias~20 as `Inflection', and WSB~82 and DoAr~33 as `Smooth.' 

The lower middle panels of Figure~\ref{fig:comp_dsharp} show the CLEAN images of the long-baseline data. 
We downloaded the original FITS images from the DSHARP data release (e.g., \cite{Andrews_2018_DSHARP}) and smoothed them using the CASA task \texttt{imsmooth} to match $\theta_{\rm eff}$. 
Hereafter, we refer to the lower middle panels as the CLEAN reference images and define their brightness distribution as $\mathbf{I}_{\rm Ref}(x, y)$. 

The bottom panels of Figure~\ref{fig:comp_dsharp} are the residual maps obtained by subtracting the brightness distributions in the CLEAN reference images from those in the SpM images, $\mathbf{I}_{\rm SpM}(x, y) - \mathbf{I}_{\rm Ref}(x, y)$, and normalizing by the peak intensities in the CLEAN reference images, ${\rm Peak}(\mathbf{I}_{\rm Ref})$.

The comparison of brightness distributions between the SpM and CLEAN reference images ($\mathbf{I}_{\rm SpM}$ and $\mathbf{I}_{\rm Ref}$) demonstrates that the SpM images successfully reveal substructures consistent with those in the CLEAN reference images from the long-baseline data, except for SR~4. 
The normalized residual maps $(\mathbf{I}_{\rm SpM}-\mathbf{I}_{\rm Ref})/{\rm Peak}(\mathbf{I}_{\rm Ref})$ show maximum differences of about 30\%.

Since the observation time per object in the short-baseline data is less than one minute, the visibility coverage in this region is sparser than that in the long-baseline data. 
Consequently, the SpM imaging cannot fully reproduce broader structures that require short baseline information.
Instead, compact structures are emphasized, resulting in stronger intensity in the SpM images, such as WSB~52 and DoAr~33.
Moreover, we assume that the visibility coverage is not filled well enough to reproduce a large brightness distribution, such as Elias~27, Elias~24, and Elias~20. 
In that case, applying the SpM algorithm to the coverage would reproduce the brightness distribution with slightly interrupted weak emissions rather than extensive ones.
We consider these distribution differences acceptable, given the limitations in data quality. 
This overall consistency suggests that the SpM algorithm can generate high-resolution images and replicate brightness distributions with a quality nearly equivalent to that of high-resolution observations.

The SpM image for SR~4 shows a ring structure with a dip in brightness distribution but does not capture the inner disk visible in the CLEAN reference image as observed by \citet{Andrews_2018_DSHARP}. 
Similarly, for WL~17, we also noted the absence of the inner disk, as described in \citet{Shoshi_2024}. 
However, this does not imply that the SpM algorithm produces artificial structures.

The left-hand panels of Figure~\ref{fig:comp_sr4} present the results obtained from the short-baseline data for SR~4.
The distribution of the observation visibility and model visibilities derived by Fourier transformation in the SpM, CLEAN, and CLEAN model images shown in Figure~\ref{fig:comp_sr4} (a) are plotted in Figure~\ref{fig:comp_sr4} (b). 
In the long-baseline region of the radial visibility profile, the real part of the observed visibility has large errors and an SNR of less than 10.
The visibility distribution modeled from the SpM images does not perfectly fit the observed visibility in the long-baseline region, where large errors are present.
Consequently, applying SpM imaging to the short-baseline data reveals only the ring structure without capturing the inner disk.

The right-hand panels of Figure~\ref{fig:comp_sr4} show the results for the same object, SR~4, but created using the data obtained with the DSHARP project.
We downloaded the final calibrated dataset for SR~4 from the DSHARP data release and limited the maximum baseline length up to that of the short-baseline data ($\sim$2\,km) using the CASA task \texttt{split}.
We applied both CLEAN and SpM imaging to the constrained long-baseline data, using the same settings described in \S\ref{subsec:method_CLEAN} and \S\ref{subsec:method_SpM}, and created the continuum images shown in Figure~\ref{fig:comp_sr4} (c).
The radial visibility profile in Figure~\ref{fig:comp_sr4}(d) shows that the observational visibility has smaller errors than those in Figure~\ref{fig:comp_sr4} (b) and an SNR greater than 10 in the long-baseline region.
The model visibility of the SpM image closely matches the observed visibility, allowing us to detect both the ring structure and the inner disk.

In summary, the SpM imaging process responds differently to disk substructure depending on the SNR in the long-baseline region. 
We note that when the number of long-baseline data points is insufficient, SpM produces a conservative model image without introducing any artificial structures. 
An SNR threshold of 10 appears useful for obtaining detailed information about disk substructures with SpM imaging. 
Therefore, the substructures we confirm are likely genuine inner disks that the substructures may also be present in ISO-Oph~51 and IRAS16201-2410, which exhibit ring-cavity structures.

\begin{figure*}[t]
    \begin{center}
    \includegraphics[width=0.95\linewidth]{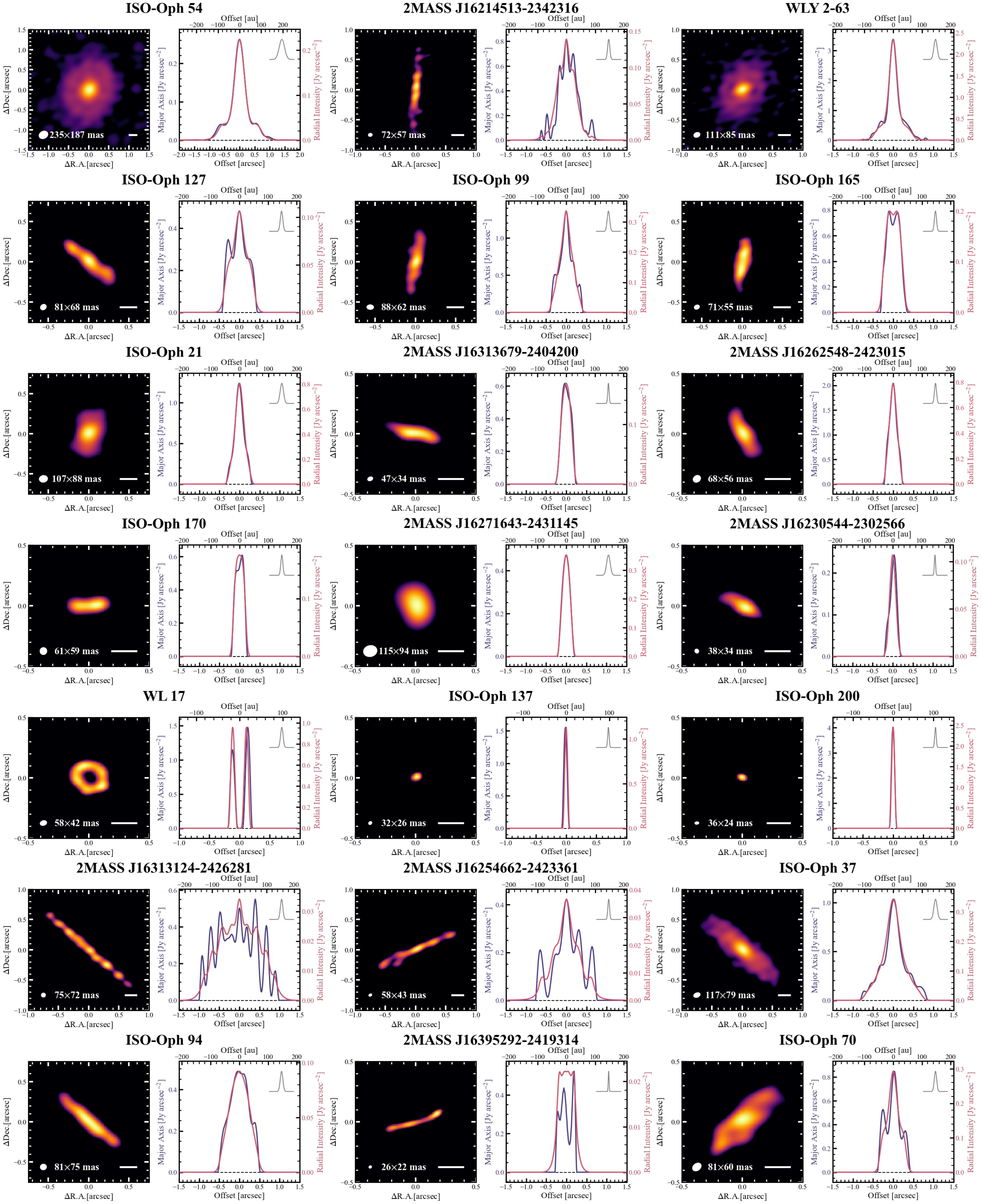}
    \end{center}
    \caption{
    Gallery of SpM images and intensity profiles for 15 Class I and 6 Class FS disks. The continuum images are identical to those in Figures~\ref{fig:spm_classi}-\ref{fig:spm_classii}. 
    Panels adjacent to the continuum images display intensity profiles: violet curves represent profiles along the major axis (aligned with the PA direction), and red curves show radial profiles averaged over all azimuthal angles. 
    Negative components of the red curves are linearly symmetrical to the positive ones.
    Gray curves in the upper right indicate the effective spatial resolution $\theta_{\rm eff}$ of the SpM images.
    }
    \label{fig:profile_vol1}
\end{figure*}

\begin{figure*}[t]
    \begin{center}
    \includegraphics[width=0.95\linewidth]{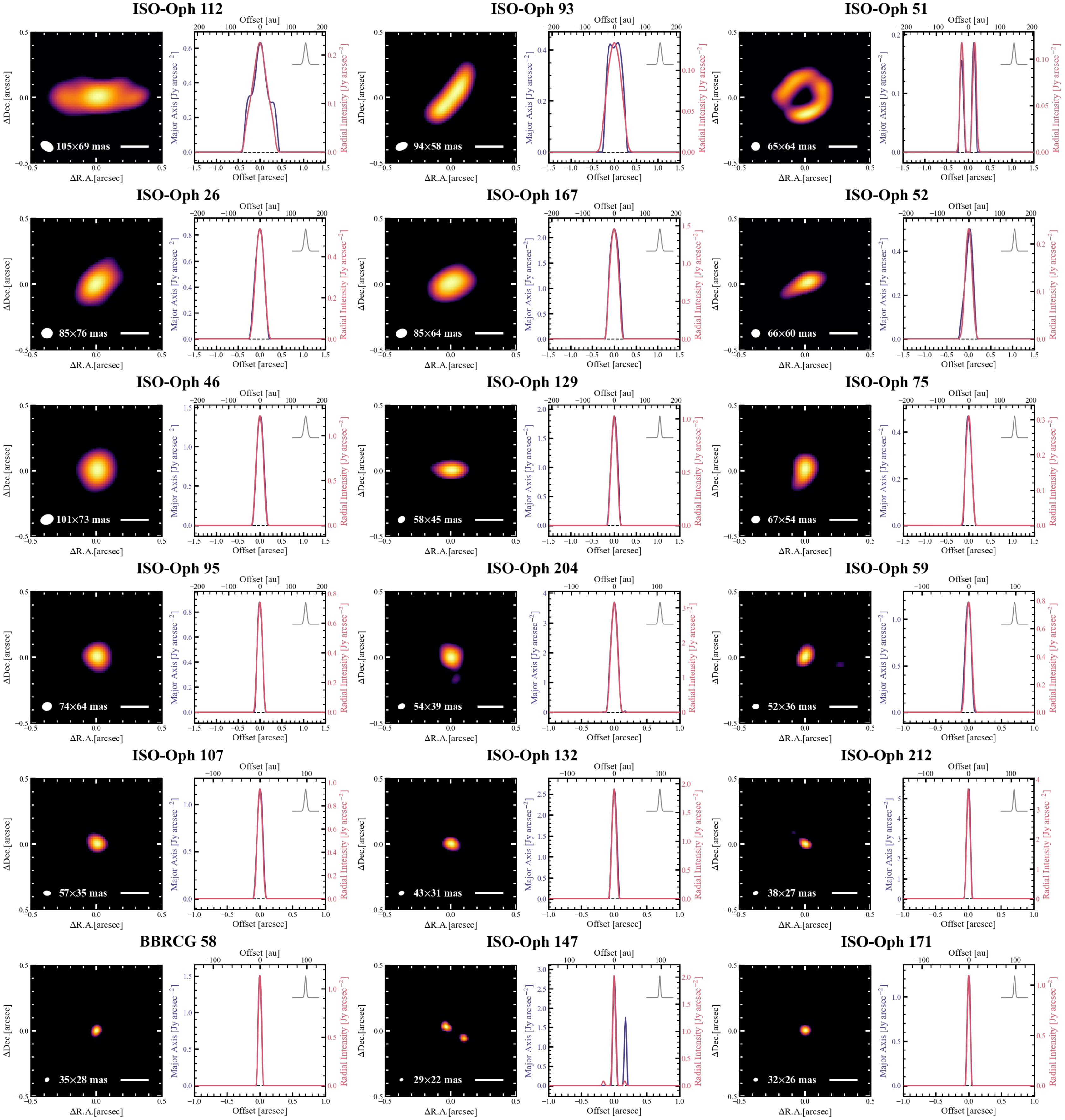}
    \end{center}
    \caption{Same as Figure~\ref{fig:profile_vol1} but for 18 different Class FS disks.
             }
    \label{fig:profile_vol2}
\end{figure*}

\begin{figure*}[t]
    \begin{center}
    \includegraphics[width=0.95\linewidth]{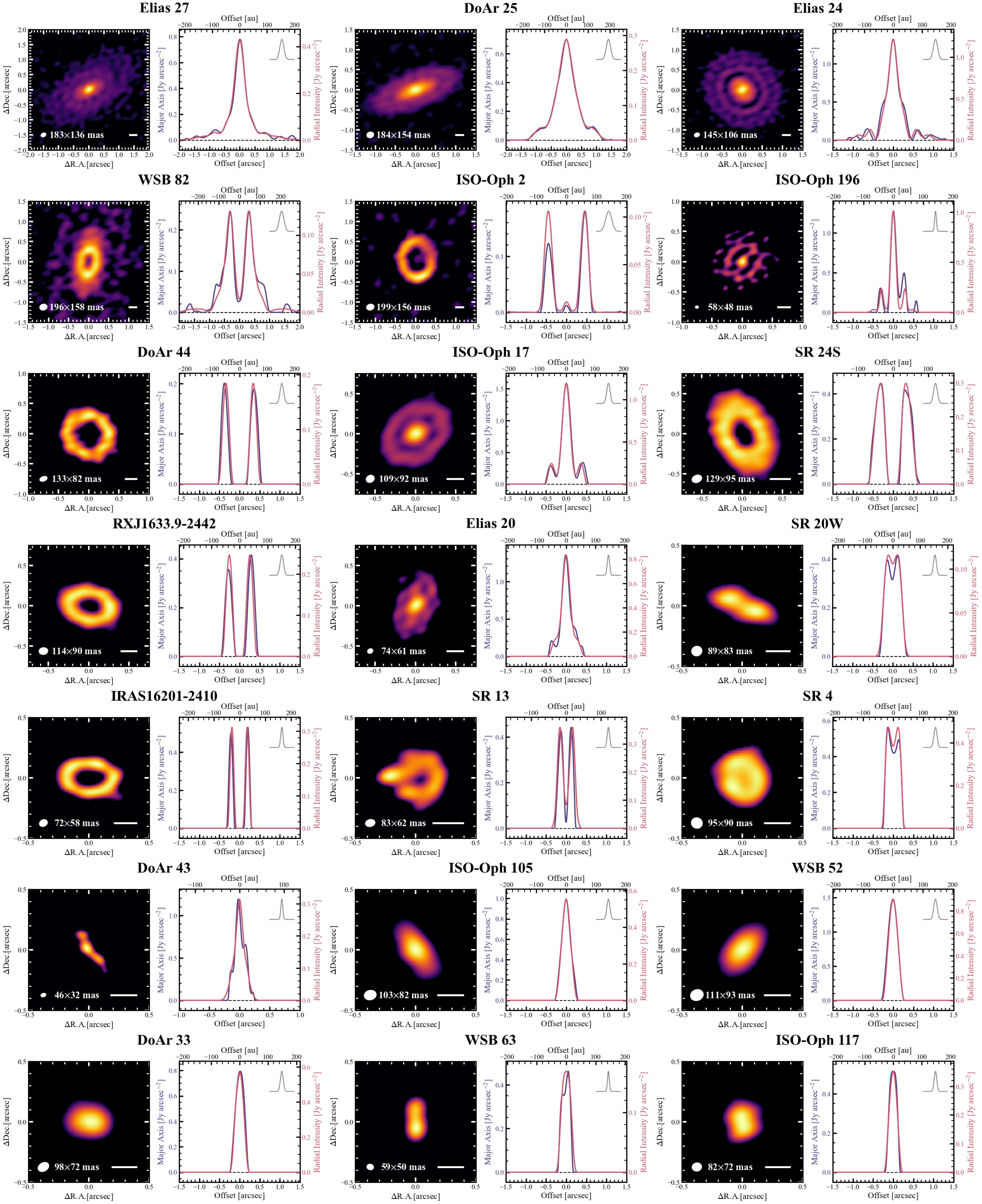}
    \end{center}
    \caption{Same as Figure~\ref{fig:profile_vol2} but for 21 Class II disks.
             }
    \label{fig:profile_vol3}
\end{figure*}

\begin{figure*}[t]
    \begin{center}
    \includegraphics[width=0.95\linewidth]{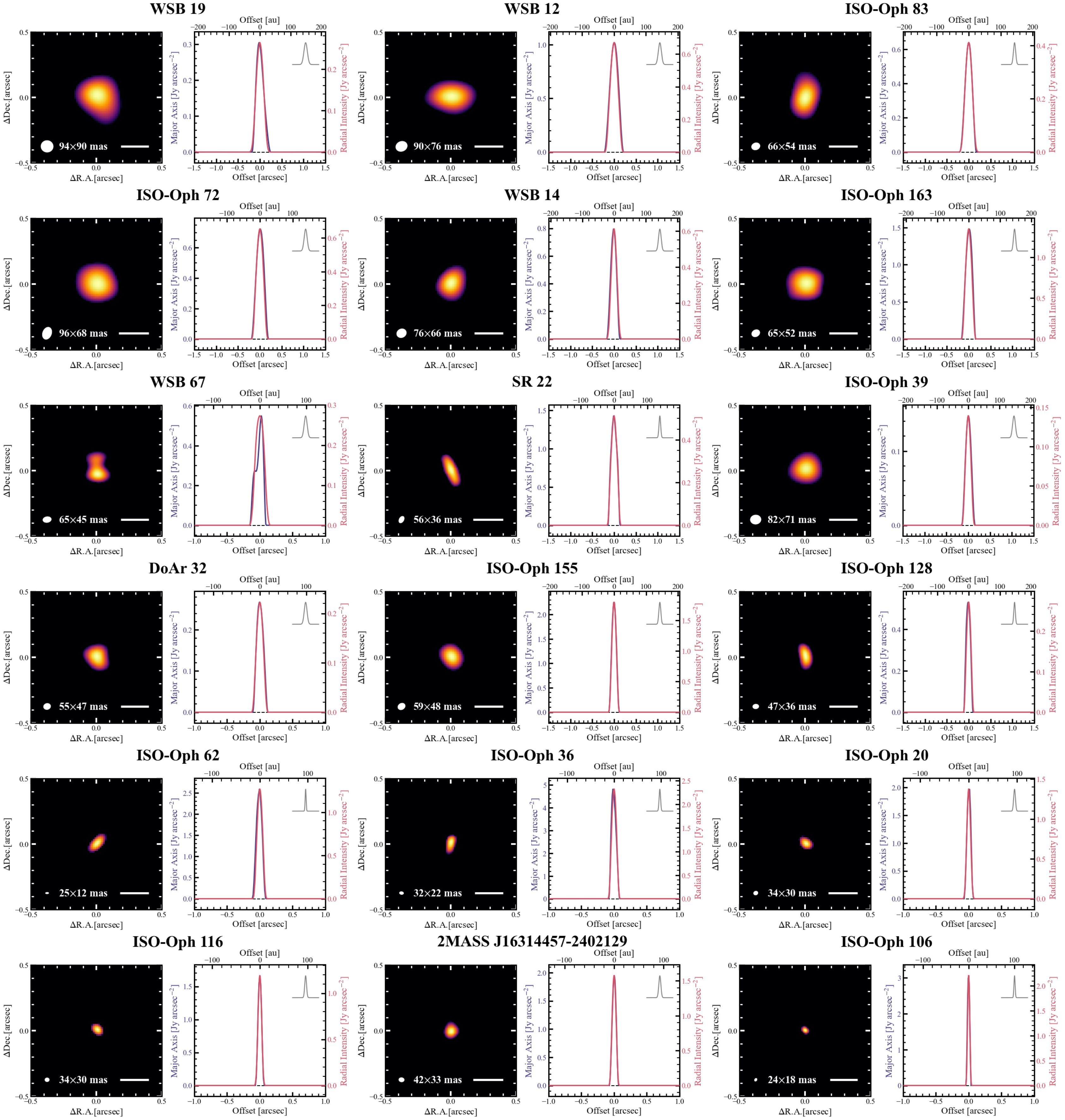}
    \end{center}
    \caption{Same as Figure~\ref{fig:profile_vol3} but for 18 different Class II disks.
             }
    \label{fig:profile_vol4}
\end{figure*}

\section{Gallery of the Intensity Profiles}\label{sec:gallery_profile}
In this section, we summarize the intensity profile.
Using $i_{\rm disk}$ and PA derived in \S\ref{subsec:disk_properities}, we deprojected the brightness distribution to a face-on view and averaged it over all azimuthal angles to obtain the radial intensity profile $I_{\rm radial}(r)$, where $r$ is the disk radius.
In addition, by setting only the PA of the disk, we determined the intensity profile along its major axis, $I_{\rm major}(r)$, using the CASA viewer.

Figures~\ref{fig:profile_vol1}-\ref{fig:profile_vol4} show the continuum images and intensity profiles of all detected disks, ordered by disk radius $R_{95\%}$. 
In each intensity profile, the red and violet curves represent the radial intensity profile $I_{\rm radial}(r)$ and the profile along the major-axis direction $I_{\rm major}(r)$, respectively. 
The negative regions of $I_{\rm radial}(r)$ are linearly symmetrical to the positive ones for comparison with $I_{\rm major}(r)$. We used these profiles to categorize the disk substructures in \S\ref{subsec:categorization}. 
Some disks can be classified as `Inflection' in $I_{\rm radial}$ and `Ring' with local maxima in $I_{\rm major}$. 
In such cases, the dust brightness distribution in the continuum image is used as a reference, and if a clear `Ring' is not identified, we conservatively categorize it as `Inflection'.

\begin{figure*}[t]
    \begin{center}
    \includegraphics[width=0.90\linewidth]{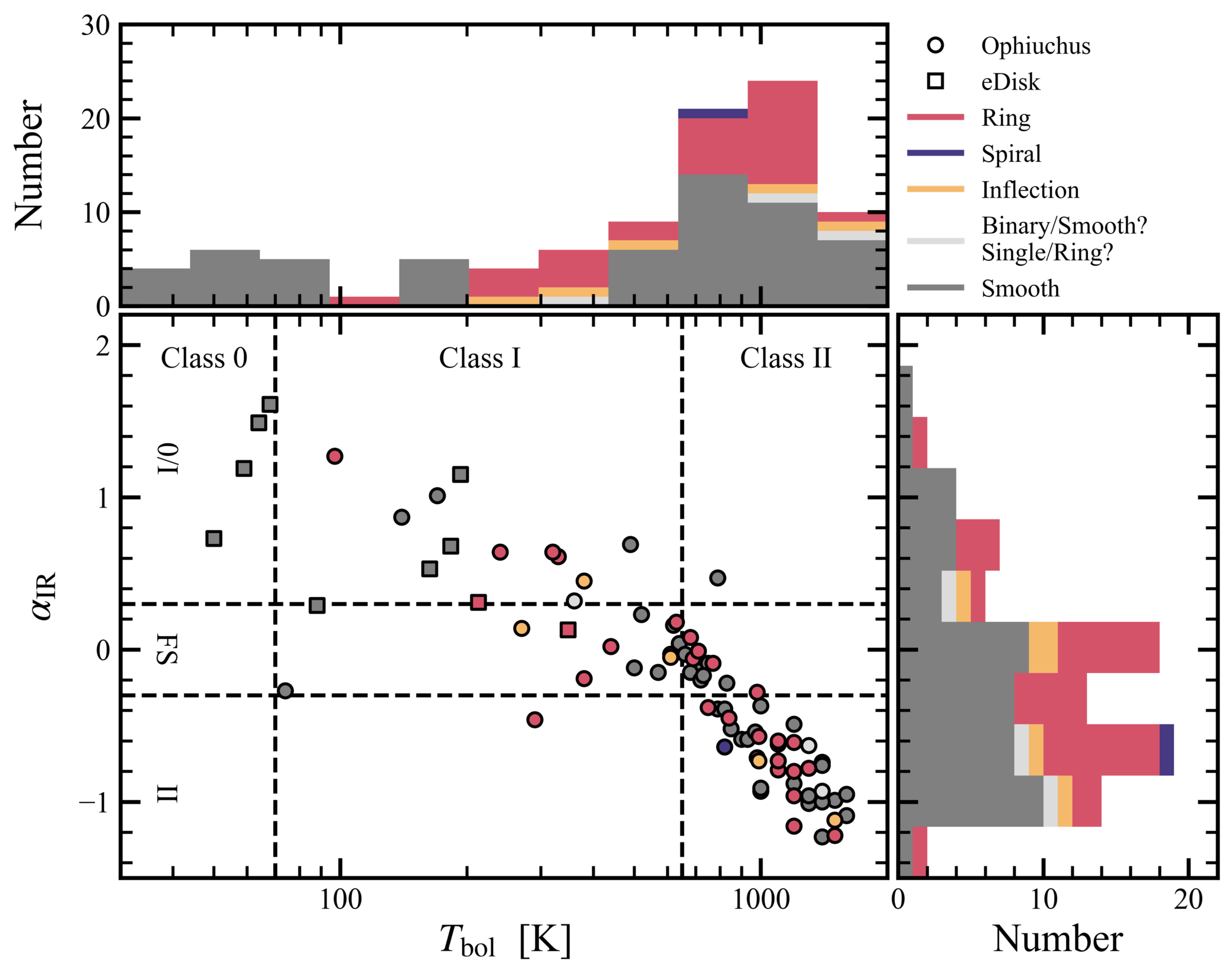}
    \end{center}
    \caption{
    Relationship between bolometric temperature $T_{\rm bol}$ and infrared spectral slope $\alpha_{\rm IR}$ of 76 Ophiuchus disks (circles) and 12 eDisk samples (squares).
    The red, violet, yellow, light gray, and dark gray colors represent disks categorized as `Ring', `Spiral’, `Inflection’ and the candidates for nearly edge-on disks with `Ring' features or circumstellar disks around binary systems, and `Smooth' brightness distributions, respectively.
    The top and right panels show the histograms of $T_{\rm bol}$ and $\alpha_{\rm IR}$ with eleven bins spanning their minimum and maximum values.
    }
    \label{fig:Tbol_IRslope}
\end{figure*}

\section{Comparison between Classification by Infrared Spectral Slope and by $T_{\rm bol}$}\label{sec:IRslope}
Figure~\ref{fig:Tbol_IRslope} compares the classification using the infrared spectral slope $\alpha_{\rm IR}$ and the bolometric temperature $T_{\rm bol}$ for our samples and the eDisk samples.
For the Ophiuchus samples and some of the eDisk samples (Oph\,IRS~43, Oph\,IRS~63, IRAS~16544-1604, R\,CrA\,IRS~7, R\,CrA\,IRS~5, and R\,CrA\,IRAS~32), we adopted $\alpha_{\rm IR}$ values from \citet{Dunham_2015}, the same as for $T_{\rm bol}$.
Additionally, we obtained $\alpha_{\rm IR}$ values for five eDisk systems (L1489~IRS, IRAS~04125+2702, TMC1A, IRAS~04302+2247, Ced\,IRS~4, and GSS~30\,IRS~3) from several literatures \citep{Connelley_2010,Luhman_2010,Manoj_2011,vanKempen_2009}.
Note that the values of $\alpha_{\rm IR}$ for the five eDisk systems are taken using the different methods for extinction corrections.
For the remaining sources (IRAS~04166+2706, L1527~IRS, BHR~71\,IRS2, BHR~71\,IRS1, IRAS~15398-3359, IRAS~16253-2499, and B335), their optically thick envelopes obscure the near-infrared radiation from the objects and make it impossible to obtain their $\alpha_{\rm IR}$ values (e.g., \cite{Andre_1994,Evans_2009}).
As reported by \citet{Ohashi_2023}, these objects lack near-infrared photometry, making it impossible to obtain their $\alpha_{\rm IR}$ values.

\bibliographystyle{forbibtex}
\bibliography{reference}

\end{document}